\documentclass[aps, 10pt, prd, longbibliography,
               notitlepage, superscriptaddress,
               nofootinbib, floatfix]{revtex4-2}
\usepackage{caption}
\usepackage{amsmath}
\usepackage{amssymb}
\usepackage{amsfonts}
\usepackage[utf8]{inputenc}
\usepackage[T1]{fontenc}
\usepackage{mathrsfs}
\usepackage{anyfontsize}
\usepackage{microtype}

\usepackage[linktocpage,breaklinks]{hyperref}
\usepackage[usenames,dvipsnames]{xcolor}

\usepackage{tensor}
\usepackage{bm}

\usepackage{graphicx}
\usepackage{epsfig}
\usepackage{epstopdf}

\usepackage{caption}
\usepackage{wrapfig}
\usepackage{xcolor}     % for colors

\usepackage{natbib}

\hypersetup{colorlinks=true,
            citecolor=NavyBlue,
            linkcolor=NavyBlue,
            urlcolor=NavyBlue}

\usepackage{float}% for subfigures
\usepackage{booktabs}             % for nice tables

\begin{document}
\title{New model of spontaneous scalarization of black holes induced by curvature and matter }

\begin{abstract}
We propose a new model of black hole spontaneous scalarization that combines a scalar-Gauss-Bonnet  interaction with a non-minimal coupling to a U(1) gauge field (a dark photon or an electromagnetic field). This construction generalizes earlier single-coupling setups and allows both curvature-induced and matter-induced scalarization within one framework, which allows us to overcome the limitations of each mechanism alone. We focus on charged, spherically symmetric black holes and demonstrate that our model substantially expands the range of black hole masses and charges that permit scalar hair. Negative Gauss-Bonnet couplings, previously associated only with near-extremal charges or rapidly spinning black holes, now trigger scalarization for much broader charge intervals. We develop a numerical procedure to solve the field equations, and investigate the various properties of these black holes. This results in new branches emerging at distinct mass thresholds, a behavior not seen in the pure Einstein-scalar-Gauss–Bonnet or Einstein-scalar-Gauss-Bonnet-Ricci models. The scalar charge depends sensitively on the coupling parameters and on the $U(1)$ charge. Our analysis also shows that these black holes have larger entropy than their Reissner–Nordström counterparts and can become overcharged, surpassing the usual extremal limit of charge-to-mass ratio. Analyzing the scalar charge behavior suggests that adding matter-coupling appears to stabilize solutions that were previously prone to higher-order instabilities in pure Gauss–Bonnet models with quadratic coupling and  broadens the range of possible configurations, making this model a promising candidate for further studies in strong gravity.

\end{abstract}

\author{Zakaria Belkhadria}
\email[Corresponding author: ]{Zakaria.belkhadria@unige.ch}
\email{Zakaria.belkhadria@gmail.com}
\affiliation{Département de Physique Théorique, Université de Genève, 24 quai Ernest Ansermet, CH-1211 Geneva 4, Switzerland}
\affiliation{Gravitational Wave Science Center (GWSC), Université de Genève, CH-1211 Geneva, Switzerland}
\affiliation{Dipartimento di Matematica, Università di Cagliari, via Ospedale 72, 09124 Cagliari, Italy}
\affiliation{INFN, Sezione di Cagliari, Cittadella Universitaria, 09042 Monserrato, Italy}

\author{Salvatore Mignemi}
\affiliation{Dipartimento di Matematica, Università di Cagliari, via Ospedale 72, 09124 Cagliari, Italy}
\affiliation{INFN, Sezione di Cagliari, Cittadella Universitaria, 09042 Monserrato, Italy}

\date{\today}

\maketitle

\setcounter{tocdepth}{2} % (optional) how many section levels to show
\begingroup              % (optional) local colour tweak for hyperlinks
  \hypersetup{linkcolor=black}
  \tableofcontents       % <<< TABLE OF CONTENTS
\endgroup
\vspace{1em}             

\section{Introduction}

Gravitational-wave observations have transformed black hole research and opened a new era of gravitational astronomy. Since 2015 the LIGO-Virgo-KAGRA collaboration has recorded  more than 200 binary black hole mergers and one binary neutron star merger, beginning with the landmark GW150914 \cite{abbott2016observation}. These signals reveal highly nonlinear spacetime dynamics and offer direct measurements of the strong-gravity regime. Complementary data from the Event Horizon Telescope enrich this picture, showing the shadow of the supermassive black hole in M87* \cite{ball2019first} and the central black hole of our galaxy Sagittarius A* \cite{akiyama2022first}. Future images will tighten constraints on horizon-scale geometry.
These data allow tests of general relativity (GR) and the Kerr hypothesis \cite{herdeiro2023black} by comparing predicted and observed waveform models and shadow profiles. Future gravitational-wave detectors such as LISA \cite{barausse2020prospects,arun2022new,auclair2023cosmology}, Einstein Telescope \cite{abac2025science,maggiore2020science} and Cosmic Explorer with improved sensitivity will enlarge the catalog of detections, and extend the mass and redshift ranges. They might probe signatures of alternative gravitational theories which were not observed before.

Black holes (BHs) in GR are characterized by only three quantities: mass, spin and electric charge. No-hair theorems reinforce this claim and exclude  extra scalar charges for some models when gravity couples to matter in the minimal way \cite{hawking1972black,bekenstein1972transcendence}. Their proofs extend to many models \cite{bekenstein1995novel}, so the metric is like GR and the scalar field is trivial. However, if one assumption of the no-hair theorems is violated, a black hole can carry scalar hair. Early attempts led to pathological, unstable solutions (e.g. the \textit{Bocharova-Bronnikov-Melnikov-Bekenstein (BBMB) solution} \cite{bocharova1970exact,bekenstein1974exact}), but later analyses found regular solutions when the coupling meets certain conditions. Identifying which couplings lead to stable, regular solutions that match observations remains essential.

GR has been extraordinarily successful in weak-field regimes, but its predictions in the strong-field regime remain largely untested. This experimental gap, now being addressed by gravitational-wave observations comes alongside theoretical puzzles that motivate extending GR. For example, phenomena like galaxy rotation curves (attributed to dark matter) and the accelerated cosmic expansion (dark energy) are not explained by GR with known fields. In addition, we cannot reconcile GR with quantum mechanics.  These issues have prompted modifications of Einstein's theory, such as adding higher-curvature terms to the action or introducing new degrees of freedom. Such extensions can emerge from fundamental physics, such as some attempts to unify gravity with other interactions naturally. A broad range of modified gravity models has been developed (see the review papers \cite{clifton2012modified,sotiriou2012black,heisenberg2019systematic,saridakis2021modified,charmousis2009higher}), and upcoming strong-field tests offer a way to distinguish these alternative theories from GR in regimes where deviations may become pronounced.

One of the principal frameworks for modifying gravity is the class of scalar-tensor theories that introduce a dynamical scalar field alongside the tensor field of GR. The prototypical example, Brans-Dicke theory, was formulated to incorporate Mach's principle by allowing the gravitational constant to vary \cite{brans1961mach}. Modern scalar-tensor models are constructed to reduce to GR in appropriate limits (ensuring consistency with solar-system experiments) while offering richer behavior in large-scale structure or in other extreme conditions. In particular, they can provide an effective explanation for cosmic acceleration without a cosmological constant, all while obeying current local gravity constraints. The most general form of these theories is Horndeski theories \cite{horndeski1974second,deffayet2011k,kobayashi2011generalized, deffayet2013formal,kobayashi2019horndeski}. They are designed so that their field equations remain second-order, avoiding ghost instabilities. No-hair theorems were developed for some classes of scalar-tensor theories. Hawking proved it for Brans-Dicke theories \cite{hawking1972black}, then \cite{sotiriou2012black} extended the proof to $f(R)$ theories. Hui and Nicolis established no-hair theorems for most interesting classes of Horndeski theories \cite{hui2013no}. Certain scalar-tensor theories allow the black hole to carry a nontrivial scalar field configuration. In other words they admit "hairy" black holes solutions that evade no-hair theorems. As proved in these theorems scalar hair is forbidden in many scalar-tensor models unless specific conditions are met, such as when the scalar is coupled to additional invariants like the Gauss-Bonnet (GB) term \cite{sotiriou2014black}. The possibility of scalar hair on black holes marks a significant departure from standard GR and motivates further empirical tests in the strong gravity regime.

Higher-curvature corrections involving the GB invariant (a particular combination of squared curvature terms) arise naturally in several fundamental contexts. Notably, a GB term appears in the low-energy effective action of heterotic string theory \cite{zwiebach1985curvature,nepomechie1985low,candelas1985vacuum,callan1986string,gross1987quartic} and in higher-dimensional generalizations of GR (Lovelock gravity) \cite{lovelock1970divergence,lovelock1971einstein}. In four dimensions the GB term by itself is a topological invariant and does not affect the field equations, but if it is coupled to a scalar field one obtains a nontrivial extension known as Einstein-scalar-Gauss-Bonnet (EsGB) gravity. EsGB models have second-order field equations and remain free of ghost instabilities, forming a subclass of Horndeski theories \cite{kobayashi2011generalized}. A well-studied example is Einstein-dilaton-Gauss-Bonnet (EdGB), which employs an exponential coupling $e^{-\alpha \phi}$ motivated by string theory .
Hairy black hole solutions in this model were first found perturbatively in \cite{mignemi1993charged} and then numerically in \cite{kanti1996dilatonic}. After that Ref \cite{sotiriou2014black} showed that shift symmetric coupling of the scalar field to GB invariant allow hairy solutions. Recently, Refs \cite{antoniou2018evasion,antoniou2018black} discovered a new violation of no-hair theorems in EsGB theories for a large class  of coupling functions including a quadratic coupling. EsGB theories provide a promising arena to explore strong-gravity phenomena and to confront theoretical predictions with observational data.

Another important category of alternative gravity involves a scalar field non-minimally coupled to the electromagnetic field. In such Einstein-Maxwell-Scalar (EMS) models, the scalar is coupled non-minimally to the Maxwell invariant through a coupling function $g(\phi)$. These models have very well motivated theoretical origins: Einstein-Maxwell-dilaton (EMD) theory for instance, emerges from Kaluza-Klein compactification and also appears as a low-energy limit of string theory, as well as in Supergravity frameworks. Another strong historical feature of such theories  is that electromagnetic fields can source the scalar, leading to black hole solutions different from the usual Reissner-Nordström (RN) solution of GR. A well-known example is the Gibbons-Maeda-Garfinkle-Horowitz-Strominger dilaton black hole in EMD gravity \cite{gibbons1988black,garfinkle1991charged}. This theory does not admit the standard RN  (bald) solution with scalar hair, any static charged BH must carry a "hair". The dilaton coupling thereby alters the black hole's charge-mass relation; BH can attain charge-to-mass ratios exceeding the extremal limit of RN black holes (they are overcharged). Models with scalar-Maxwell couplings, often inspired by high energy theory, thus provide a rich framework for how additional fields might influence electromagnetic interactions in strong gravity, offering distinctive black hole signatures that can be probed observationally.

One of the most recent and actively studied mechanisms that allow compact objects to develop scalar hair  only under some extreme conditions is \textit{spontaneous scalarization} (see \cite{doneva2024spontaneous} for a review).
Damour and Esposito-Farèse first uncovered the mechanism for neutron stars \cite{damour1993nonperturbative}.
In their model a massless scalar field is coupled to the trace of matter stress-energy tensor. When the stellar compactness reaches a critical value of the coupling, it drives a tachyonic instability. The scalar grows until nonlinear terms stop the process, leaving the star with a non-trivial configuration of scalar field "scalar hair". The scalar field is dormant in weak fields and still agrees with solar-system tests but emerges for dense configurations. Later it was shown \cite{doneva2018new,Silva2018SpontaneousCoupling} that spontaneous scalarization can apply to vacuum spacetimes (black holes) if the scalar couples to curvature or other matter invariants.

Two broad classes of models realize this phenomenon: one is the curvature-induced scalarization through Gauss-Bonnet coupling \cite{doneva2018new,Silva2018SpontaneousCoupling}, and the other is matter-induced scalarization through couplings to the field such as electromagnetism \cite{herdeiro2018spontaneous, myung2019instability}. In both cases, a nonminimal coupling of a scalar field  $\phi$ to an additional invariant (geometric invariant such as the GB term, the Ricci scalar or the Chern-Simons term \cite{doneva2021spontaneously, gao2019scalar,myung2021onset}, or matter invariant such as the electromagnetic tensor $F_{\mu\nu}F^{\mu\nu}$) triggers a tachyonic instability when the invariant is sufficiently large, causing the GR solution (with $\phi=0$) to become unstable, inducing a new ground state with $\phi\neq0$. This provides a natural way to circumvent no-hair theorems by introducing scalar hair only beyond a certain threshold of curvature or field strength.
In the curvature-induced case, represented by EsGB theories, the scalar field is coupled to the Gauss–Bonnet invariant ${\cal G}$ (a quadratic curvature term). A representative action is 

\begin{equation*}
S_{\rm EsGB} =\frac{1}{16\pi } \int d^4x\sqrt{-g}\Big[R - 2\nabla_\mu\phi\nabla^\mu\phi + f(\phi){\cal G}\Big]~
\end{equation*}
with $f(\phi)$ being a coupling function. The Schwarzschild black hole is a solution with $\phi=0$; however, if $f(\phi)$ is chosen such that $f''(0)\,{\cal G}>0$ induces a tachyonic mass for $\phi$, then above some coupling strength or below a certain mass scale the Schwarzschild solution becomes unstable. Indeed, it was shown that a static vacuum black hole can spontaneously scalarize when its curvature (characterized by $M^{-2}$ for a black hole of mass $M$) exceeds a critical value \cite{Silva2018SpontaneousCoupling,doneva2018new}. The end point is a “hairy” black hole with a nontrivial $\phi(r)$ profile. A key feature of these Gauss–Bonnet models is that they preserve the GR solutions for weak curvature (large $M$), activating scalar hair only in the high-curvature regime. This behavior is motivated by the idea that deviations from GR appear only in extreme gravity environments (as might be expected from quantum gravity or high-energy effective corrections), thereby satisfying observational constraints in the weak-field regime, while providing new physics for compact objects. This is why curvature-induced scalarization of black holes has attracted a lot of interest \cite{minamitsuji2019scalarized,silva2021dynamical,andreou2019spontaneous,doneva2019gauss,blazquez2020axial,blazquez2020polar,brihaye2019hairy,julie2022black,doneva2022dynamical,minamitsuji2019spontaneous,wong2022constraining,brihaye2020black, doneva2022beyond, antoniou2021compact, ventagli2020onset, cunha2019spontaneously,collodel2020spinning,kuan2021dynamical, danchev2022constraining}. Another attractive aspect is that the Gauss–Bonnet term is a natural curvature correction arising in effective descriptions (it appears in low-energy expansions of string theory), so EsGB models introduce scalar hair through a well-motivated extension of the gravitational action. 

\bigskip

Likewise, spontaneous scalarization can be induced by couplings to matter fields. An important example is the EMS class of models \cite{herdeiro2018spontaneous, myung2019instability, fernandes2019spontaneous,astefanesei2019einstein}, where the scalar field couples to the electromagnetic field invariant $F_{\mu\nu}F^{\mu\nu}$. In these models a charged Reissner–Nordström black hole (which in electrovacuum has $\phi=0$) can become unstable if the electric field is strong enough, leading to a new branch of scalarized charged black holes. For instance, with an action term of the form $f(\phi)F_{\mu\nu}F^{\mu\nu}$, the Maxwell field can trigger a tachyonic growth of $\phi$ when the charge $Q$ of the black hole exceeds a certain threshold. This matter-induced scalarization is analogous to the Gauss–Bonnet case: the scalar hair develops only when the electric field (or charge-to-mass ratio $Q/M$) is sufficiently large, ensuring that minimally coupled GR solutions remain stable for weakly charged black holes. Such EMS models have been extensively studied \cite{belkhadria2024mixed, konoplya2019analytical,gan2021photon,gan2021photon,myung2019quasinormal,myung2019stability,brihaye2020black,zou2019scalarized,blazquez2020einstein,blazquez2021quasinormal,guo2022quasinormal,fernandes2020einstein,zhang2022critical,zhang2021dynamical,khalil2019theory} as a way to generate hairy black holes that are smooth deformations of the Reissner–Nordström solution, again obeying the principle that hair appears only in regimes where the Maxwell field is above some threshold.

\bigskip

There have been some studies of charged black holes with Gauss-Bonnet-induced scalarization, without any direct scalar-Maxwell coupling \cite{doneva2018charged,brihaye2019spontaneous, herdeiro2021aspects}. These studies found that the inclusion of charge can enrich the scalarization behavior. For example, the Gauss–Bonnet term admits two types of scalarization (depending on the sign of the coupling) $GB^+$ or $GB^-$, and while a spherical symmetric neutral black hole only scalarizes in the positive-coupling case, adding charge allows scalarization with negative coupling ($GB^-$), but only near extremality. The electromagnetic field in those models influences the curvature-induced mechanism indirectly by modifying the black hole spacetime, but it does not act as an independent source of scalarization because $\phi$ does not couple to $F_{\mu\nu}F^{\mu\nu}$. Thus, the curvature-driven instability remains the primary route to hair in these models. 

\bigskip

A crucial challenge for the viability of EsGB scalarization is the stability of the scalarized solutions. It has become evident that many black holes with Gauss-Bonnet-induced hair suffer from perturbative instabilities \cite{blazquez2024instabilities}. In particular, for the simple choice of a quadratic coupling function $f(\phi)\propto \phi^2$, all the resulting scalarized black hole solutions are unstable under radial (spherically symmetric) perturbations \cite{blazquez2018radial,silva2019stability}. This means that although the hair can form, it does not represent a stable end state for evolution, and small perturbations will drive the solution away (often back to the GR state). More sophisticated coupling functions (for example, exponential \cite{doneva2018new,blazquez2018radial} or higher-order forms \cite{macedo2019self,silva2019stability}) can produce branches of solutions that are radially stable. However, even those “stable” branches have been shown to be vulnerable to non-radial (angular) instabilities. In fact, modes with higher angular momentum  can become unstable on hairy solutions that are stable to monopole perturbations. For instance, in theories of EsGB with Ricci coupled quadratically to $\phi$ referred to as (EsGBR) \cite{antoniou2021black, antoniou2022stable}, a quadrupole ($l=2$) instability was found by \cite{kleihaus2023quadrupole}, which can be extended to EsGB theories. Another instability was reported in EsGB and in EsGBR, called gradient instability, where the sound speed of angular propagation becomes imaginary for even perturbations in the case of large harmonic index $l$ \cite{minamitsuji2023instability}. The same authors later mentioned a mistake in their previous calculation, and found that the instability occurs for only part of the parameter space \cite{minamitsuji2024angular}. Many references reported a hyperbolicity loss in EsGB and EsGBR scalarization models and analyzed its threshold using numerical simulations \cite{east2021dynamics,doneva2023testing,doneva20243+,hegade2023and,thaalba2024spherical,franchini2022fixing}.

These instability issues imply that achieving a truly stable scalarized black hole in Gauss–Bonnet models is nontrivial. The existence of such instabilities motivates searching for other alternative models that can stabilize the solutions.

\bigskip

Motivated by these developments, in this work we propose a new model of black hole spontaneous scalarization induced by both curvature and matter, aiming to capture both curvature-induced and matter-induced scalarization within a single framework. The model introduces a simultaneous nonminimal coupling of the scalar field to the Gauss–Bonnet invariant and to $U(1)$ gauge field (a dark photon or an electromagnetic field), incorporating the two mechanisms within one framework. One should have in mind that a theory which contains a nonminimal coupling to both Gauss-Bonnet and Maxwell terms is very well motivated. Earlier works have consider it as a low energy limit of string actions \cite{gross1987quartic,zwiebach1985curvature,candelas1985vacuum,mignemi1993charged}, and black hole solutions were already investigated in this effective string theory \cite{mignemi1993charged, mignemi1995dyonic,torii1997dilatonic,alexeyev1997singular,torii1998stability}.

Our aim is to leverage the two sources of scalarization to try to overcome some of the difficulties faced by the pure EsGB case like instabilities. By including the matter-coupling directly, the scalar field has an additional channel to become tachyonic, which may widen the parameter space for stable scalarized solutions or provide new branches of hair that are absent in single-coupling models. Intuitively, the curvature coupling ensures that strong-gravity environments can trigger the scalar, while the matter coupling can also trigger scalarization and adds an extra contribution that  modify the properties of the resulting black holes. This unified model is a logical next step because it reduces to the known cases in the appropriate limits (recovering pure Gauss-Bonnet scalarization when charge is zero, and pure Maxwell-induced scalarization when the Gauss-Bonnet coupling is turned off). At the same time, it offers the prospect of richer phenomenology by allowing a broader scalarization threshold for different masses and, potentially, more robust hairy black hole solutions. 
 
  Another motivation, comes from the fact that this model can be seen as a subset of Maxwell-Horndeski theories, whose linear stability has already been studied \cite{kase2023black}. This will allow us to easily check  if the model suffers from the same instabilities reported previously.

 Finally, let us clarify the interpretation of the $U(1)$ gauge field. We treat this field and the associated charge in a broad sense without fixing its physical interpretation. 
The standard view is  to consider this $U(1)$ field as the usual electromagnetic field. If the black hole is electrically charged, astrophysical arguments imply that net electric charge stays negligible when the black hole interacts with ordinary plasmas and light charged particles. In that setting the charge-to-mass ratio is very small, or the theory serves as a toy model to study the impact of additional interactions with EsGB models of scalarization.  

Another possibility is a magnetically charged black hole \footnote{We consider only electrically charged BHs in this paper, but our  calculations can be extended to magnetic BHs and dyonic BHs}, motivated by hypothetical magnetic monopoles if magnetic monopoles exist, either primordial or formed through other early-Universe processes or motivated  by GUT-based arguments. In that case, black holes might preserve large magnetic charge over long timescales because there is no abundant light magnetic counterpart to neutralize it. This has attracted much interest recently \cite{bai2020phenomenology,ghosh2021astrophysical,diamond2022constraints,gervalle2024black,pereniguez2024superradiant,dyson2023magnetic}, especially after the paper of Maldacena \cite{maldacena2021comments}.

 Alternatively, one may view  the $U(1)$ field as living in a hidden sector decoupled from the standard model (dark photon) \cite{alexander2016dark,ackerman2009dark,nelson2011dark,fabbrichesi2021physics,curtin2015illuminating,abel2008kinetic,foot2015dissipative,an2013new,caputo2021dark,agrawal2020relic,chaudhuri2015radio,mirizzi2009microwave,an2013dark,pierce2018searching,caputo2020dark,feng2016detecting,villas2025bright,an2025situ,long2019dark}, with black holes carrying a dark charge that evades standard discharge arguments because the charge-to-mass ratio of dark particles can be much larger or smaller. We can identify the gauge field with a small fractional charge (minicharged dark matter) that might endow black holes with a net hidden charge \cite{cardoso2016black}. Various other references studied black holes involving dark photons \cite{cardoso2017superradiance,cardoso2018constraining,alexander2018hidden,caputo2021electromagnetic,gupta2021bounding,east2022dark,cannizzaro2022dark,east2022vortex,siemonsen2023dark,xin2025dark,ovgun2025black,bhattacharyya2023worldline}.
 
 Consequently, the model remains agnostic about how the black hole acquired its charge, whether it is electrical, magnetic, or hidden, and accommodates several theoretical and phenomenological contexts under a single framework.

From a theoretical point of view, RN black holes are interesting because they admit an extremal limit, so they can also serve as a toy model for understanding rotating solutions. Among four-dimensional Einstein-Maxwell solutions \cite{zilhao2012collisions}, RN black holes have a unique property: in the extremal limit, they admit a static, regular multi-black-hole configuration (the Majumdar-Papapetrou solution  \cite{majumdar1947class,papapetrou1948einstein} that remains regular on and outside the event horizon). In that scenario, gravitational attraction and electromagnetic repulsion cancel exactly at every point. This condition often occurs in supersymmetric contexts. The extremal RN solution is the only four-dimensional Einstein-Maxwell black hole with Killing spinors when viewed as the bosonic sector of $\mathcal{N}=2$ Supergravity \cite{gibbons1982bogomolny,tod1983all}.

 In the following sections, we detail the construction of this  model with two sources and explore its black hole solutions and their properties. Sec \ref{sec2} introduces the model and field equations, then analyzes the onset of scalarization, Sec  \ref{sec3} presents numerical solutions of the scalarized black holes their physical properties like the scalar charge and thermodynamic properties, and Sec \ref{sec4} discusses  conclusions and possible future directions.

In this study, we adopt the convention $4\pi G = 1 = 4\pi \epsilon_0$ for convenience. The spacetime signature is $(-,+,+,+)$. Our analysis is limited to spherically symmetric solutions, meaning that all metric and matter fields are functions of the radial coordinate only. To simplify our notation, any function initially introduced as dependent on the radial coordinate, like $X(r)$, will henceforth be referred to simply as $X$, keeping in mind its radial dependence. Derivatives are denoted as $X' \equiv \frac{dX}{dr}$ for radial derivatives and $X_{,\phi} \equiv \frac{dX}{d\phi}$ for derivatives with respect to the scalar field $\phi$.

\section{Setup : Theoretical framework} \label{sec2}

\subsection{Description of the model}
We consider a family of theories\footnote{These theories can be considered as a generalization of Einstein-Maxwell-scalar-Gauss-Bonnet theories (EMsGB), and a  subclass of Maxwell-Horndeski theories.} where a real scalar field couples non-minimally both to the Gauss-Bonnet invariant and to a $U(1)$ gauge field (e.g. an electromagnetic field). The action is  

 \begin{equation}
 \label{eqaction}
        S=\frac{1}{16\pi}\int d^4x \, \sqrt{-g} \Big[R-2\nabla _\mu \phi \nabla ^\mu \phi + f(\phi)\mathcal{G} -g(\phi) F_{\mu \nu} F^{\mu \nu}\Big]\ ,
    \end{equation}
with $R$ the Ricci scalar and $F_{\mu \nu}=\partial_\mu A_\nu-\partial_\nu A_\mu$  where $A_\mu$ is a $U(1)$ gauge field, $F_{\mu \nu}$ is its strength tensor (that can be identified as a Maxwell tensor), $\phi$ is a scalar field non-minimally coupled to gravity and the $U(1)$ gauge field  through the functions $f(\phi)$ and $g(\phi)$ respectively, and $\mathcal{G}$ is the Gauss-Bonnet invariant which is defined as follows:   \begin{equation}
        \mathcal{G} = R^2 -4R_{\mu \nu}R^{\mu \nu}+R_{\mu \nu \alpha \beta}R^{\mu \nu \alpha \beta}\ .
    \end{equation}

Variation of the action with respect to the metric and the different fields leads to the following equations: 

\bigskip

\begin{itemize}
    \item  Einstein equations:
\begin{equation}
\label{EOM-G}
G_{\mu \nu} = T_{\mu\nu}^{\rm (\phi)}+ T_{ \mu\nu}^{\rm (EM)} 
\end{equation}
with
\begin{eqnarray}
T_{\mu \nu }^{\rm (\phi)}= 2 ( \partial_\mu \phi \partial_\nu \phi -\frac{1}{2} g_{\mu\nu} \partial_\alpha \phi\partial^\alpha \phi \ - 2  P_{\mu\gamma \nu \alpha}\nabla^\alpha \nabla^\gamma f(\phi) )\ ,
\\
T_{\mu \nu }^{\rm (EM)}= 2 g(\phi) (F_{\mu \alpha}F_{\nu}^{\ \alpha}-\frac{1}{4} g_{\mu \nu }F_{\alpha\beta}F^{\alpha\beta}) \ ,
\end{eqnarray}
where
 \begin{eqnarray}
\nonumber
		P_{\alpha\beta\mu\nu}  = -\frac14 \varepsilon_{\alpha\beta\rho\sigma} R^{\rho\sigma\gamma\delta} \varepsilon_{\mu\nu\gamma\delta} 
		 &= R_{\alpha\beta\mu\nu}+ g_{\alpha\nu} R_{\beta\mu} - g_{\alpha\mu} R_{\beta\nu} + g_{\beta\mu} R_{\alpha\nu}-g_{\beta\nu} R_{\alpha\mu} \\
			&+\frac12 \left( g_{\alpha\mu}g_{\beta\nu} - g_{\alpha\nu}g_{\beta\mu}\right) R \ ,
 \end{eqnarray}   
and $ \varepsilon_{\alpha\beta\rho\sigma}$ is the Levi-Civita tensor.
     \item  Klein-Gordon (KG) equation: 
\begin{equation}
\label{KG-eq}
\Box \phi +\frac{1}{4} f_{, \phi}(\phi) \mathcal{G}-\frac{1}{4} g_{, \phi}(\phi)F_{\mu \nu} F^{\mu \nu}=0 \ 
\end{equation}
    \item  Maxwell's equations: 
\begin{equation}
g(\phi)\nabla_\nu F^{\mu \nu}+g_{, \phi}(\phi) F^{\mu \nu} \nabla_\nu \phi =0
\end{equation}
\end{itemize}

\subsection{Field equations and ansatz}

We focus on spherically symmetric models. A convenient metric ansatz is 
    \begin{equation}\label{ansatz}
	 ds^2 = - N(r) \sigma(r)^2 dt^2+\frac{dr^2}{N(r)}+r^2 \big(d\theta ^2 +\sin ^2 \theta\,  d \varphi ^2\big)\ ,~~{\rm with}~~N(r)\equiv 1-\frac{2\, m(r) }{r},
    \end{equation}
    where $m(r)$ is the Misner-Sharp mass function~\cite{misner1964relativistic} and $\sigma (r)$ is a second metric function. Spherical symmetry, in the absence of a magnetic charge\footnote{A magnetic charge would also be compatible with spherical symmetry, but it shall not be considered here -  see~\cite{astefanesei2019einstein} for dyonic charged BHs in the context of EMS theories.}, imposes an electrostatic 4-vector potential, $A=V(r)\,dt$, and a scalar field that depends only on $r$, $\phi=\phi (r) $.

Substituting this ansatz into the field equations leads to a set of three coupled differential equations for the radial functions $N$, $\sigma$, and $\phi$, with an additional constraint. This system of equations can be put in a form suitable for numerical integration as

\begin{equation}\label{eqform}
    \begin{aligned}
        & N'= F_{1}(N,\sigma,\phi,\phi',\phi''), \qquad &\sigma'=F_{2}(N,\sigma,\phi,\phi',\phi''), \qquad 
        & \phi '' = F_{3}(N,\sigma,\phi,\phi') \\
      &\qquad \text{and} \quad(\frac{V'g(\phi) r^2}{\sigma})'=0,
    \end{aligned}
\end{equation}

In more detail, these equations can be written as

\begin{equation}\label{ODE}
    \begin{aligned}
       N' &= \frac{-r^2 g(\phi) \left(N \left(\phi'^2 \left(r^2-4 (N-1) f_{, \phi \phi}(\phi)\right)-4 (N-1) \phi'' f_{, \phi}(\phi)+1\right)-1\right)-Q^2}{r^2 g(\phi) \left(2 (1-3 N) \phi' f_{, \phi}(\phi)+r\right)}, \\
        \sigma' &= \frac{\sigma \left(\phi'^2 \left(r^2-8 (N-1) f_{, \phi \phi}(\phi)\right)-8  (N-1) \phi'' f_{, \phi}(\phi)\right)}{2  (1-3 N) \phi' f_{, \phi}(\phi)+r}, \\
        V' &= -\frac{Q \sigma}{g(\phi) r^2}, \\
        \phi'' &= \frac{W}{Z},
    \end{aligned}
\end{equation}
The electrostatic potential $V$ gives rise to a first integral, simplifying the remaining equations. After integration of the last equation of \eqref{eqform}, the integration constant $Q$ will be interpreted as the electric charge.

The detailed expressions for $W$ and $Z$ are reported in Appendix~\ref{sec:appendixA}.

\subsection{Boundary conditions and physical quantities of interest}

To solve the set of four coupled ordinary differential equations Eq \eqref{ODE}, one must implement the appropriate boundary conditions. We impose regularity at the horizon, $r=r_H$, and the field equations can be approximated by a power series expansion in \( r - r_H \) as
    \begin{equation}\label{BC}
        \begin{aligned}
         & N = N_0 + N_1 (r-r_H) +\cdots\ \ , \qquad
        \sigma   = \sigma _0 + \sigma_1 (r-r_H)+\cdots\ ,	\\
        & \phi =  \phi _0 + \phi_1 (r-r_H)+\cdots\ ,\qquad  \ \ V = V_1 (r-r_H)+\cdots\ ,
        \end{aligned}
    \end{equation}
where the coefficients are obtained by using the expansion into the field equations:
\begin{equation}
    \begin{aligned}
        N_0 &= 0, &  N_1 &= \frac{r_H^2 g(\phi_0) - Q^2}{r_H^2 g(\phi_0) \left(2  \phi_1 f_{, \phi}(\phi_0) + r_H\right)}, \\
        \sigma_1 &= \frac{\sigma_0 \phi_1^2 \left(2  f_{, \phi \phi}(\phi_0) + r_H^2\right)}{2  \phi_1 f_{, \phi}(\phi_0) + r_H}, &
        \phi_1 &= \frac{B \pm \sqrt{\Delta}}{A},
    \end{aligned}
\end{equation}

with
\begin{align*}
    B &= Q^2 r_H^2 \left(g(\phi_0) \left(4 f_{, \phi}(\phi_0)^2 + r_H^4\right) - 2  r_H^2 f_{, \phi}(\phi_0) g_{, \phi}(\phi_0)\right) + 4  Q^4 g_{, \phi}(\phi_0)^2 - r_H^8 g(\phi_0)^2, \\
    \Delta &= \left(Q^2 - r_H^2 g(\phi_0)\right)^2 \Big(8 Q^2 f_{, \phi}(\phi_0)^3 \left(2  Q^2 f_{, \phi}(\phi_0) - 3 r_H^4 g_{, \phi}(\phi_0)\right) \\
    &\quad + 16  Q^2 r_H^2 g(\phi_0) f_{, \phi}(\phi_0)^2 \left(6 f_{, \phi}(\phi_0)^2 + r_H^4\right) + r_H^8 g(\phi_0)^2 \left(r_H^4 - 24  f_{, \phi}(\phi_0)^2\right) \Big), \\
    A  &= 4  r_H f_{, \phi}(\phi_0) \left( Q^2 r_H^2 f_{, \phi}(\phi_0) g_{, \phi}(\phi_0) - Q^2 g(\phi_0) \left(4  f_{, \phi}(\phi_0)^2 + r_H^4\right) + r_H^6 g(\phi_0)^2\right).
\end{align*}
The coefficients depend on the two  parameters $\phi_0$ and $\sigma_0$, where the subscript $0$ denotes functions evaluated at the horizon $r_H$.

The equation determining $\phi_1$ is obtained from the scalar field equation near the horizon and has two solutions, but we choose the only one that gives a black hole solution  $\phi_1 = \frac{B+\sqrt{\Delta}}{A}$, while the other solution $\phi_1 = \frac{B-\sqrt{\Delta}}{A}$ does not lead to an asymptotically flat BH \cite{kanti1996dilatonic}. To obtain a physical solution we need to ensure regularity by imposing that $\Delta \geq 0$ which keeps $\phi_1$ real and $\phi''_0$ finite \cite{kanti1996dilatonic}.

   \bigskip
   
   At spatial infinity we expand in power series of $1/r$, so the leading-order expansion for far fields is
   
\begin{equation}\label{BCinf}
        \begin{aligned}
         & N \approx 1- \frac{2M}{r}  , \qquad
        \sigma   \approx 1 - \frac{D^2}{2r^2},	\qquad
        & \phi \approx \frac{D}{r},\qquad  \ \ V \approx \psi_{e}+\frac{Q}{r}
        \end{aligned}
    \end{equation}

where $M$ is the Arnowitt-Deser-Misner (ADM) mass,  $D$ is the scalar charge and $\psi_{e}$ is the electrostatic potential at infinity.
\bigskip

We can check the regularity of the solutions and the presence of singularities in some cases by checking the divergence of the Ricci scalar $R$, and the Kretschmann scalar $K\equiv R_{\mu \nu \delta \lambda} R^{\mu \nu \delta \lambda}$,

 \begin{equation}
    R = -\frac{\sigma (r) \left(-2 + 2 N(r) + 4 r N'(r) + r^2 N''(r) \right) + r \left(3 r N'(r) \sigma '(r) + 2 N(r) \left(2 \sigma '(r) + r \sigma ''(r)\right)\right)}{r^2 \sigma (r)}.
\end{equation}

\begin{equation}
\begin{aligned}
K &= \frac{4\,(N(r)-1)^2\,\sigma(r)^2 + 2 r^2 \sigma(r)^2 N'(r)^2
          + 2 r^2\!\bigl(\sigma(r) N'(r) + 2 N(r) \sigma'(r)\bigr)^2}
        {r^4 \sigma(r)^2} \\
  &+ \frac{r^4\!\bigl(3 N'(r) \sigma'(r) + \sigma(r) N''(r)
          + 2 N(r) \sigma''(r)\bigr)^2}
        {r^4 \sigma(r)^2}.
\end{aligned}
\end{equation}

\subsection{Scaling symmetry}

The field equations of the model exhibit a scaling symmetry \cite{herdeiro2021aspects, Silva2018SpontaneousCoupling, brihaye2019hairy}, so that the solutions are invariant under the rescaling transformations 
\begin{equation}
r \to \beta r\,, \quad M \to \beta M\,, \quad \eta \to \beta^2 \eta\,, \quad Q \;\to\; \beta\, Q,
\end{equation}
where $\beta>0$ is an arbitrary scaling parameter, and $\eta$ is the  coupling constant of the GB term, defined so that $f=\eta\, \tilde f$, with dimensionless $\tilde f$, so that $\eta$ has the dimension of length square.

This symmetry and freedom in choosing the length scale  implies that physical quantities of the black hole solution should be expressed in terms of dimensionless combinations that remain invariant under these transformations.
To characterize the black hole solutions we introduce the reduced quantities 
\begin{equation}
q=\frac{Q}{M}\ ,\quad \frac{\eta}{M^2},\quad \frac{D}{\sqrt{\eta}}, \quad \frac{Q}{\sqrt{\eta}}, \quad \frac{M}{\sqrt{\eta}},\quad \frac{\eta}{r_H^2}
\end{equation}

 For numerical analysis, we exploit this symmetry to fix one parameter, typically either the horizon radius, $r_H = 1$, or the mass, $M=1$, or the coupling parameter $\eta$ when appropriate, reducing the solution space to a family characterized by dimensionless quantities like $\eta/M^2$, without loss of generality.

All dimensional quantities presented in the results are normalized accordingly, either by appropriate powers of $M$ or $\sqrt{\eta}$ or $r_H$.

\subsection{Conditions for scalarization: Scalarization threshold and coupling functions}

The scalar field $\phi$ obeys the Klein-Gordon equation 
      \begin{equation}\label{KG}
        \Box \phi = \frac{-\mathcal{I}_{GB}}{4} f_{,\phi} + \frac{\mathcal{I}_M}{4} g_{,\phi},
      \end{equation}
where we have renamed
\begin{eqnarray}
\mathcal{I}_{M}=F_{\mu\nu}F^{\mu\nu}\ , \quad
\mathcal{I}_{GB}= \mathcal{G}.
\end{eqnarray}

Two conditions are required for spontaneous scalarization to happen, as follows: 
\begin{itemize}
    \item When the theory admits solutions with \( \phi = 0 \). This demands 

\begin{equation}
\frac{df}{d\phi}\Big |_{\phi=0} = 0, \quad \frac{dg}{d\phi}\Big |_{\phi=0} = 0.
\end{equation}

These conditions are generally met in models with \(\mathbb{Z}_2\) symmetry (under \(\phi \rightarrow -\phi\)) and allow the existence of standard electro-vacuum GR solutions. 
\item If the second derivatives of \( f(\phi) \) and \( g(\phi) \) near $\phi=0$ create a tachyonic instability.
 Small-\(\phi\) expansions for both coupling functions show

\begin{equation}\label{expanfg}
f(\phi) = f(0) + \frac{1}{2}\frac{d^2 f}{d \phi^2}\Big |_{\phi=0}\phi^2 + \cdots, \quad g(\phi) = g(0) + \frac{1}{2}\frac{d^2 g}{d \phi^2}\Big |_{\phi=0}\phi^2 + \cdots,
\end{equation}

The linearized Klein-Gordon equation then becomes 
\begin{equation} \label{scalarpert}
(\Box - \mu_{\rm eff}^2)\delta \phi = 0, \quad \text{where} \quad \mu_{\rm eff}^2 = \mu_{\rm eff,f}^2 + \mu_{\rm eff,g}^2
\end{equation}
with
\begin{equation}
\mu_{\rm eff,f}^2 = \frac{-1}{4}\frac{d^2 f}{d \phi^2}\Big |_{\phi=0} \mathcal{I}_{GB}, \quad \mu_{\rm eff,g}^2 = \frac{1}{4}\frac{d^2 g}{d \phi^2}\Big |_{\phi=0} \mathcal{I}_M
\end{equation}
and $\delta \phi $ is the scalar field perturbation, while $\mu_{\rm eff,f}^2$ and $\mu_{\rm eff,g}^2$ are the effective mass squares.

A tachyonic instability appears when  $\mu_{\rm eff}^2$ is negative, and 

\begin{equation}
\frac{d^2 f(\phi)}{d \phi^2}\Bigg |_{\phi=0} \neq 0   \quad \text{or} \quad \frac{d^2 g(\phi)}{d \phi^2}\Bigg |_{\phi=0} \neq 0.
\end{equation}

\end{itemize}

 When $\mu_{\rm eff}^2<0$ a new branch of scalarized solutions may bifurcate from the original scalar-free solutions (GR black holes). These scalarized solutions form a family of solutions  which are continuously connected to the scalar-free solutions and characterized by additional parameters (e.g., the scalar field's value at the horizon).
The quantities $\mathcal{I}_{GB}$ and $\mathcal{I}_{M}$ influence both the onset of stability and the resulting scalarized solutions. 

The choice of the coupling functions  $f(\phi)$ and $g(\phi)$ plays an important role. Various forms have been proposed in the literature, which satisfy the condition that $f(0)=0$, $g(0)=1$, so that for the vanishing scalar field one recovers the Einstein-Maxwell action, whose unique spherically symmetric solution is the Reissner-Nordstr\"om (RN) metric. 
\bigskip

\begin{minipage}[t]{0.45\linewidth}
\textbf{Functions for $f(\phi)$ \cite{Silva2018SpontaneousCoupling,doneva2018charged}:}
\begin{enumerate}
  \item $\frac{\eta \phi^{2}}{2}$
  \item $\frac{\eta}{2\beta}(1 - e^{-\beta\,\phi^{2}})$
     \item $\frac{\eta}{\beta^{2}}(1 - \frac{1}{\cosh(\beta\,\phi)})$
  \item $\frac{\eta \phi^{2}}{2\,(1+\beta^{2}\,\phi^{2})}$

\end{enumerate}
\end{minipage}%
\hfill
\begin{minipage}[t]{0.45\linewidth}
\textbf{Functions for $g(\phi)$ \cite{herdeiro2018spontaneous,fernandes2019spontaneous,konoplya2019analytical}:}
\begin{enumerate}
  \item $1-\alpha\,\phi^{2}$
  \item $e^{-\alpha\,\phi^{2}}$
  \item $\cosh(\sqrt{2|\alpha|}\,\phi)$
  \item $\frac{1}{1+\alpha \phi ^2 }$
\end{enumerate}
\end{minipage}
\bigskip

\noindent For small $\phi$ they all have the same behavior, so for simplicity
here we adopt the simplest quadratic form : 
\begin{eqnarray}
    f(\phi) = \frac{\eta}{2}\phi^2, \qquad   g(\phi) = 1-\alpha \phi^2
\end{eqnarray}
where  $\alpha$ is a dimensionless coupling constant. \footnote{$\alpha$ is often negative in the pure EMS scalarization, as in the coupling functions $1-\alpha\,\phi^{2}$ or $e^{-\alpha\,\phi^{2}}$. The required overall positive sign ensures a positive second derivative, so the negative sign in the effective mass arises from the chosen invariants. Typically  $\eta>0$ in the pure EsGB scalarization, but Refs \cite{brihaye2019spontaneous, herdeiro2021aspects} show that scalarization can happen with a negative $\eta$ in a theory of EMsGB (without coupling between $\phi$ and Maxwell invariant) in the case of near extremality, when the Gauss-Bonnet invariant changes sign from positive to negative. In the present setup, the situation is very different, since having both couplings allows wider sign choices, with the key element being the sign of the total effective mass squared}

If we take the ansatz \eqref{ansatz} and put 

\begin{eqnarray}
N(r)=1-\frac{2M}{r}+\frac{Q^2}{r^2}\ \ , \qquad  \sigma(r)=1   , \qquad  V(r)=\frac{Q}{r} \ .
\label{RNn2} 
\end{eqnarray}
we get the metric of the RN black hole, which has an event horizon located at
\begin{equation}   
r_H = M + \sqrt{M^2 - Q^2}.
\end{equation}

For this RN background, the matter and the Gauss-Bonnet invariants are \begin{eqnarray}
\mathcal{I}_{M}=-\frac{2 Q^2}{r^4}<0 \ , \quad \text{and} \\
\mathcal{I}_{GB}= \frac{8}{r^8}\left[6M^2r^2-12Q^2Mr+5 Q^4\right] \ 
\end{eqnarray}

The total effective mass becomes 

\begin{eqnarray}
\mu_{\rm eff}^2=\frac{- 2\eta}{ r^8}\left(6M^2r^2-12Q^2Mr+5 Q^4\right) \ +\frac{\alpha Q^2}{r^4}
\end{eqnarray}

 We will analyze the stability of the RN solution within this theory (we neglect back-reaction). This approach provides scalarization thresholds and identifies the onset of tachyonic instability associated with spontaneous scalarization.

Let us write the linearized KG equation (scalar field perturbation equation) on a fixed RN background, then decompose the scalar field perturbations into real spherical harmonics
\begin{equation}
\label{harmo}
\delta \phi=U(r) Y_{\ell m}(\theta,\varphi) \ 
\end{equation}
This allows the scalar field perturbation equation (linearized KG equation) to be only radially dependent and written in the form 
\begin{eqnarray}
\label{KgeRN}
\left[r^2 N(r)U'(r)\right]'=U(r)\left(\ell(\ell+1)- \frac{\eta}{r^2} 
(
\frac{12 M^2}{r^2}+\frac{10Q^4}{r^4}-\frac{24 MQ^2}{r^3}
)+\alpha \frac{Q^2}{r^2}\right).
\end{eqnarray}

In the analysis of the KG equation in the theory (\ref{eqaction}), the spherical symmetry make it easy to separate  the wave equation (characterized by $\ell = 0$).

For $\ell=0$ this equation can be written as\begin{equation}
(Q^2 + r (-2 M + r)) (U^{\prime\prime})(r) = \frac{\left(-10 Q^4 \eta - 12 M^2 r^2 \eta + Q^2 (r^4 \alpha + 24 M r \eta)\right) U(r)}{r^6} + 2 (M - r) U^\prime(r).
\end{equation}
To facilitate our investigation, we introduce dimensionless variables 
\begin{equation}
    q \equiv \frac{Q}{M}, \quad  u \equiv \frac{U}{M}, \quad   \rho \equiv \frac{r}{M},\quad  
    \gamma \equiv \frac{\eta}{M^2}
\end{equation}
leading to the dimensionless form of the equation

\begin{equation}\label{dimensionless lin KG}
\rho^6 ((q^2 + (-2 + \rho) \rho) u''(\rho) + 2 (-1 + \rho) u'(\rho)) = (q^2 \rho^4 \alpha - 2 (5 q^4 - 12 q^2 \rho + 6 \rho^2) \gamma) u(\rho).
\end{equation}

The differential equation Eq \eqref{dimensionless lin KG} can be solved numerically. We start our numerical integration near the event horizon (at $\rho = \rho_H + 10^{-5}$), with $\rho_H=1+\sqrt{1-q^2}$ \footnote{We will integrate the equation for the range of values $0\leqslant q\leqslant 1$, where $q=1$ is the extremal RN black hole.} using a power series solution for $u$ valid in that region. The leading-order behavior near the horizon is given by:

\begin{equation}
u = u_H + \frac{(q^2\rho_H^4 \alpha+\gamma(-10 q^4 - 12 \rho_H^2 +q^2 24 \rho_H )) u_H}{2(\rho_H-1)\rho_H^6} (\rho -\rho_H)
\end{equation}

We extend the integration to a large value of $\rho$, typically around $10^{5}$, for specific values of the coupling constants $(\gamma,\, \alpha)$ and an arbitrary horizon value $u_H$. For every set of these parameters, Eq \eqref{dimensionless lin KG} yields a distinct solution. Our focus is on solutions $u$ that vanish at spatial infinity.

Our exploration covers a parameter space defined by the reduced charge $q$ and the coupling parameters $(\gamma,\, \alpha)$. This results in a three-dimensional domain characterizing the model in question. To simplify our analysis, we fix one of the coupling constants, for instance $\alpha$, and vary the other, $\gamma$.

 We varied these parameters systematically to map out the region in the parameter space where the GR is unstable in order to identify where scalarization can happen. 
  This region is characterized by the onset of tachyonic instability, indicated by the contribution of effective mass squared, $\mu_{\rm eff}^2$, as outlined in Eq. \eqref{KgeRN}.

\begin{figure}[ht]
    \centering
    \includegraphics[width=0.45\textwidth]{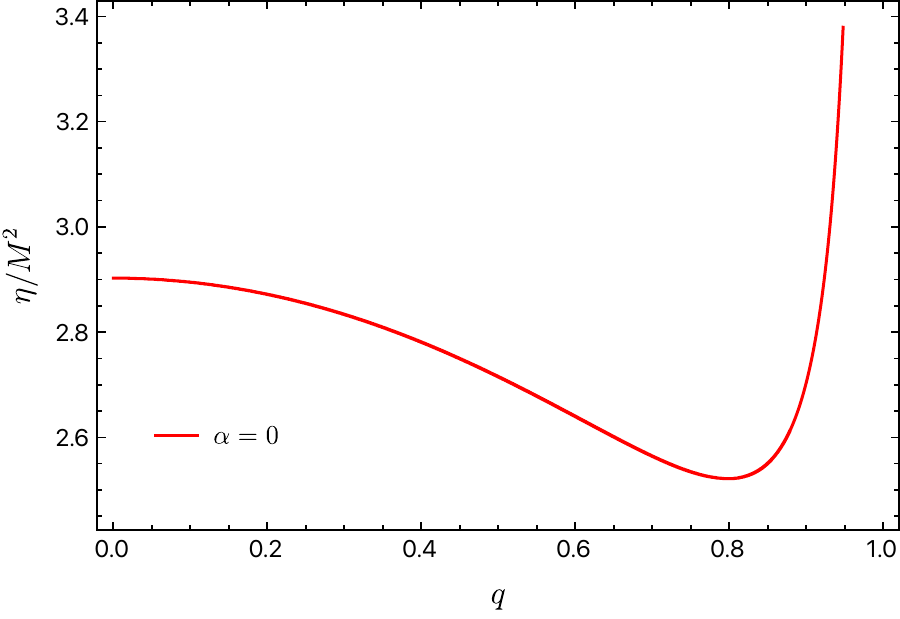}
    \includegraphics[width=0.45\textwidth]{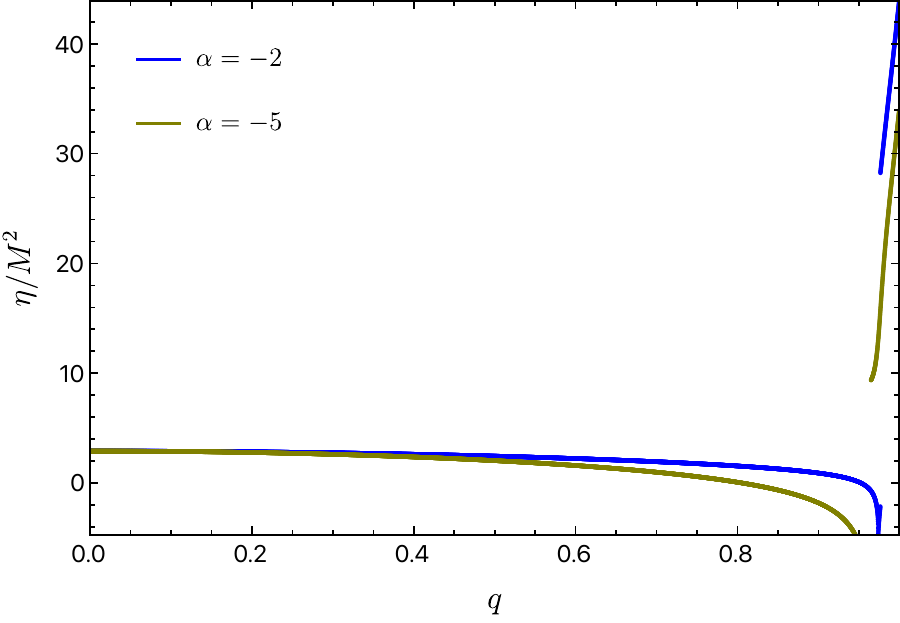}
    \includegraphics[width=0.45\textwidth]{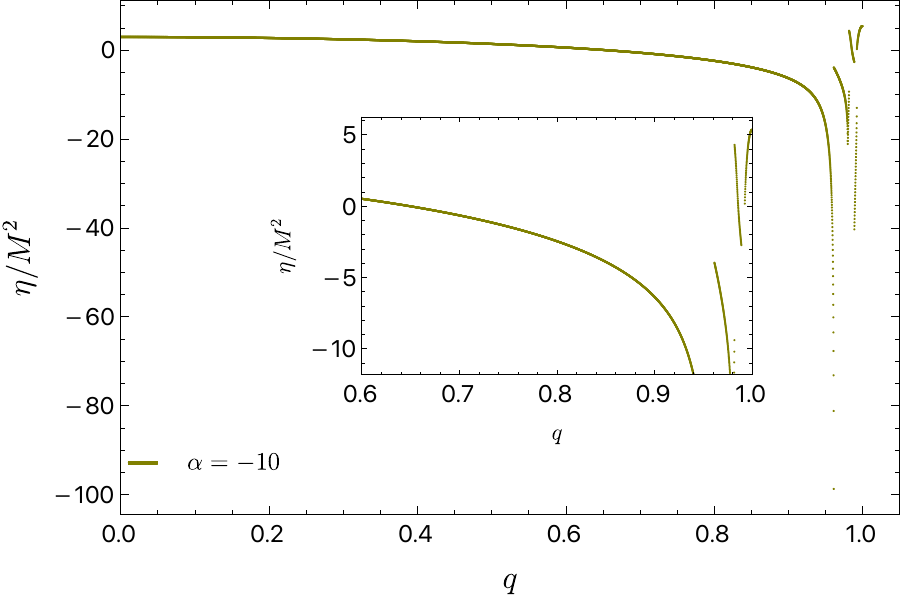}
    \includegraphics[width=0.45\textwidth]{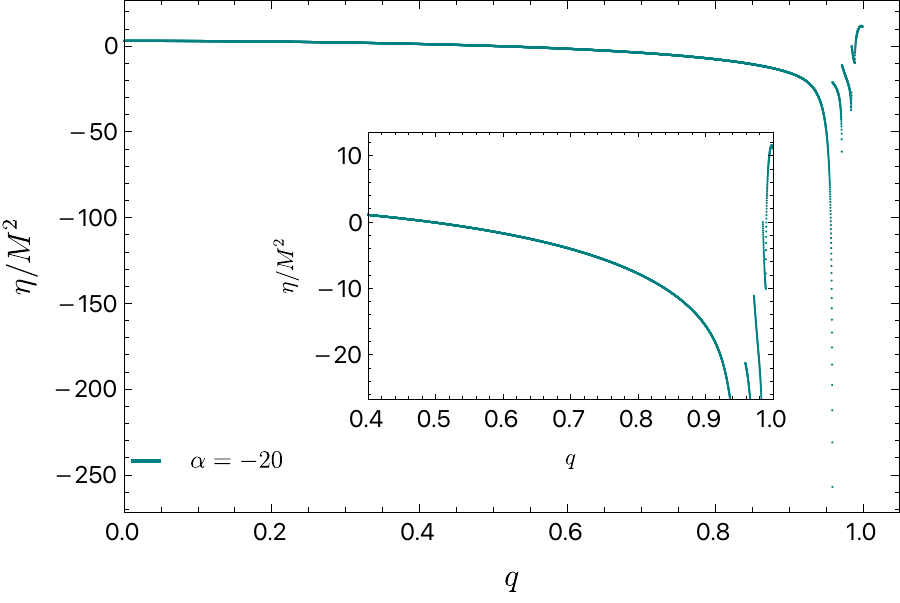}
    \includegraphics[width=0.45\textwidth]{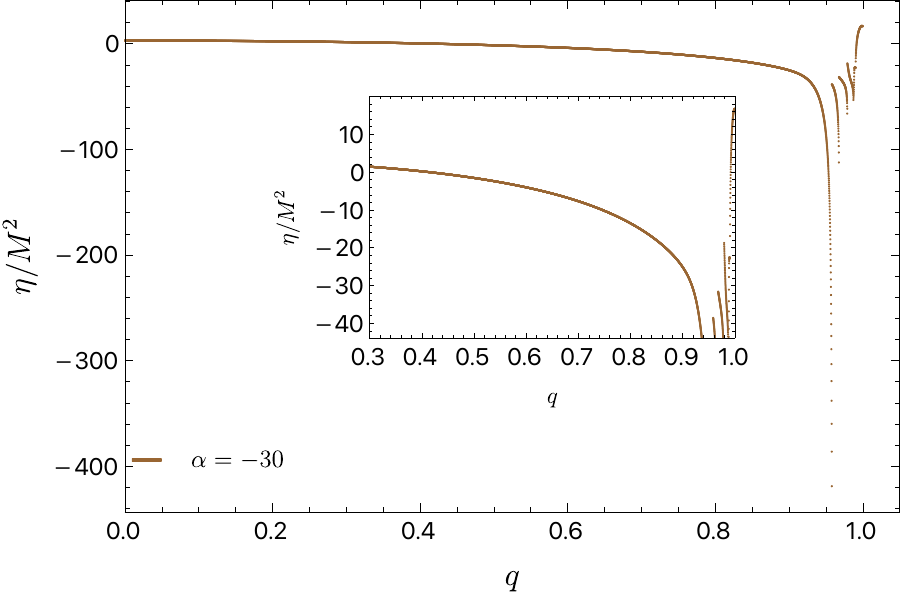}
    \includegraphics[width=0.45\textwidth]{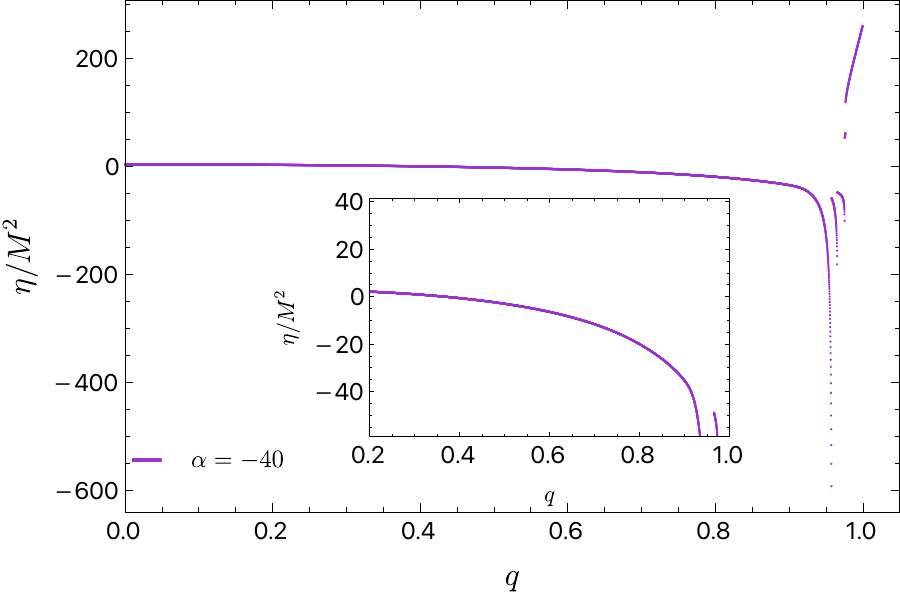}
    
    \caption{This figure illustrates the behavior of the lower limit of the dimensionless Gauss-Bonnet coupling constant $\gamma$, which is necessary to guarantee the instability of the Reissner-Nordström black hole. It showcases the dimensionless ratio $\gamma=\eta/M^2$ for a set of Reissner-Nordström solutions that support the $\ell=0$ scalar cloud, plotted against the reduced electric charge $q=Q/M$.The inset is a magnification of the main plot near the region where $\eta$ becomes negative}
    \label{plotbif}
\end{figure}

In Fig.~\ref{plotbif} we plot, for six fixed values of the matter coupling $\alpha$, the minimum dimensionless Gauss-Bonnet strength $\gamma=\eta/M^{2}$ that triggers the instability of a Reissner–Nordström black hole as a function of the charge ratio $q=Q/M$. The six panels are arranged by $\alpha$ from the top left $(\alpha=0)$ to the bottom right $(\alpha=-40)$. Each curve, therefore, marks the bifurcation line where a non‑trivial scalar configuration first appears. The bullet list that follows comments on the behavior in every panel:

\begin{itemize}
  \item \textbf{Top‑left, $\alpha=0$.}  The curve reproduces the results already discussed in \cite{doneva2018charged,herdeiro2021aspects}.
  
  \item \textbf{Top‑right, $\alpha=-2$ and $\alpha=-5$.}  The blue $(\alpha=-2)$ and orange $(\alpha=-5)$ curves bend below the $\gamma=0$ axis at high charge.  The Maxwell coupling, therefore, allows scalarization with $\eta<0$ once $q$ is large.
  
  \item \textbf{Middle‑left, $\alpha=-10$.}  The fundamental branch crosses $\gamma=0$ at $q\approx0.6$, extending the $GB^{-}$ window that earlier work found only near $q\approx0.96$ \cite{brihaye2019spontaneous,herdeiro2021aspects}. Other branches appear, and numerical checks confirm that the curve that appears just above the first fundamental branch is the first excited scalar mode with one radial node $(n=1)$.
  
  \item \textbf{Middle‑right, $\alpha=-20$.}  The sign flip now occurs near $q\approx0.4$.  Three excited branches $(n=1,\;2, \;3)$ are visible.  Larger $|\alpha|$, therefore, lowers the charge threshold and creates more bifurcation lines.
  
  \item \textbf{Bottom‑left, $\alpha=-30$.}  The fundamental branch turns negative already around $q\approx0.3$.  Three excited branches track higher‑node clouds.
  
  \item \textbf{Bottom‑right, $\alpha=-40$.}  Almost the entire plot lies below $\gamma=0$.  Scalarization with negative $\eta$ is now possible for moderate charges $(q\gtrsim0.2)$ and several higher‑node branches follow the same trend.
\end{itemize}

Overall, increasing $|\alpha|$ widens the parameter space in which the instability occurs and introduces additional node branches. It permits $\eta<0$ solutions with less restrictive conditions. The critical value $q^{(-)}$ at which $\eta$ flips sign decreases monotonically with $|\alpha|$, and larger couplings push the $\eta<0$ scalarization window down to $q \approx 0.2$, making negative-$\eta$ scalarization accessible at a much broader class of RN BH.

Earlier studies located this so-called $GB^-$ branch in narrow regimes. In case of spherical symmetry studied in \cite{brihaye2019spontaneous, herdeiro2021aspects} RN clouds with $\eta<0$ were found only near extremality $q \approx 0.957058$. For rotating black holes, Refs \cite{dima2020spin,herdeiro2021spin,berti2021spin}  reported the so-called \textit{spin-induced scalarization} with $\eta<0$ scalarization when the spin is high ($J/M^2 \gtrsim 0.5$. In both cases the Gauss-Bonnet invariant turns negative.

\section{Numerical scalarized black hole solutions and their properties}\label{sec3}

\subsection{Numerical procedure}

We solve the system of ordinary differential equations in \eqref{ODE} with the boundary conditions in \eqref{BC} and \eqref{BCinf}. These equations describe a boundary value problem with conditions at the event horizon and at large distance. Standard scaling arguments allow us to set the horizon radius to $r_H = 1$.

We work with three parameters: $\eta$, $\alpha$, and $Q$. We fix $\alpha$ and load from the bifurcation data an initial list of values of ($q,\eta/M^2$) that satisfy a certain ratio $Q/\sqrt{\eta}$, which defines the onset of the instability of the RN background. Starting from the bifurcation (GR) value, we vary $\eta$ in small increments or decrements while keeping $Q/\sqrt{\eta}$ constant. For each step, we solve for the scalar field value at the horizon so that the field vanishes at large distance. This strategy leads to families of asymptotically flat black holes with scalar hair.

We adopt a shooting method combined with a stiff solver. We start the integration near $r=r_H + 10^{-4}$, using the near-horizon expansions in \eqref{BC}. We move outward to $r \sim 10^{3}$ and compare the result to the far-field expansions in \eqref{BCinf}. We adjust the shooting parameters, which include $\phi_0$ and $\sigma_0$, until the scalar field approaches zero at large $r$ and the metric converges to a RN form. We then rescale $\sigma$ at large $r$ to fix the normalization. In practice, we set $\sigma_0=1$ at the start, then adjust it to $\sigma_0 = 1/\sigma_\infty$.

When the step in $\eta$ fails to produce a well-behaved solution, we revert to the previous successful solution and reduce the step size. We require $\Delta>0$ and the solutions for mass and charge to remain real. The local integration error is set close to $10^{-10}$. 

This procedure generates black hole solutions labeled by $\alpha$, $\eta$, and $Q/\sqrt{\eta}$. They arise from the Reissner--Nordstr\"om configuration through a tachyonic instability.

\subsection{Scalarized black hole solutions}
In the following, we focus only on the case where $\eta$ is positive and $\alpha$ is negative (which is the usual choice of sign for scalarization both in EsGB and EMS).
 One of the most important quantities to study for scalarized black holes is the scalar charge, which gives a clear idea of how scalarized black holes bifurcate from the GR ones and how they evolve.

\begin{figure}[ht]
    \centering
    \includegraphics[width=0.45\textwidth]{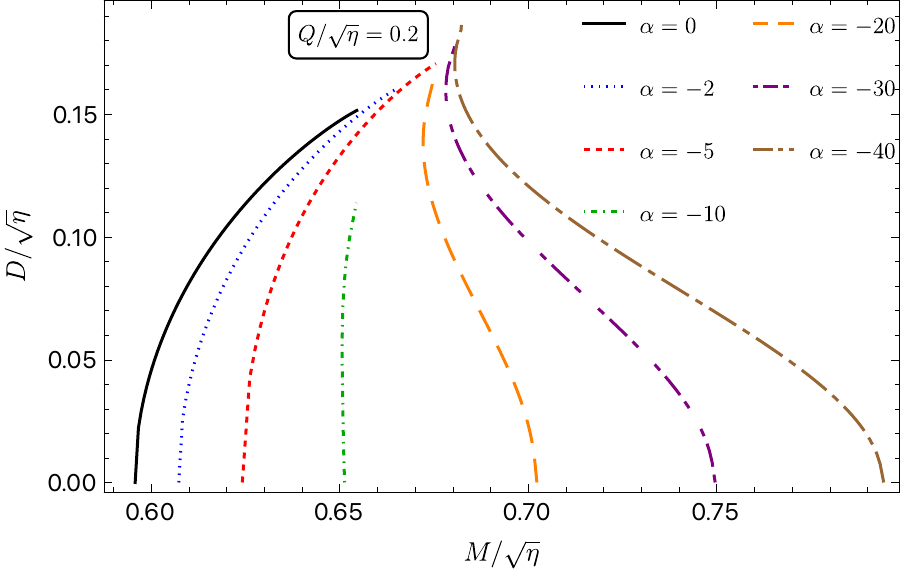}
    \includegraphics[width=0.45\textwidth]{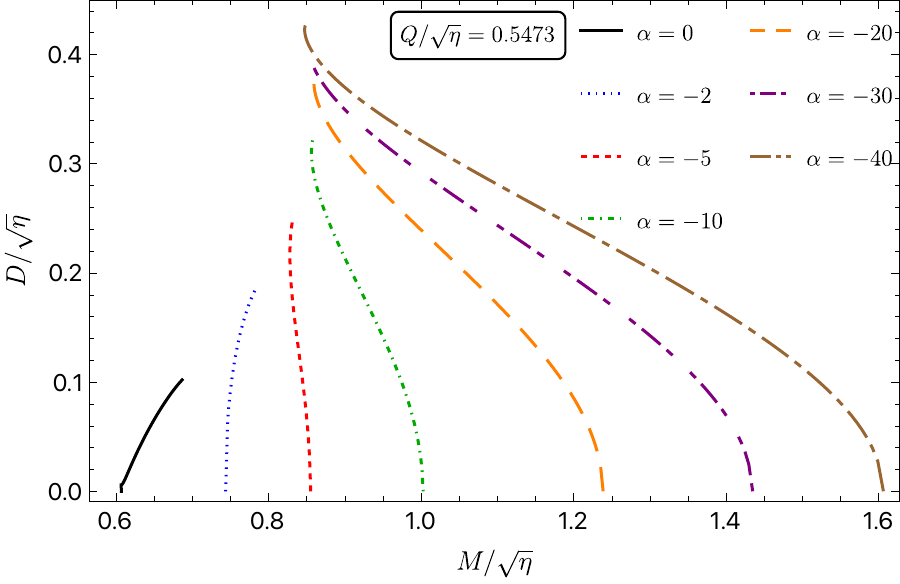}
    \includegraphics[width=0.45\textwidth]{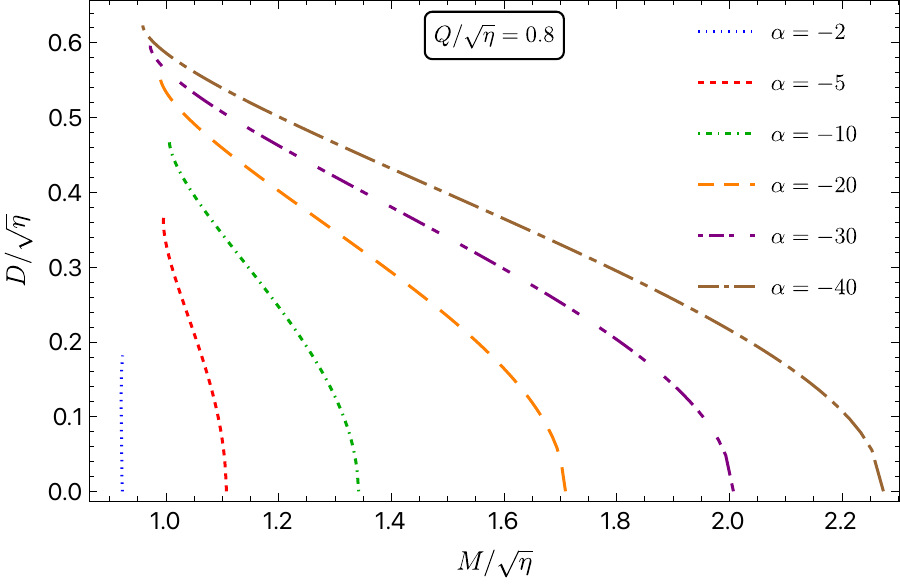}
    \includegraphics[width=0.45\textwidth]{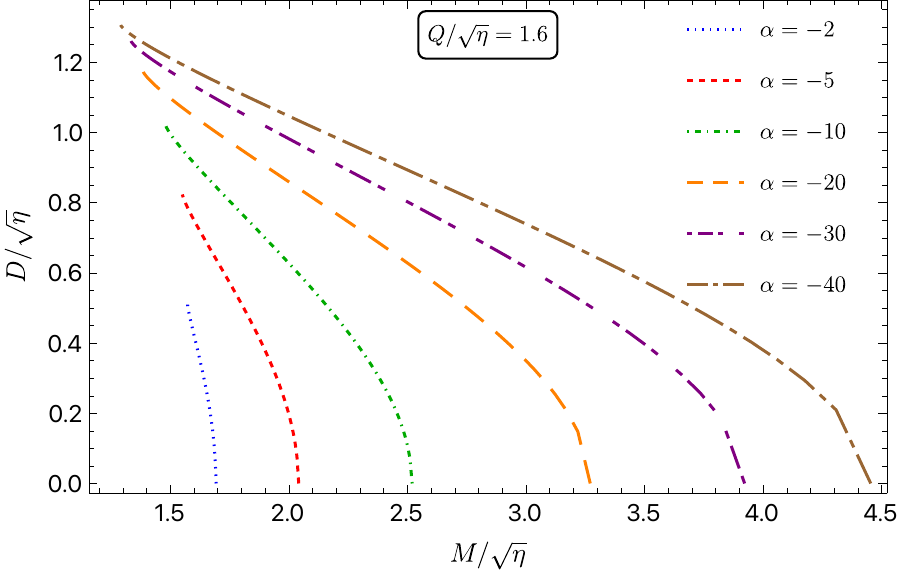}
    
\includegraphics[width=0.45\textwidth]{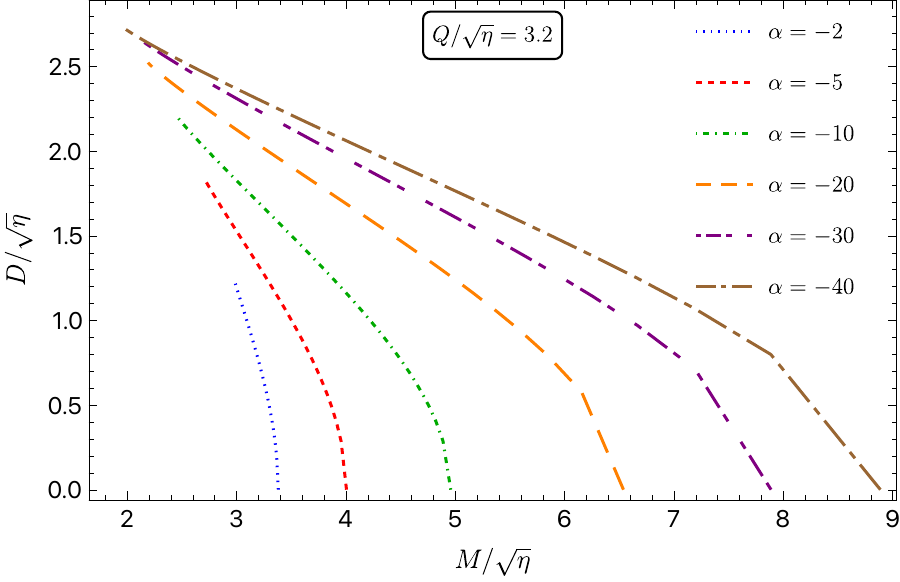}
    
    \caption{Normalized scalar charge $D/\sqrt{\eta}$ with respect to the normalized mass $M/\sqrt{\eta}$ for a fixed normalized electric charge $Q/\sqrt{\eta}$, comparing different values of $\alpha$,  for fixed values  of $Q/\sqrt{\eta}$ }
    \label{plotD}
\end{figure}

Figure \ref{plotD} shows the plots of the scalar charge $\frac{D}{\sqrt{\eta}}$ with respect to the mass $\frac{M}{\sqrt{\eta}}$
 for fixed ratio $Q/\sqrt{\eta}$, comparing several solutions with different $\alpha$. The scalar charge $\frac{D}{\sqrt{\eta}}$ versus mass $\frac{M}{\sqrt{\eta}}$ plots for scalarized black holes show that each choice of the coupling parameter $\alpha$ yields a distinct branch of solutions bifurcating from RN at different threshold masses. In single-coupling models (pure EsGB or EsGB plus a Ricci term) those mass thresholds are identical. In the present model, the threshold mass depends strongly on $\alpha$ and on $Q/\sqrt{\eta}$ . This allows a shift of the bifurcation mass $ \tilde{M}_{th}$ over a wide interval: the largest onset masses are up to three to four times the smallest ones. Furthermore after the bifurcation point, some scalarized branches extend toward higher masses while others extend toward lower masses. The direction of this extension appears to correlate with stability, as outlined below:
 
\begin{itemize}

\item \textbf{Varying threshold masses:} scalarization in this model sets in at different threshold masses $\tilde{M}_{th}=\frac{M}{\sqrt{\eta}}$ for different $\alpha$ values. This is in contrast with simpler cases of pure EsGB or EsGB with Ricci coupling, where all the branches share the same onset mass. For $\alpha=0$ (no Maxwell coupling), the threshold masses $\tilde{M}_{th}$ are very close and almost identical (small difference), as previously found in \cite{doneva2018charged}. By introducing a Maxwell coupling, the scalarization onset spans a much wider range of masses (some black holes have $\tilde{M}_{th}$ four times the lightest one), with each $\alpha$  producing its own distinct threshold.

\item\leavevmode\begin{minipage}[t]{0.62\linewidth}%
\textbf{Branch behavior at low charge:}
at low charge fraction $Q/\sqrt{\eta}=0.2$, the direction in which
scalarized branches extend after bifurcation depends on $\alpha$ .
Coupling values $\alpha=0,-2,-5$ produce branches that extend toward
higher mass values beyond the threshold (rightward on the plot).  In
contrast $\alpha=-20,-30,-40$ yield branches extending toward lower
masses.  The case $\alpha=-10$ shows a critical turning point where the branch has a curve that changes direction when analyzed individually in a small interval of $M/\sqrt{\eta}$ (see Fig. 3).
\end{minipage}%
\hfill
\begin{minipage}[t]{0.34\linewidth}%
  \centering
\includegraphics[width=0.94\linewidth]{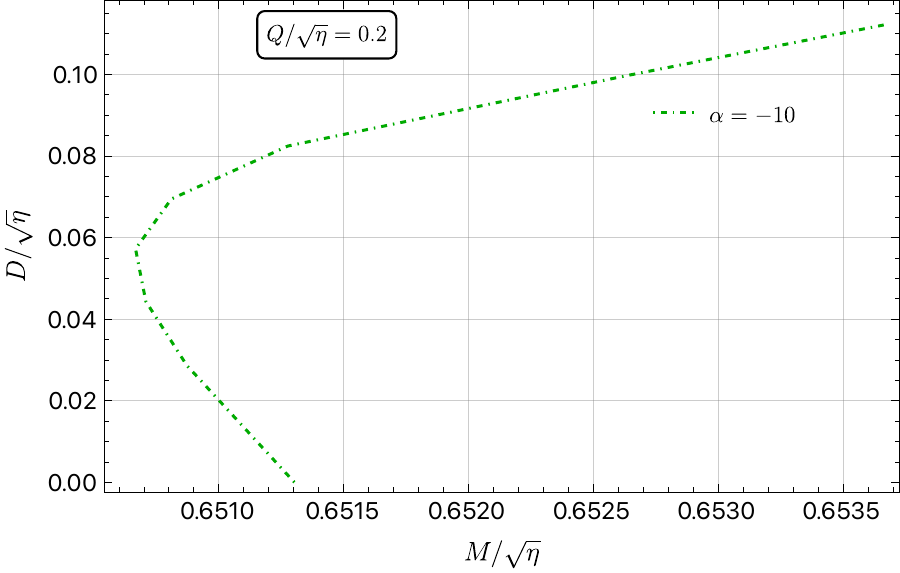}  % was 1.00
  \captionsetup{type=figure}
  \captionof{figure}{Scalar‑charge branch for $\alpha=-10$ and $Q/\sqrt{\eta}=0.2$.}
\end{minipage}\par

\item \textbf{Branch behavior at higher charge:} for larger charge to coupling ratio (e.g. $Q/\sqrt{\eta}=1.6$ or above), all the scalarized solution branches tend to extend toward smaller masses.
\item \textbf{Stability implications:} the direction in which a scalarized branch evolves is believed to indicate its stability. Branches that extended toward smaller masses (decreasing $M/\sqrt{\eta}$ as the scalar charge grows) are likely stable solutions, whereas those that extend toward larger masses beyond the bifurcation point may be unstable. This interpretation is in line with the suggestions of \cite{macedo2019self,antoniou2021black}, which means that introducing the matter coupling may have a stabilizing effect.
\item \textbf{Convergent end points:} several of the scalar charge curves for different values of $\alpha$ converge toward similar end points in the plane. Despite their different onset masses and trajectories, many branches approach nearly the same final scalar charge (and associated mass) at their termination. This suggests that different coupling strengths $\alpha$ can lead to scalarized black holes with comparable limiting properties. The behavior of the curves suggests maybe the existence of an attractor. 
\end{itemize}

\subsection{Thermodynamic properties }

We analyze the thermodynamic properties of the scalarized black hole solutions. Key quantities include the horizon area and the Hawking temperature. For the metric ansatz \eqref{ansatz}, the area and the temperature are :
\begin{eqnarray}
A_H = 4\pi r_H^2, \qquad T_H = \frac{\sigma(r_H) N'(r_H)}{4\pi}.
\end{eqnarray}

The black hole entropy $S_H$ in this model presents corrections to the Bekenstein-Hawking law, due to the non-minimal coupling with the Gauss-Bonnet term. We adopt the Wald–Iyer approach \cite{wald1993black,iyer1994some}, which associates a Noether charge to the black hole entropy in theories with diffeomorphism-invariant Lagrangian: 
Then, the entropy $S_H$ is defined as
\begin{equation}
    S_H = 2\pi \int_H \frac{\partial \mathcal{L}}{\partial R_{\mu\nu\alpha\beta}} \epsilon_{\mu\nu} \epsilon_{\alpha\beta},
\end{equation}
where $\mathcal{L}$ is the Lagrangian density and $\epsilon_{\alpha\beta}$ is the horizon's volume form.

For the action in Eq.~\eqref{eqaction}, this reduces to:
\begin{equation}
    S_H = \frac{1}{4} A_H + 4\pi f(\phi_0),
\end{equation}
where $\phi_0$ is the scalar field value at the horizon. This expression holds for both static and rotating black holes .

For instance scalarized black holes in EMsGB theories satisfy the Smarr relation \cite{iyer1994some,liberati2016smarr,herdeiro2021aspects}. For the static solutions considered here ($J=0, \Omega_H=0$), this relation is
\begin{equation} \label{eq:smarr}
M = 2 T_H S_H + \psi_e Q + M_s \ ,
\end{equation}
where $M_s$  is a contribution of the scalar field \cite{fernandes2022exploring},
\begin{equation*}
 M_s = \frac{1}{4\pi} \int d^3x \sqrt{-g} \frac{f(\phi)}{f_{, \phi}(\phi)} \Box \phi.
 \end{equation*}

 The first law of black hole thermodynamics in many of these models retains the form
\begin{equation}
    dM = T_H\,dS_H + \psi_e\,dQ.
\end{equation}
We can see that there is no direct contribution to this equation from the scalar charge $D$.

The most useful quantities to plot for studying black hole thermodynamic properties are the horizon area, the Hawking temperature, and the entropy.
The entropy can be plotted as $S_H/\eta$ or as $S_{H}/S_{GR}$. The temperature and area can be represented either by normalizing them with respect to the coupling constant, for instance $A_H/\eta$ and $T_H/\eta$, or by using the reduced quantities  
\begin{eqnarray}
\label{scale1}
a_H\equiv \frac{A_H}{16\pi M^2}\ , \qquad t_H\equiv 8\pi T_H M = 2M  N'(r_H)\sigma(r_H)\ ,  
\end{eqnarray}

\begin{figure}[ht]
    \centering
    \includegraphics[width=0.45\textwidth]{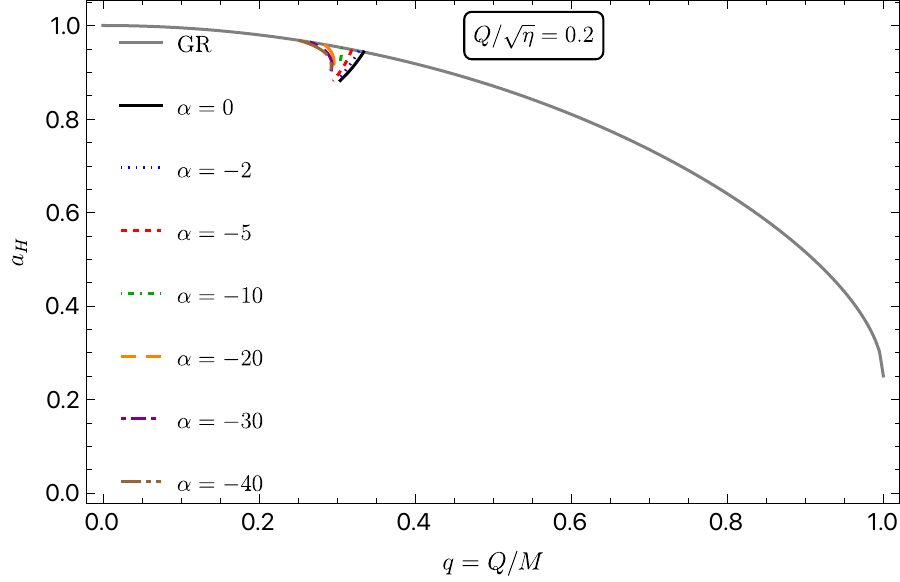}
    \includegraphics[width=0.45\textwidth]{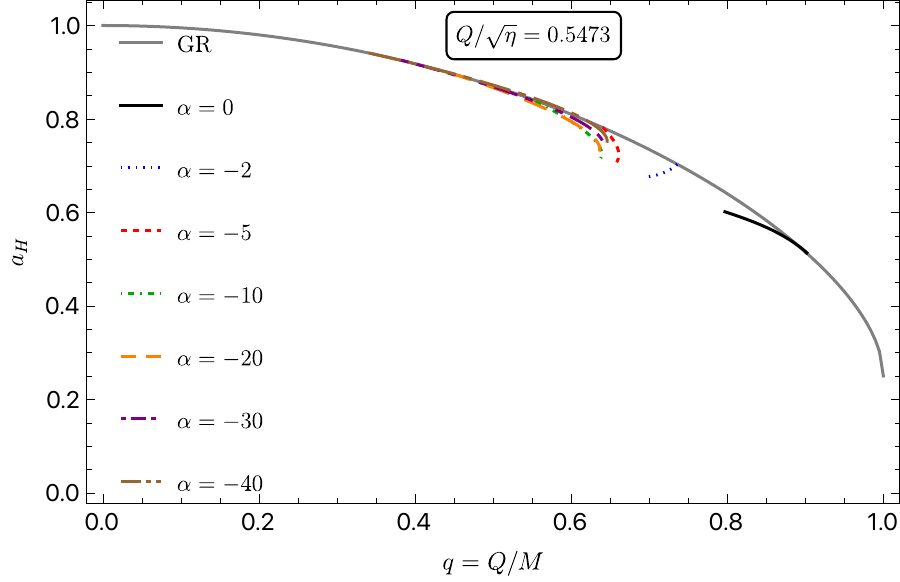}
    \includegraphics[width=0.45\textwidth]{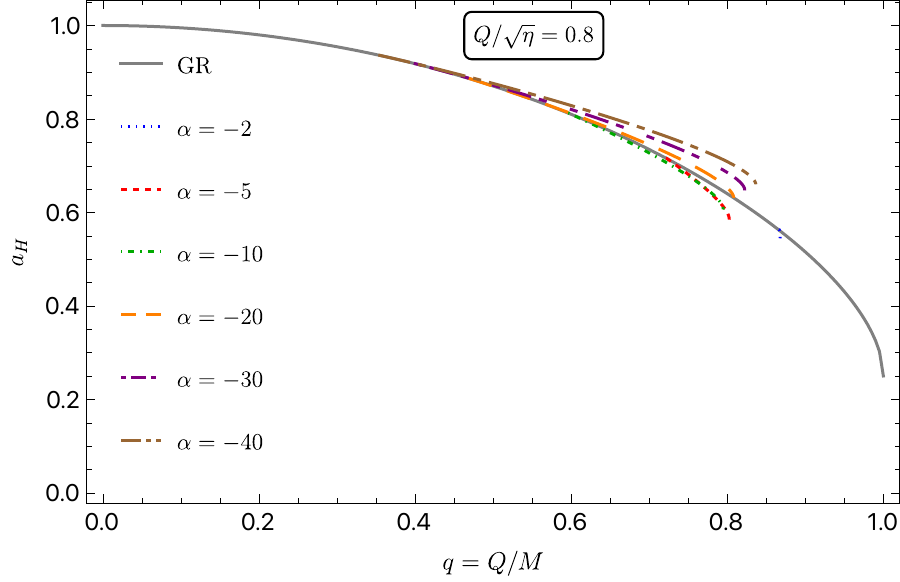}
    \includegraphics[width=0.45\textwidth]{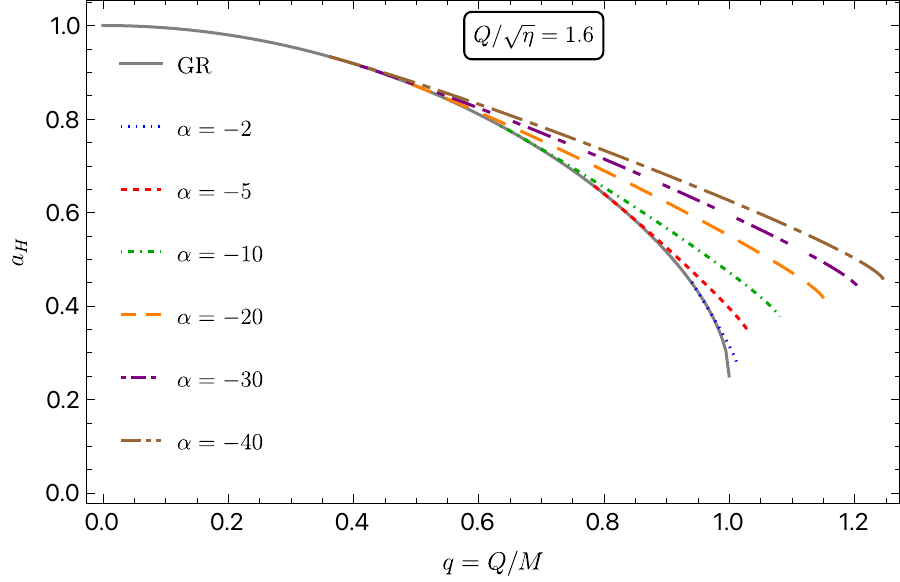}
    
\includegraphics[width=0.45\textwidth]{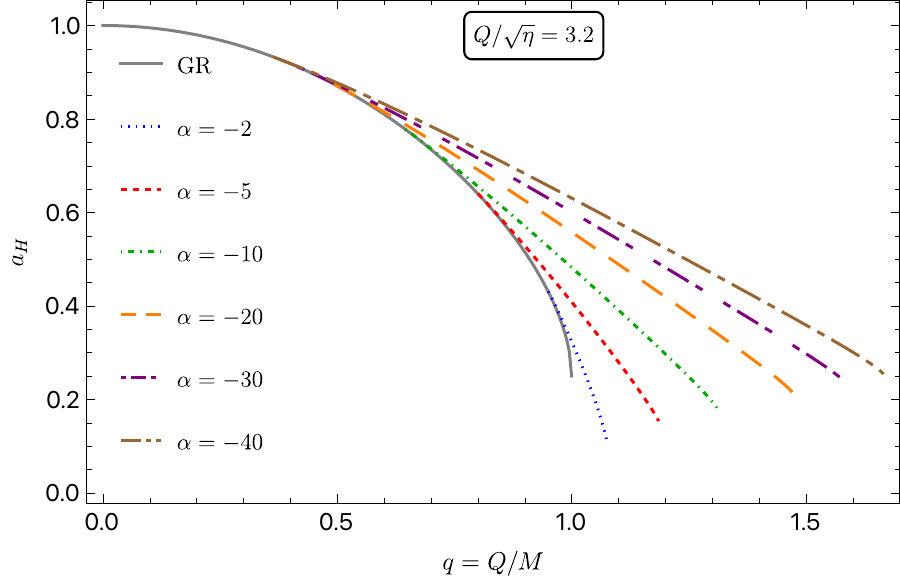}
    
    \caption{The figures present several plots of the reduced area $a_H=\frac{A_H}{16\pi M^2}$ with respect to the reduced charge $q=Q/M$, for different fixed values of $Q/\sqrt{\eta}$, and each of them shows many solutions with different values of $\alpha$.}
    \label{plotah}
\end{figure}
In Fig. \ref{plotah} the reduced area is reported with respect to the reduced charge for several values of the other parameters. We can notice that the first two plots, corresponding to $Q/\sqrt{\eta}=0.2$ and 0.5473 show that the area of all the scalarized solutions is smaller than the RN one, while in the third plot, with $Q/\sqrt{\eta}=0.8$, the behavior of the solutions with $\alpha=-20, -30, -40$ is opposite, showing scalarized black holes with higher area than the RN one, which is in agreement with the fact that matter-induced scalarization becomes dominant  due to the high value of the coupling to Maxwell ($\alpha$). From the value $Q/\sqrt{\eta}=1.6$ we clearly see that all the scalarized solutions with different values of $\alpha$ show higher area with respect to the RN one. This is expected even for the smallest value of $\alpha$ because of the high value of $Q/\sqrt{\eta}$ which enhances the contribution of scalarization induced by Maxwell in comparison with  the one induced by the GB invariant.

\begin{figure}[H]
    \centering
    \includegraphics[width=0.45\textwidth]{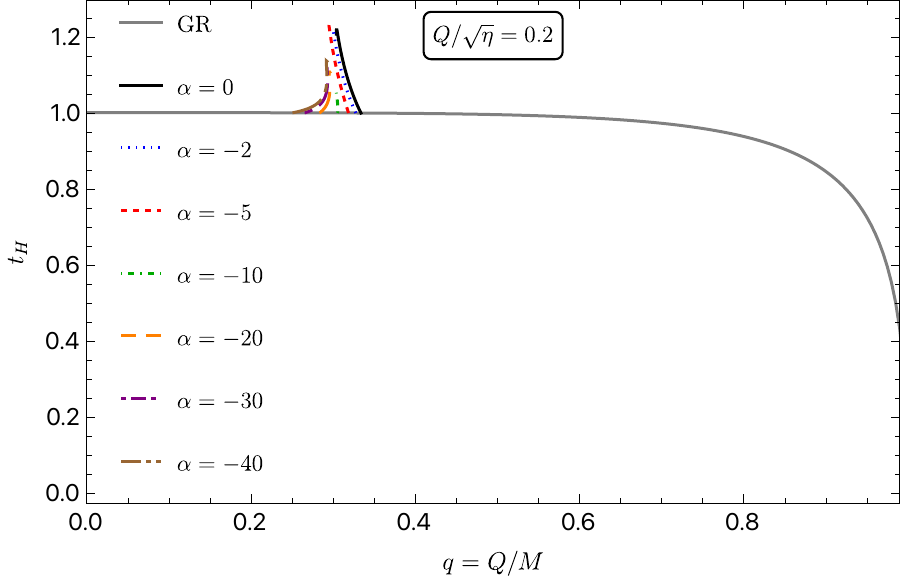}
    \includegraphics[width=0.45\textwidth]{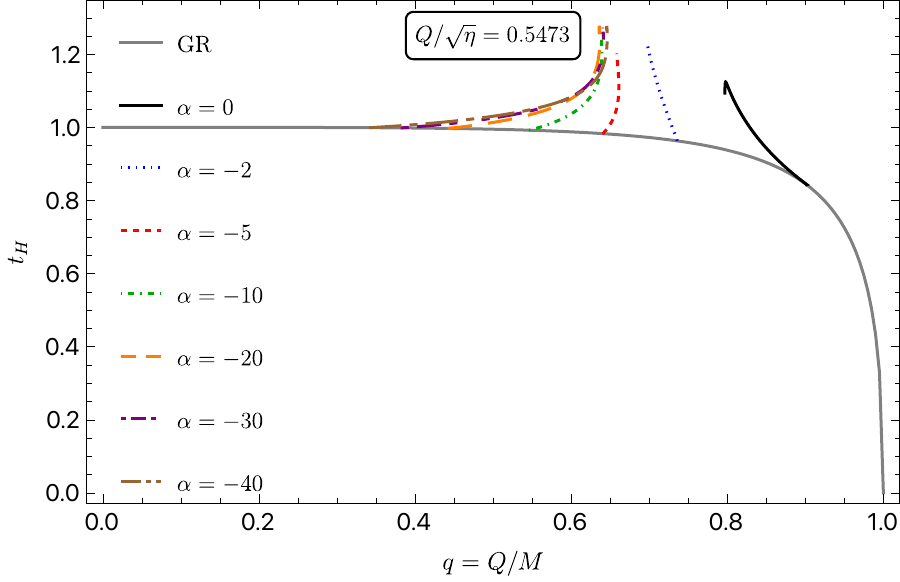}
    \includegraphics[width=0.45\textwidth]{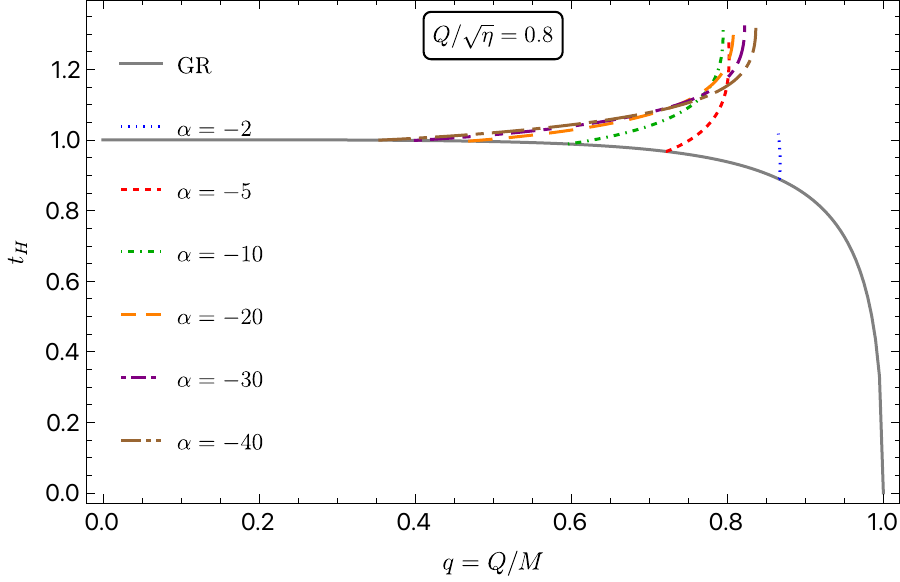}
    \includegraphics[width=0.45\textwidth]{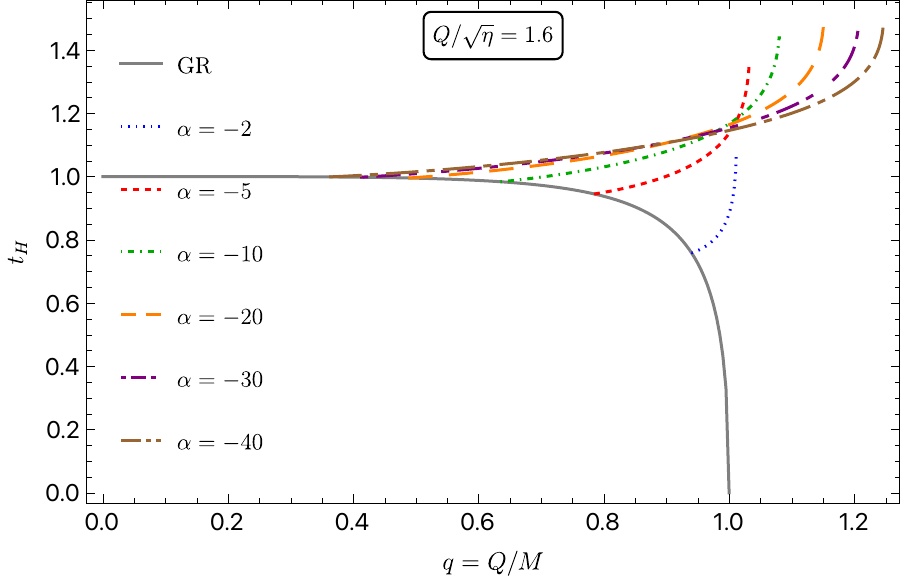}
      \includegraphics[width=0.45\textwidth]{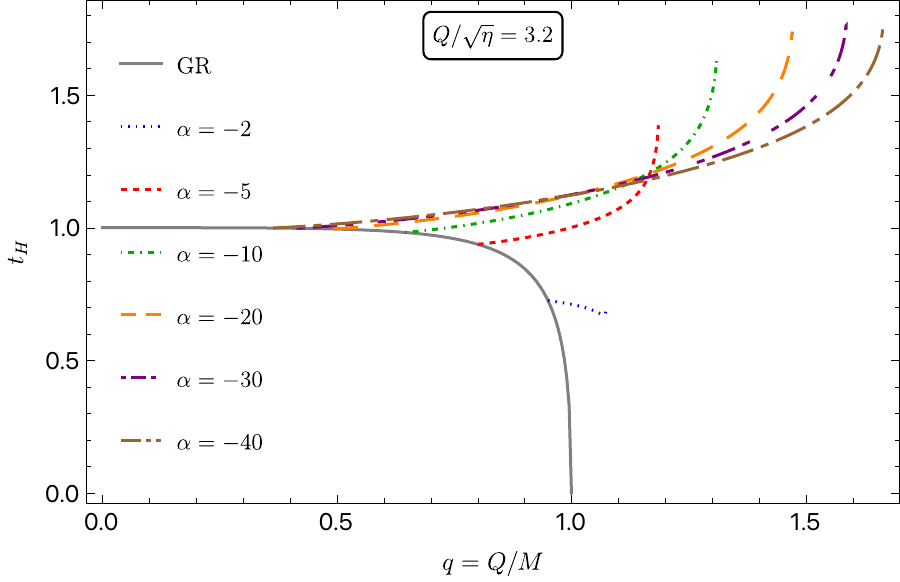}
    \caption{The figures present several plots of the reduced temperature $t_H=8 \pi M T_H$ with respect to the reduced charge $q=Q/M$, for different fixed values of $Q/\sqrt{\eta}$, and each one shows many solutions with different values of $\alpha$. }
    \label{plotth}
\end{figure}
In Fig. \ref{plotth} the reduced temperature of the scalarized solutions is plotted against the reduced charge. One can see that all scalarized solutions bifurcate from the GR branch, and have higher temperatures than the RN black hole. 
Another important property is that they never reach extremality ($t_H$ never vanishes). Finally, it turns out that, starting from $Q/\sqrt{\eta}=1.6$, the scalarized black hole can become overcharged ($q>1$). This fact was never observed for $GB^+$ ($\eta>0$) scalarization before, but only in the case of $GB^-$ ($\eta<0$) as pointed out in \cite{herdeiro2021aspects}.

\begin{figure}[H]
    \centering
    \includegraphics[width=0.45\textwidth]{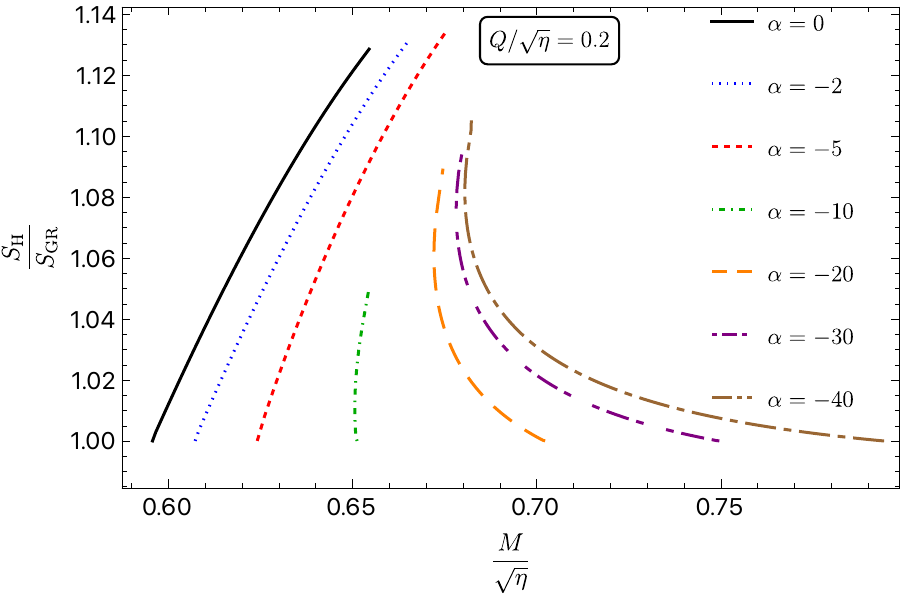}
    \includegraphics[width=0.45\textwidth]{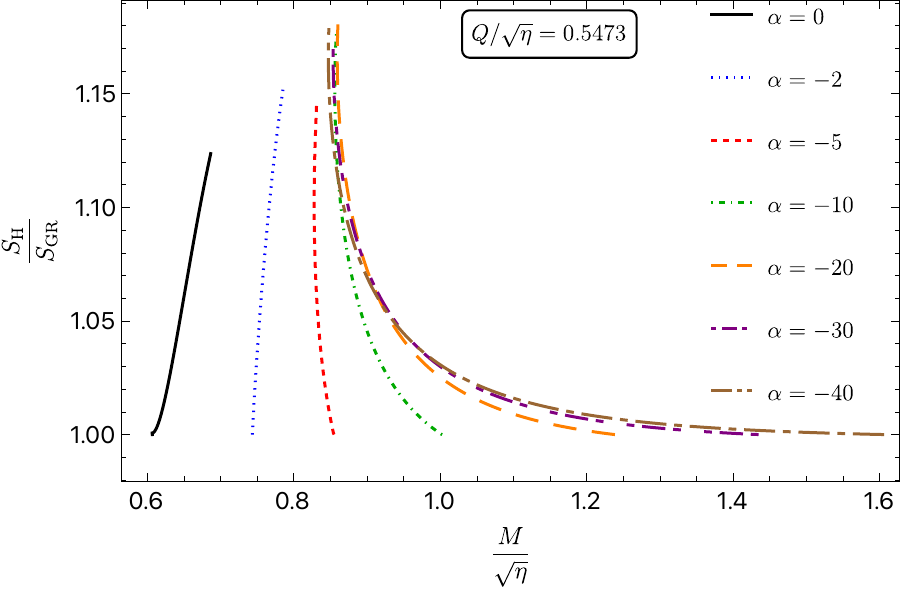}
    \includegraphics[width=0.45\textwidth]{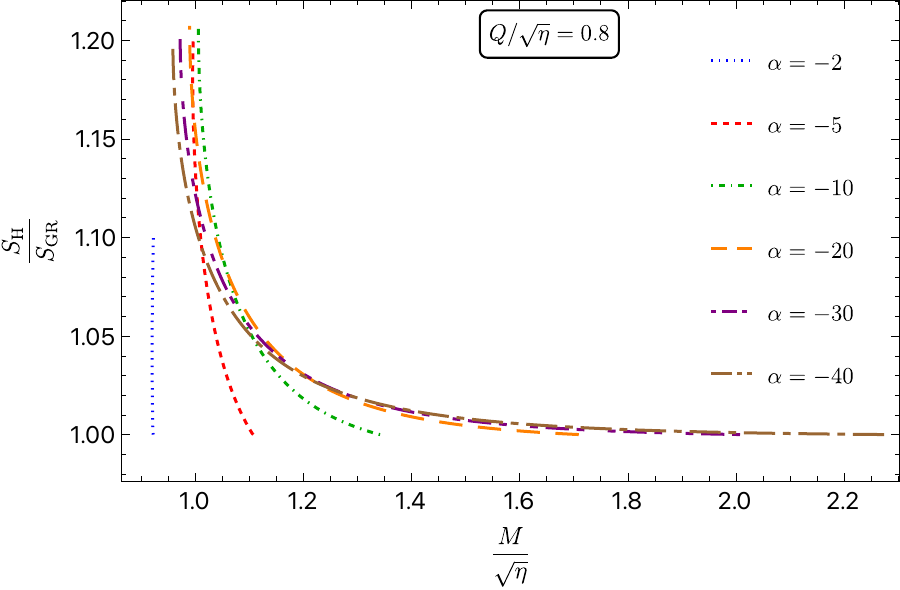}
    \includegraphics[width=0.45\textwidth]{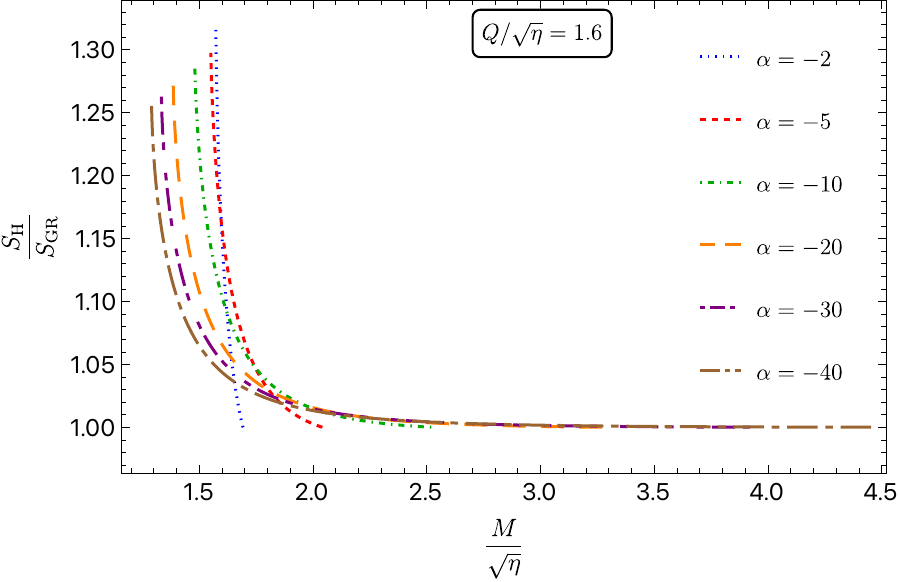}
      \includegraphics[width=0.45\textwidth]{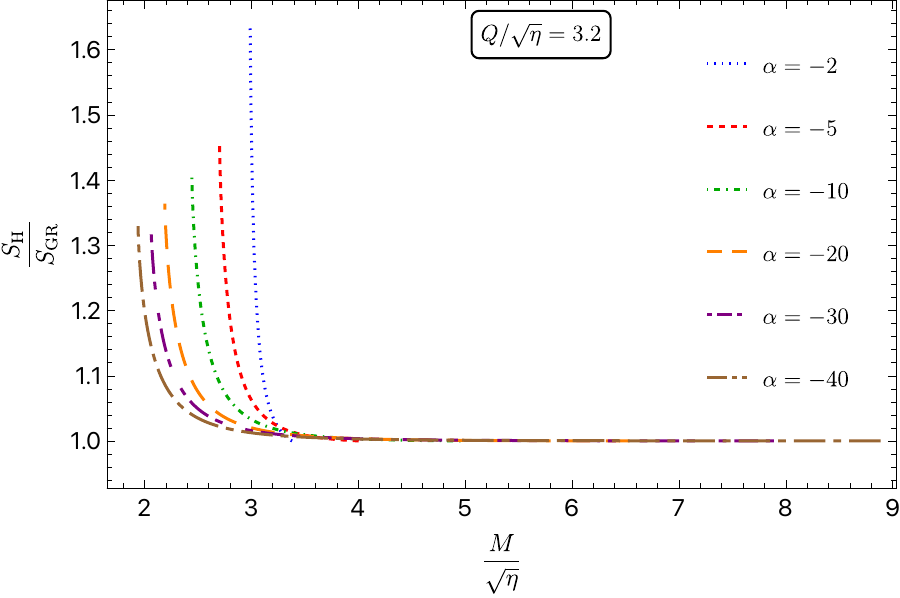}
    \caption{The figures show the entropy of scalarized black holes normalized by the entropy of GR black holes with the same parameters ($S_{H}/S_{GR}$), with respect to the normalized mass $M/\sqrt{\eta}$, and each one shows many solutions with different values of $\alpha$. }
    \label{plotsh}
\end{figure}
\bigskip

In Fig. \ref{plotsh}, the entropy of scalarized black holes is depicted with respect to their GR counterparts. We see clearly that the entropy of all scalarized black holes is larger than the entropy of RN black holes, so they are thermodynamically preferred. 
For large values of $Q/\sqrt{\eta}$, the entropy is almost the same as in GR, especially for larger masses $M/\sqrt{\eta}$, while for small values of $Q/\sqrt{\eta}$ the entropy of scalarized black holes deviates from its GR value.

\section{Conclusions}\label{sec4}
In this work, we introduced a new model of spontaneous black hole scalarization that includes couplings to both the Gauss-Bonnet curvature term and a $U(1)$ gauge field that can be taken as a Maxwell field type or a dark photon. By combining these curvature and matter couplings, we obtained charged black hole solutions with scalar hair that are not present in models with a single coupling. Our analysis revealed that the matter coupling extends the regime of scalarization even into regions with negative Gauss-Bonnet coupling, which was not possible with only curvature coupling. In practice, this means the scalar field can arise around a black hole under less restrictive conditions, thanks to the additional influence of the matter coupling.

The bifurcation plots in Fig. \ref{plotbif} show that the matter coupling allows scalarization with negative $\eta$ to occur with sub-extremal charges, which relaxes the near-extremality requirement observed in pure Gauss-Bonnet models associated with a Maxwell term. 

Our numerical results further quantify how the black hole properties depend on the coupling constants. We examined the behavior of the scalar charge $D/\sqrt{\eta}$ as a function of the black hole mass $M/\sqrt{\eta}$. Generally, the scalar charge grows from zero at the threshold and then increases with $M$ for the scalarized solutions. The solutions show several key features. The presence of two coupling parameters leads to different mass thresholds for the onset of scalarization. We found that the critical mass  at which the scalar field appears can vary widely: for certain values of the Gauss-Bonnet coupling $\eta$ and matter coupling $\alpha$, scalarization starts at much higher black hole masses than in the single-coupling cases, while for others it occurs earlier.

For different choices of $\alpha$ and $\eta$, the curves   can follow different trajectories, but interestingly many branches tend toward similar values of scalar charge as they approach their end points.

By comparing the branches, we found indications of which Maxwell coupling may stabilize solutions with quadratic coupling as the Ricci coupling and self-interactions did.  

Finally, we found that our model permits overcharged black hole solutions (with charge $Q$ exceeding the extremal value $M$). We observed that beyond a certain charge (for example, $Q/\sqrt{\eta} \gtrsim 1.6$ in our examples), the scalarized branches can exceed the $q = 1$ limit, yielding black holes with $q > 1$. Such overcharged scalarized black holes were not seen in earlier Gauss-Bonnet-only models (except in the special case of negative $\eta$) and thus represent a novel feature introduced by the inclusion of the matter coupling. 

 Many future directions to extend this work are possible. One first step should derive the black hole solutions for $\eta<0$ ($GB^-$). The next one will be to study the stability of these solutions (radial and angular), where the work of Ref \cite{kase2023black} will make it simple to do the analysis. 
 Future work can examine different coupling functions in our model. For instance, using an exponential form of coupling for one of the invariants or both of them instead of the quadratic forms.
 
  One can also compute the quasi-normal modes of these black holes in order to test them with future detectors (LISA and Einstein Telescope), and study their shadow in light of the observations by the Event Horizon Telescope.
  
 An ambitious direction should also target dynamical studies through numerical simulations for this model and see the impact of adding this new interaction to the hyperbolicity loss.

Another path that we suggest, is examining the implications of our findings on dark matter in the context of dark photon models. In a dark photon scenario, the Maxwell field in our model may represent a hidden $U(1)$ charge (a “dark” charge) carried by the black hole. Studying scalarization in this context may link our theoretical results to dark matter or dark sector phenomenology, and it could offer potential avenues to detect or constrain these exotic forms of black hole hair through their subtle gravitational or astrophysical signatures. A possible way is studying dark photon superradiance in this model \cite{cardoso2017superradiance,cardoso2018constraining,caputo2021electromagnetic,east2022dark,east2022vortex,cannizzaro2022dark,siemonsen2023dark}.

Finally, it will be very interesting to verify if these theories can have any impact for explaining late time acceleration without breaking GW170817 constraints on the speed of gravitational waves \cite{abbott2017gw170817,ezquiaga2017dark,creminelli2017dark,baker2017strong,sakstein2017implications}.

Our new model with both curvature and matter couplings broadens the domain of spontaneous black hole scalarization and reveals a much richer structure of solutions. We have shown that scalarized black holes in this theory can exist under more flexible conditions, where they appear to be more stable, possess higher entropy, and even achieve charge-to-mass ratios beyond the traditional extremal limit. These findings deepen our understanding of scalarization phenomena by introducing a new type of scalarization induced by two different sources and set the stage for further investigations into the stability and astrophysical implications of scalarized black holes.

\section*{Acknowledgments}
We are very grateful to Pedro G.S. Fernandes, Hector O. Silva and Caio F.B. Macedo for discussions and comments. 

\newpage

\appendix
\section{Field equations of motion for the model } 
\label{sec:appendixA}
The expressions for $W$ and $Z$ are as follows:

\begin{align*}
    W &= 16 r (1-3 N)^2 f_{, \phi}(\phi)^3 \phi'^2 \left(g(\phi)^2 N (5 N-1) \phi'^2 r^4 + Q^2 N g_{, \phi}(\phi) \phi' r - Q^2 g(\phi) (N-1)\right)  \\
    &\quad - 4 f_{, \phi}(\phi)^2 \phi' \left(-2 Q^4 + r^3 (3 N-1) (7 N-1) g_{, \phi}(\phi) \phi' Q^2 \right)  \\
    &\quad - 4 f_{, \phi}(\phi)^2 \phi' \left(2 r^2 g(\phi) \left(N \left(2 r^2 \phi'^2 - 6 r^2 N \phi'^2 + 2  (N-1) (3 N+1) f_{, \phi \phi}(\phi) \phi'^2 - 9 N + 10\right) - 1\right) Q^2 \right)  \\
    &\quad - 4 f_{, \phi}(\phi)^2 \phi' \left(2 r^4 g(\phi)^2 N \left(2 r^2 \phi'^2 + N \left(r^4 \phi'^4 - 20 r^2 \phi'^2 + 42 r^2 N \phi'^2 + 6  (N-1)^2 \left(r^2 \phi'^2 + 2\right) f_{, \phi \phi}(\phi) \phi'^2 - 3 N + 6\right) - 3\right) \right)  \\
    &\quad - 2 r f_{, \phi}(\phi) \left(Q^4 + 2 r^3 (1-4 N) g_{, \phi}(\phi) \phi' Q^2 + 2 r^2 g(\phi) \left(-r^2 \phi'^2 + N \left(\left(5 r^2 - 4  (N-1) f_{, \phi \phi}(\phi)\right) \phi'^2 + 3\right) - 3\right) Q^2 \right)  \\
    &\quad - 2 r f_{, \phi}(\phi) \left(r^4 g(\phi)^2 \left(2 r^2 \phi'^2 \right. \right. \\
    &\quad \quad \left. \left. + N \left(-4  (N-1) \left(r^2 (3 N-1) \phi'^2 + 3 (N-1)\right) f_{, \phi \phi}(\phi) \phi'^2 - 2 \left(r^2 \phi'^2 + 3\right) - N \left(r^4 \phi'^4 + 24 r^2 \phi'^2 - 3\right)\right) + 3\right) \right) \\
    &\quad - r^5 \left(g_{, \phi}(\phi) Q^2 + 2 r g(\phi) \phi' \left(r^2 g(\phi) \left(2  (N-1) N f_{, \phi \phi}(\phi) \phi'^2 + N + 1\right) - Q^2\right)\right), \\
    Z &= -2 r^2 g(\phi) N \left(8 Q^2 (N-1) f_{, \phi}(\phi)^2 \left( (3 N+1) f_{, \phi}(\phi) \phi' - r\right)  + r^2 g(\phi) \left(r^5 + 2  f_{, \phi}(\phi) \left((1-7 N) \phi' r^4 \right. \right. \right. \\
    &\quad \left. \left. \left. + 6  f_{, \phi}(\phi) \left(2 r^2 N (3 N-1) \phi'^2 - (N-1)^2\right) r + 4  N f_{, \phi}(\phi)^2 \phi' \left(6 (N-1)^2 + r^2 (3 N (2-5 N) + 1) \phi'^2\right)\right)\right)\right).
\end{align*}

\bibliography{biblio}

\begin{thebibliography}{163}%
\makeatletter
\providecommand \@ifxundefined [1]{%
 \@ifx{#1\undefined}
}%
\providecommand \@ifnum [1]{%
 \ifnum #1\expandafter \@firstoftwo
 \else \expandafter \@secondoftwo
 \fi
}%
\providecommand \@ifx [1]{%
 \ifx #1\expandafter \@firstoftwo
 \else \expandafter \@secondoftwo
 \fi
}%
\providecommand \natexlab [1]{#1}%
\providecommand \enquote  [1]{``#1''}%
\providecommand \bibnamefont  [1]{#1}%
\providecommand \bibfnamefont [1]{#1}%
\providecommand \citenamefont [1]{#1}%
\providecommand \href@noop [0]{\@secondoftwo}%
\providecommand \href [0]{\begingroup \@sanitize@url \@href}%
\providecommand \@href[1]{\@@startlink{#1}\@@href}%
\providecommand \@@href[1]{\endgroup#1\@@endlink}%
\providecommand \@sanitize@url [0]{\catcode `\\12\catcode `\$12\catcode `\&12\catcode `\#12\catcode `\^12\catcode `\_12\catcode `\%12\relax}%
\providecommand \@@startlink[1]{}%
\providecommand \@@endlink[0]{}%
\providecommand \url  [0]{\begingroup\@sanitize@url \@url }%
\providecommand \@url [1]{\endgroup\@href {#1}{\urlprefix }}%
\providecommand \urlprefix  [0]{URL }%
\providecommand \Eprint [0]{\href }%
\providecommand \doibase [0]{https://doi.org/}%
\providecommand \selectlanguage [0]{\@gobble}%
\providecommand \bibinfo  [0]{\@secondoftwo}%
\providecommand \bibfield  [0]{\@secondoftwo}%
\providecommand \translation [1]{[#1]}%
\providecommand \BibitemOpen [0]{}%
\providecommand \bibitemStop [0]{}%
\providecommand \bibitemNoStop [0]{.\EOS\space}%
\providecommand \EOS [0]{\spacefactor3000\relax}%
\providecommand \BibitemShut  [1]{\csname bibitem#1\endcsname}%
\let\auto@bib@innerbib\@empty
%</preamble>
\bibitem [{\citenamefont {Abbott}\ \emph {et~al.}(2016)\citenamefont {Abbott}, \citenamefont {Abbott}, \citenamefont {Abbott}, \citenamefont {Abernathy}, \citenamefont {Acernese}, \citenamefont {Ackley}, \citenamefont {Adams}, \citenamefont {Adams}, \citenamefont {Addesso}, \citenamefont {Adhikari} \emph {et~al.}}]{abbott2016observation}%
  \BibitemOpen
  \bibfield  {author} {\bibinfo {author} {\bibfnamefont {B.~P.}\ \bibnamefont {Abbott}}, \bibinfo {author} {\bibfnamefont {R.}~\bibnamefont {Abbott}}, \bibinfo {author} {\bibfnamefont {T.}~\bibnamefont {Abbott}}, \bibinfo {author} {\bibfnamefont {M.}~\bibnamefont {Abernathy}}, \bibinfo {author} {\bibfnamefont {F.}~\bibnamefont {Acernese}}, \bibinfo {author} {\bibfnamefont {K.}~\bibnamefont {Ackley}}, \bibinfo {author} {\bibfnamefont {C.}~\bibnamefont {Adams}}, \bibinfo {author} {\bibfnamefont {T.}~\bibnamefont {Adams}}, \bibinfo {author} {\bibfnamefont {P.}~\bibnamefont {Addesso}}, \bibinfo {author} {\bibfnamefont {R.}~\bibnamefont {Adhikari}}, \emph {et~al.},\ }\bibfield  {title} {\bibinfo {title} {Observation of gravitational waves from a binary black hole merger},\ }\href@noop {} {\bibfield  {journal} {\bibinfo  {journal} {Physical review letters}\ }\textbf {\bibinfo {volume} {116}},\ \bibinfo {pages} {061102} (\bibinfo {year} {2016})}\BibitemShut {NoStop}%
\bibitem [{\citenamefont {Ball}\ \emph {et~al.}(2019)\citenamefont {Ball}, \citenamefont {Chan}, \citenamefont {Christian}, \citenamefont {Jannuzi}, \citenamefont {Kim}, \citenamefont {Marrone}, \citenamefont {Medeiros}, \citenamefont {Ozel}, \citenamefont {Psaltis}, \citenamefont {Rose} \emph {et~al.}}]{ball2019first}%
  \BibitemOpen
  \bibfield  {author} {\bibinfo {author} {\bibfnamefont {D.}~\bibnamefont {Ball}}, \bibinfo {author} {\bibfnamefont {C.-k.}\ \bibnamefont {Chan}}, \bibinfo {author} {\bibfnamefont {P.}~\bibnamefont {Christian}}, \bibinfo {author} {\bibfnamefont {B.~T.}\ \bibnamefont {Jannuzi}}, \bibinfo {author} {\bibfnamefont {J.}~\bibnamefont {Kim}}, \bibinfo {author} {\bibfnamefont {D.~P.}\ \bibnamefont {Marrone}}, \bibinfo {author} {\bibfnamefont {L.}~\bibnamefont {Medeiros}}, \bibinfo {author} {\bibfnamefont {F.}~\bibnamefont {Ozel}}, \bibinfo {author} {\bibfnamefont {D.}~\bibnamefont {Psaltis}}, \bibinfo {author} {\bibfnamefont {M.}~\bibnamefont {Rose}}, \emph {et~al.},\ }\bibfield  {title} {\bibinfo {title} {First m87 event horizon telescope results. i. the shadow of the supermassive black hole},\ }\href@noop {} {\bibfield  {journal} {\bibinfo  {journal} {IOP PUBLISHING LTD}\ } (\bibinfo {year} {2019})}\BibitemShut {NoStop}%
\bibitem [{\citenamefont {Akiyama}\ \emph {et~al.}(2022)\citenamefont {Akiyama}, \citenamefont {Alberdi}, \citenamefont {Alef}, \citenamefont {Algaba}, \citenamefont {Anantua}, \citenamefont {Asada}, \citenamefont {Azulay}, \citenamefont {Bach}, \citenamefont {Baczko}, \citenamefont {Ball} \emph {et~al.}}]{akiyama2022first}%
  \BibitemOpen
  \bibfield  {author} {\bibinfo {author} {\bibfnamefont {K.}~\bibnamefont {Akiyama}}, \bibinfo {author} {\bibfnamefont {A.}~\bibnamefont {Alberdi}}, \bibinfo {author} {\bibfnamefont {W.}~\bibnamefont {Alef}}, \bibinfo {author} {\bibfnamefont {J.~C.}\ \bibnamefont {Algaba}}, \bibinfo {author} {\bibfnamefont {R.}~\bibnamefont {Anantua}}, \bibinfo {author} {\bibfnamefont {K.}~\bibnamefont {Asada}}, \bibinfo {author} {\bibfnamefont {R.}~\bibnamefont {Azulay}}, \bibinfo {author} {\bibfnamefont {U.}~\bibnamefont {Bach}}, \bibinfo {author} {\bibfnamefont {A.-K.}\ \bibnamefont {Baczko}}, \bibinfo {author} {\bibfnamefont {D.}~\bibnamefont {Ball}}, \emph {et~al.},\ }\bibfield  {title} {\bibinfo {title} {First sagittarius a* event horizon telescope results. i. the shadow of the supermassive black hole in the center of the milky way},\ }\href@noop {} {\bibfield  {journal} {\bibinfo  {journal} {The Astrophysical Journal Letters}\ }\textbf {\bibinfo {volume} {930}},\ \bibinfo {pages} {L12} (\bibinfo {year}
  {2022})}\BibitemShut {NoStop}%
\bibitem [{\citenamefont {Herdeiro}(2023)}]{herdeiro2023black}%
  \BibitemOpen
  \bibfield  {author} {\bibinfo {author} {\bibfnamefont {C.~A.}\ \bibnamefont {Herdeiro}},\ }\bibfield  {title} {\bibinfo {title} {Black holes: on the universality of the kerr hypothesis},\ }in\ \href@noop {} {\emph {\bibinfo {booktitle} {Modified and Quantum Gravity: From Theory to Experimental Searches on All Scales}}}\ (\bibinfo  {publisher} {Springer},\ \bibinfo {year} {2023})\ pp.\ \bibinfo {pages} {315--331}\BibitemShut {NoStop}%
\bibitem [{\citenamefont {Barausse}\ \emph {et~al.}(2020)\citenamefont {Barausse}, \citenamefont {Berti}, \citenamefont {Hertog}, \citenamefont {Hughes}, \citenamefont {Jetzer}, \citenamefont {Pani}, \citenamefont {Sotiriou}, \citenamefont {Tamanini}, \citenamefont {Witek}, \citenamefont {Yagi} \emph {et~al.}}]{barausse2020prospects}%
  \BibitemOpen
  \bibfield  {author} {\bibinfo {author} {\bibfnamefont {E.}~\bibnamefont {Barausse}}, \bibinfo {author} {\bibfnamefont {E.}~\bibnamefont {Berti}}, \bibinfo {author} {\bibfnamefont {T.}~\bibnamefont {Hertog}}, \bibinfo {author} {\bibfnamefont {S.~A.}\ \bibnamefont {Hughes}}, \bibinfo {author} {\bibfnamefont {P.}~\bibnamefont {Jetzer}}, \bibinfo {author} {\bibfnamefont {P.}~\bibnamefont {Pani}}, \bibinfo {author} {\bibfnamefont {T.~P.}\ \bibnamefont {Sotiriou}}, \bibinfo {author} {\bibfnamefont {N.}~\bibnamefont {Tamanini}}, \bibinfo {author} {\bibfnamefont {H.}~\bibnamefont {Witek}}, \bibinfo {author} {\bibfnamefont {K.}~\bibnamefont {Yagi}}, \emph {et~al.},\ }\bibfield  {title} {\bibinfo {title} {Prospects for fundamental physics with lisa},\ }\href@noop {} {\bibfield  {journal} {\bibinfo  {journal} {General Relativity and Gravitation}\ }\textbf {\bibinfo {volume} {52}},\ \bibinfo {pages} {1} (\bibinfo {year} {2020})}\BibitemShut {NoStop}%
\bibitem [{\citenamefont {Arun}\ \emph {et~al.}(2022)\citenamefont {Arun}, \citenamefont {Belgacem}, \citenamefont {Benkel}, \citenamefont {Bernard}, \citenamefont {Berti}, \citenamefont {Bertone}, \citenamefont {Besancon}, \citenamefont {Blas}, \citenamefont {B{\"o}hmer}, \citenamefont {Brito} \emph {et~al.}}]{arun2022new}%
  \BibitemOpen
  \bibfield  {author} {\bibinfo {author} {\bibfnamefont {K.}~\bibnamefont {Arun}}, \bibinfo {author} {\bibfnamefont {E.}~\bibnamefont {Belgacem}}, \bibinfo {author} {\bibfnamefont {R.}~\bibnamefont {Benkel}}, \bibinfo {author} {\bibfnamefont {L.}~\bibnamefont {Bernard}}, \bibinfo {author} {\bibfnamefont {E.}~\bibnamefont {Berti}}, \bibinfo {author} {\bibfnamefont {G.}~\bibnamefont {Bertone}}, \bibinfo {author} {\bibfnamefont {M.}~\bibnamefont {Besancon}}, \bibinfo {author} {\bibfnamefont {D.}~\bibnamefont {Blas}}, \bibinfo {author} {\bibfnamefont {C.~G.}\ \bibnamefont {B{\"o}hmer}}, \bibinfo {author} {\bibfnamefont {R.}~\bibnamefont {Brito}}, \emph {et~al.},\ }\bibfield  {title} {\bibinfo {title} {New horizons for fundamental physics with lisa},\ }\href@noop {} {\bibfield  {journal} {\bibinfo  {journal} {Living Reviews in Relativity}\ }\textbf {\bibinfo {volume} {25}},\ \bibinfo {pages} {4} (\bibinfo {year} {2022})}\BibitemShut {NoStop}%
\bibitem [{\citenamefont {Auclair}\ \emph {et~al.}(2023)\citenamefont {Auclair}, \citenamefont {Bacon}, \citenamefont {Baker}, \citenamefont {Barreiro}, \citenamefont {Bartolo}, \citenamefont {Belgacem}, \citenamefont {Bellomo}, \citenamefont {Ben-Dayan}, \citenamefont {Bertacca}, \citenamefont {Besancon} \emph {et~al.}}]{auclair2023cosmology}%
  \BibitemOpen
  \bibfield  {author} {\bibinfo {author} {\bibfnamefont {P.}~\bibnamefont {Auclair}}, \bibinfo {author} {\bibfnamefont {D.}~\bibnamefont {Bacon}}, \bibinfo {author} {\bibfnamefont {T.}~\bibnamefont {Baker}}, \bibinfo {author} {\bibfnamefont {T.}~\bibnamefont {Barreiro}}, \bibinfo {author} {\bibfnamefont {N.}~\bibnamefont {Bartolo}}, \bibinfo {author} {\bibfnamefont {E.}~\bibnamefont {Belgacem}}, \bibinfo {author} {\bibfnamefont {N.}~\bibnamefont {Bellomo}}, \bibinfo {author} {\bibfnamefont {I.}~\bibnamefont {Ben-Dayan}}, \bibinfo {author} {\bibfnamefont {D.}~\bibnamefont {Bertacca}}, \bibinfo {author} {\bibfnamefont {M.}~\bibnamefont {Besancon}}, \emph {et~al.},\ }\bibfield  {title} {\bibinfo {title} {Cosmology with the laser interferometer space antenna},\ }\href@noop {} {\bibfield  {journal} {\bibinfo  {journal} {Living Reviews in Relativity}\ }\textbf {\bibinfo {volume} {26}},\ \bibinfo {pages} {5} (\bibinfo {year} {2023})}\BibitemShut {NoStop}%
\bibitem [{\citenamefont {Abac}\ \emph {et~al.}(2025)\citenamefont {Abac}, \citenamefont {Abramo}, \citenamefont {Albanesi}, \citenamefont {Albertini}, \citenamefont {Agapito}, \citenamefont {Agathos}, \citenamefont {Albertus}, \citenamefont {Andersson}, \citenamefont {Andrade}, \citenamefont {Andreoni} \emph {et~al.}}]{abac2025science}%
  \BibitemOpen
  \bibfield  {author} {\bibinfo {author} {\bibfnamefont {A.}~\bibnamefont {Abac}}, \bibinfo {author} {\bibfnamefont {R.}~\bibnamefont {Abramo}}, \bibinfo {author} {\bibfnamefont {S.}~\bibnamefont {Albanesi}}, \bibinfo {author} {\bibfnamefont {A.}~\bibnamefont {Albertini}}, \bibinfo {author} {\bibfnamefont {A.}~\bibnamefont {Agapito}}, \bibinfo {author} {\bibfnamefont {M.}~\bibnamefont {Agathos}}, \bibinfo {author} {\bibfnamefont {C.}~\bibnamefont {Albertus}}, \bibinfo {author} {\bibfnamefont {N.}~\bibnamefont {Andersson}}, \bibinfo {author} {\bibfnamefont {T.}~\bibnamefont {Andrade}}, \bibinfo {author} {\bibfnamefont {I.}~\bibnamefont {Andreoni}}, \emph {et~al.},\ }\bibfield  {title} {\bibinfo {title} {The science of the einstein telescope},\ }\href@noop {} {\bibfield  {journal} {\bibinfo  {journal} {arXiv preprint arXiv:2503.12263}\ } (\bibinfo {year} {2025})}\BibitemShut {NoStop}%
\bibitem [{\citenamefont {Maggiore}\ \emph {et~al.}(2020)\citenamefont {Maggiore}, \citenamefont {Van Den~Broeck}, \citenamefont {Bartolo}, \citenamefont {Belgacem}, \citenamefont {Bertacca}, \citenamefont {Bizouard}, \citenamefont {Branchesi}, \citenamefont {Clesse}, \citenamefont {Foffa}, \citenamefont {Garc{\'\i}a-Bellido} \emph {et~al.}}]{maggiore2020science}%
  \BibitemOpen
  \bibfield  {author} {\bibinfo {author} {\bibfnamefont {M.}~\bibnamefont {Maggiore}}, \bibinfo {author} {\bibfnamefont {C.}~\bibnamefont {Van Den~Broeck}}, \bibinfo {author} {\bibfnamefont {N.}~\bibnamefont {Bartolo}}, \bibinfo {author} {\bibfnamefont {E.}~\bibnamefont {Belgacem}}, \bibinfo {author} {\bibfnamefont {D.}~\bibnamefont {Bertacca}}, \bibinfo {author} {\bibfnamefont {M.~A.}\ \bibnamefont {Bizouard}}, \bibinfo {author} {\bibfnamefont {M.}~\bibnamefont {Branchesi}}, \bibinfo {author} {\bibfnamefont {S.}~\bibnamefont {Clesse}}, \bibinfo {author} {\bibfnamefont {S.}~\bibnamefont {Foffa}}, \bibinfo {author} {\bibfnamefont {J.}~\bibnamefont {Garc{\'\i}a-Bellido}}, \emph {et~al.},\ }\bibfield  {title} {\bibinfo {title} {Science case for the einstein telescope},\ }\href@noop {} {\bibfield  {journal} {\bibinfo  {journal} {Journal of Cosmology and Astroparticle Physics}\ }\textbf {\bibinfo {volume} {2020}}\bibinfo  {number} { (03)},\ \bibinfo {pages} {050}}\BibitemShut {NoStop}%
\bibitem [{\citenamefont {Hawking}(1972)}]{hawking1972black}%
  \BibitemOpen
\bibfield  {number} {  }\bibfield  {author} {\bibinfo {author} {\bibfnamefont {S.}~\bibnamefont {Hawking}},\ }\bibfield  {title} {\bibinfo {title} {Black holes in the brans-dicke: Theory of gravitation},\ }\href@noop {} {\bibfield  {journal} {\bibinfo  {journal} {Communications in Mathematical Physics}\ }\textbf {\bibinfo {volume} {25}},\ \bibinfo {pages} {167} (\bibinfo {year} {1972})}\BibitemShut {NoStop}%
\bibitem [{\citenamefont {Bekenstein}(1972)}]{bekenstein1972transcendence}%
  \BibitemOpen
  \bibfield  {author} {\bibinfo {author} {\bibfnamefont {J.~D.}\ \bibnamefont {Bekenstein}},\ }\bibfield  {title} {\bibinfo {title} {Transcendence of the law of baryon-number conservation in black-hole physics},\ }\href@noop {} {\bibfield  {journal} {\bibinfo  {journal} {Physical Review Letters}\ }\textbf {\bibinfo {volume} {28}},\ \bibinfo {pages} {452} (\bibinfo {year} {1972})}\BibitemShut {NoStop}%
\bibitem [{\citenamefont {Bekenstein}(1995)}]{bekenstein1995novel}%
  \BibitemOpen
  \bibfield  {author} {\bibinfo {author} {\bibfnamefont {J.~D.}\ \bibnamefont {Bekenstein}},\ }\bibfield  {title} {\bibinfo {title} {Novel ‘‘no-scalar-hair’’theorem for black holes},\ }\href@noop {} {\bibfield  {journal} {\bibinfo  {journal} {Physical Review D}\ }\textbf {\bibinfo {volume} {51}},\ \bibinfo {pages} {R6608} (\bibinfo {year} {1995})}\BibitemShut {NoStop}%
\bibitem [{\citenamefont {Bocharova}\ \emph {et~al.}(1970)\citenamefont {Bocharova}, \citenamefont {Bronnikov},\ and\ \citenamefont {Melnikov}}]{bocharova1970exact}%
  \BibitemOpen
  \bibfield  {author} {\bibinfo {author} {\bibfnamefont {N.}~\bibnamefont {Bocharova}}, \bibinfo {author} {\bibfnamefont {K.}~\bibnamefont {Bronnikov}},\ and\ \bibinfo {author} {\bibfnamefont {V.}~\bibnamefont {Melnikov}},\ }\bibfield  {title} {\bibinfo {title} {An exact solution of the system of einstein equations and mass-free scalar field},\ }\href@noop {} {\bibfield  {journal} {\bibinfo  {journal} {Vestn. Mosk. Univ. Fiz. Astro}\ }\textbf {\bibinfo {volume} {6}},\ \bibinfo {pages} {706} (\bibinfo {year} {1970})}\BibitemShut {NoStop}%
\bibitem [{\citenamefont {Bekenstein}(1974)}]{bekenstein1974exact}%
  \BibitemOpen
  \bibfield  {author} {\bibinfo {author} {\bibfnamefont {J.~D.}\ \bibnamefont {Bekenstein}},\ }\bibfield  {title} {\bibinfo {title} {Exact solutions of einstein-conformal scalar equations},\ }\href@noop {} {\bibfield  {journal} {\bibinfo  {journal} {Annals of Physics}\ }\textbf {\bibinfo {volume} {82}},\ \bibinfo {pages} {535} (\bibinfo {year} {1974})}\BibitemShut {NoStop}%
\bibitem [{\citenamefont {Clifton}\ \emph {et~al.}(2012)\citenamefont {Clifton}, \citenamefont {Ferreira}, \citenamefont {Padilla},\ and\ \citenamefont {Skordis}}]{clifton2012modified}%
  \BibitemOpen
  \bibfield  {author} {\bibinfo {author} {\bibfnamefont {T.}~\bibnamefont {Clifton}}, \bibinfo {author} {\bibfnamefont {P.~G.}\ \bibnamefont {Ferreira}}, \bibinfo {author} {\bibfnamefont {A.}~\bibnamefont {Padilla}},\ and\ \bibinfo {author} {\bibfnamefont {C.}~\bibnamefont {Skordis}},\ }\bibfield  {title} {\bibinfo {title} {Modified gravity and cosmology},\ }\href@noop {} {\bibfield  {journal} {\bibinfo  {journal} {Physics reports}\ }\textbf {\bibinfo {volume} {513}},\ \bibinfo {pages} {1} (\bibinfo {year} {2012})}\BibitemShut {NoStop}%
\bibitem [{\citenamefont {Sotiriou}\ and\ \citenamefont {Faraoni}(2012)}]{sotiriou2012black}%
  \BibitemOpen
  \bibfield  {author} {\bibinfo {author} {\bibfnamefont {T.~P.}\ \bibnamefont {Sotiriou}}\ and\ \bibinfo {author} {\bibfnamefont {V.}~\bibnamefont {Faraoni}},\ }\bibfield  {title} {\bibinfo {title} {Black holes in scalar-tensor gravity},\ }\href@noop {} {\bibfield  {journal} {\bibinfo  {journal} {Physical Review Letters}\ }\textbf {\bibinfo {volume} {108}},\ \bibinfo {pages} {081103} (\bibinfo {year} {2012})}\BibitemShut {NoStop}%
\bibitem [{\citenamefont {Heisenberg}(2019)}]{heisenberg2019systematic}%
  \BibitemOpen
  \bibfield  {author} {\bibinfo {author} {\bibfnamefont {L.}~\bibnamefont {Heisenberg}},\ }\bibfield  {title} {\bibinfo {title} {A systematic approach to generalisations of general relativity and their cosmological implications},\ }\href@noop {} {\bibfield  {journal} {\bibinfo  {journal} {Physics Reports}\ }\textbf {\bibinfo {volume} {796}},\ \bibinfo {pages} {1} (\bibinfo {year} {2019})}\BibitemShut {NoStop}%
\bibitem [{\citenamefont {Saridakis}\ \emph {et~al.}(2021)\citenamefont {Saridakis}, \citenamefont {Lazkoz}, \citenamefont {Salzano}, \citenamefont {Moniz}, \citenamefont {Capozziello}, \citenamefont {Jim{\'e}nez}, \citenamefont {De~Laurentis},\ and\ \citenamefont {Olmo}}]{saridakis2021modified}%
  \BibitemOpen
  \bibfield  {author} {\bibinfo {author} {\bibfnamefont {E.~N.}\ \bibnamefont {Saridakis}}, \bibinfo {author} {\bibfnamefont {R.}~\bibnamefont {Lazkoz}}, \bibinfo {author} {\bibfnamefont {V.}~\bibnamefont {Salzano}}, \bibinfo {author} {\bibfnamefont {P.~V.}\ \bibnamefont {Moniz}}, \bibinfo {author} {\bibfnamefont {S.}~\bibnamefont {Capozziello}}, \bibinfo {author} {\bibfnamefont {J.~B.}\ \bibnamefont {Jim{\'e}nez}}, \bibinfo {author} {\bibfnamefont {M.}~\bibnamefont {De~Laurentis}},\ and\ \bibinfo {author} {\bibfnamefont {G.~J.}\ \bibnamefont {Olmo}},\ }\href@noop {} {\emph {\bibinfo {title} {Modified gravity and cosmology}}},\ \bibinfo {type} {Tech. Rep.}\ (\bibinfo  {institution} {Springer},\ \bibinfo {year} {2021})\BibitemShut {NoStop}%
\bibitem [{\citenamefont {Charmousis}(2009)}]{charmousis2009higher}%
  \BibitemOpen
  \bibfield  {author} {\bibinfo {author} {\bibfnamefont {C.}~\bibnamefont {Charmousis}},\ }\bibfield  {title} {\bibinfo {title} {Higher order gravity theories and their black hole solutions},\ }in\ \href@noop {} {\emph {\bibinfo {booktitle} {Physics of Black Holes: A Guided Tour}}}\ (\bibinfo  {publisher} {Springer},\ \bibinfo {year} {2009})\ pp.\ \bibinfo {pages} {299--346}\BibitemShut {NoStop}%
\bibitem [{\citenamefont {Brans}\ and\ \citenamefont {Dicke}(1961)}]{brans1961mach}%
  \BibitemOpen
  \bibfield  {author} {\bibinfo {author} {\bibfnamefont {C.}~\bibnamefont {Brans}}\ and\ \bibinfo {author} {\bibfnamefont {R.~H.}\ \bibnamefont {Dicke}},\ }\bibfield  {title} {\bibinfo {title} {Mach's principle and a relativistic theory of gravitation},\ }\href@noop {} {\bibfield  {journal} {\bibinfo  {journal} {Physical review}\ }\textbf {\bibinfo {volume} {124}},\ \bibinfo {pages} {925} (\bibinfo {year} {1961})}\BibitemShut {NoStop}%
\bibitem [{\citenamefont {Horndeski}(1974)}]{horndeski1974second}%
  \BibitemOpen
  \bibfield  {author} {\bibinfo {author} {\bibfnamefont {G.~W.}\ \bibnamefont {Horndeski}},\ }\bibfield  {title} {\bibinfo {title} {Second-order scalar-tensor field equations in a four-dimensional space},\ }\href@noop {} {\bibfield  {journal} {\bibinfo  {journal} {International Journal of Theoretical Physics}\ }\textbf {\bibinfo {volume} {10}},\ \bibinfo {pages} {363} (\bibinfo {year} {1974})}\BibitemShut {NoStop}%
\bibitem [{\citenamefont {Deffayet}\ \emph {et~al.}(2011)\citenamefont {Deffayet}, \citenamefont {Gao}, \citenamefont {Steer},\ and\ \citenamefont {Zahariade}}]{deffayet2011k}%
  \BibitemOpen
  \bibfield  {author} {\bibinfo {author} {\bibfnamefont {C.}~\bibnamefont {Deffayet}}, \bibinfo {author} {\bibfnamefont {X.}~\bibnamefont {Gao}}, \bibinfo {author} {\bibfnamefont {D.~A.}\ \bibnamefont {Steer}},\ and\ \bibinfo {author} {\bibfnamefont {G.}~\bibnamefont {Zahariade}},\ }\bibfield  {title} {\bibinfo {title} {From k-essence to generalized galileons},\ }\href@noop {} {\bibfield  {journal} {\bibinfo  {journal} {Physical Review D—Particles, Fields, Gravitation, and Cosmology}\ }\textbf {\bibinfo {volume} {84}},\ \bibinfo {pages} {064039} (\bibinfo {year} {2011})}\BibitemShut {NoStop}%
\bibitem [{\citenamefont {Kobayashi}\ \emph {et~al.}(2011)\citenamefont {Kobayashi}, \citenamefont {Yamaguchi},\ and\ \citenamefont {Yokoyama}}]{kobayashi2011generalized}%
  \BibitemOpen
  \bibfield  {author} {\bibinfo {author} {\bibfnamefont {T.}~\bibnamefont {Kobayashi}}, \bibinfo {author} {\bibfnamefont {M.}~\bibnamefont {Yamaguchi}},\ and\ \bibinfo {author} {\bibfnamefont {J.}~\bibnamefont {Yokoyama}},\ }\bibfield  {title} {\bibinfo {title} {Generalized g-inflation: —inflation with the most general second-order field equations—},\ }\href@noop {} {\bibfield  {journal} {\bibinfo  {journal} {Progress of Theoretical Physics}\ }\textbf {\bibinfo {volume} {126}},\ \bibinfo {pages} {511} (\bibinfo {year} {2011})}\BibitemShut {NoStop}%
\bibitem [{\citenamefont {Deffayet}\ and\ \citenamefont {Steer}(2013)}]{deffayet2013formal}%
  \BibitemOpen
  \bibfield  {author} {\bibinfo {author} {\bibfnamefont {C.}~\bibnamefont {Deffayet}}\ and\ \bibinfo {author} {\bibfnamefont {D.~A.}\ \bibnamefont {Steer}},\ }\bibfield  {title} {\bibinfo {title} {A formal introduction to horndeski and galileon theories and their generalizations},\ }\href@noop {} {\bibfield  {journal} {\bibinfo  {journal} {Classical and Quantum Gravity}\ }\textbf {\bibinfo {volume} {30}},\ \bibinfo {pages} {214006} (\bibinfo {year} {2013})}\BibitemShut {NoStop}%
\bibitem [{\citenamefont {Kobayashi}(2019)}]{kobayashi2019horndeski}%
  \BibitemOpen
  \bibfield  {author} {\bibinfo {author} {\bibfnamefont {T.}~\bibnamefont {Kobayashi}},\ }\bibfield  {title} {\bibinfo {title} {Horndeski theory and beyond: a review},\ }\href@noop {} {\bibfield  {journal} {\bibinfo  {journal} {Reports on Progress in Physics}\ }\textbf {\bibinfo {volume} {82}},\ \bibinfo {pages} {086901} (\bibinfo {year} {2019})}\BibitemShut {NoStop}%
\bibitem [{\citenamefont {Hui}\ and\ \citenamefont {Nicolis}(2013)}]{hui2013no}%
  \BibitemOpen
  \bibfield  {author} {\bibinfo {author} {\bibfnamefont {L.}~\bibnamefont {Hui}}\ and\ \bibinfo {author} {\bibfnamefont {A.}~\bibnamefont {Nicolis}},\ }\bibfield  {title} {\bibinfo {title} {No-hair theorem for the galileon},\ }\href@noop {} {\bibfield  {journal} {\bibinfo  {journal} {Physical Review Letters}\ }\textbf {\bibinfo {volume} {110}},\ \bibinfo {pages} {241104} (\bibinfo {year} {2013})}\BibitemShut {NoStop}%
\bibitem [{\citenamefont {Sotiriou}\ and\ \citenamefont {Zhou}(2014)}]{sotiriou2014black}%
  \BibitemOpen
  \bibfield  {author} {\bibinfo {author} {\bibfnamefont {T.~P.}\ \bibnamefont {Sotiriou}}\ and\ \bibinfo {author} {\bibfnamefont {S.-Y.}\ \bibnamefont {Zhou}},\ }\bibfield  {title} {\bibinfo {title} {Black hole hair in generalized scalar-tensor gravity},\ }\href@noop {} {\bibfield  {journal} {\bibinfo  {journal} {Physical Review Letters}\ }\textbf {\bibinfo {volume} {112}},\ \bibinfo {pages} {251102} (\bibinfo {year} {2014})}\BibitemShut {NoStop}%
\bibitem [{\citenamefont {Zwiebach}(1985)}]{zwiebach1985curvature}%
  \BibitemOpen
  \bibfield  {author} {\bibinfo {author} {\bibfnamefont {B.}~\bibnamefont {Zwiebach}},\ }\bibfield  {title} {\bibinfo {title} {Curvature squared terms and string theories},\ }\href@noop {} {\bibfield  {journal} {\bibinfo  {journal} {Physics Letters B}\ }\textbf {\bibinfo {volume} {156}},\ \bibinfo {pages} {315} (\bibinfo {year} {1985})}\BibitemShut {NoStop}%
\bibitem [{\citenamefont {Nepomechie}(1985)}]{nepomechie1985low}%
  \BibitemOpen
  \bibfield  {author} {\bibinfo {author} {\bibfnamefont {R.~I.}\ \bibnamefont {Nepomechie}},\ }\bibfield  {title} {\bibinfo {title} {Low-energy limit of strings},\ }\href@noop {} {\bibfield  {journal} {\bibinfo  {journal} {Physical Review D}\ }\textbf {\bibinfo {volume} {32}},\ \bibinfo {pages} {3201} (\bibinfo {year} {1985})}\BibitemShut {NoStop}%
\bibitem [{\citenamefont {Candelas}\ \emph {et~al.}(1985)\citenamefont {Candelas}, \citenamefont {Horowitz}, \citenamefont {Strominger},\ and\ \citenamefont {Witten}}]{candelas1985vacuum}%
  \BibitemOpen
  \bibfield  {author} {\bibinfo {author} {\bibfnamefont {P.}~\bibnamefont {Candelas}}, \bibinfo {author} {\bibfnamefont {G.~T.}\ \bibnamefont {Horowitz}}, \bibinfo {author} {\bibfnamefont {A.}~\bibnamefont {Strominger}},\ and\ \bibinfo {author} {\bibfnamefont {E.}~\bibnamefont {Witten}},\ }\bibfield  {title} {\bibinfo {title} {Vacuum configurations for superstrings},\ }\href@noop {} {\bibfield  {journal} {\bibinfo  {journal} {Nuclear Physics B}\ }\textbf {\bibinfo {volume} {258}},\ \bibinfo {pages} {46} (\bibinfo {year} {1985})}\BibitemShut {NoStop}%
\bibitem [{\citenamefont {Callan}\ \emph {et~al.}(1986)\citenamefont {Callan}, \citenamefont {Klebanov},\ and\ \citenamefont {Perry}}]{callan1986string}%
  \BibitemOpen
  \bibfield  {author} {\bibinfo {author} {\bibfnamefont {C.~G.}\ \bibnamefont {Callan}}, \bibinfo {author} {\bibfnamefont {I.~R.}\ \bibnamefont {Klebanov}},\ and\ \bibinfo {author} {\bibfnamefont {M.}~\bibnamefont {Perry}},\ }\bibfield  {title} {\bibinfo {title} {String theory effective actions},\ }\href@noop {} {\bibfield  {journal} {\bibinfo  {journal} {Nuclear Physics B}\ }\textbf {\bibinfo {volume} {278}},\ \bibinfo {pages} {78} (\bibinfo {year} {1986})}\BibitemShut {NoStop}%
\bibitem [{\citenamefont {Gross}\ and\ \citenamefont {Sloan}(1987)}]{gross1987quartic}%
  \BibitemOpen
  \bibfield  {author} {\bibinfo {author} {\bibfnamefont {D.~J.}\ \bibnamefont {Gross}}\ and\ \bibinfo {author} {\bibfnamefont {J.~H.}\ \bibnamefont {Sloan}},\ }\bibfield  {title} {\bibinfo {title} {The quartic effective action for the heterotic string},\ }\href@noop {} {\bibfield  {journal} {\bibinfo  {journal} {Nuclear Physics B}\ }\textbf {\bibinfo {volume} {291}},\ \bibinfo {pages} {41} (\bibinfo {year} {1987})}\BibitemShut {NoStop}%
\bibitem [{\citenamefont {Lovelock}(1970)}]{lovelock1970divergence}%
  \BibitemOpen
  \bibfield  {author} {\bibinfo {author} {\bibfnamefont {D.}~\bibnamefont {Lovelock}},\ }\bibfield  {title} {\bibinfo {title} {Divergence-free tensorial concomitants},\ }\href@noop {} {\bibfield  {journal} {\bibinfo  {journal} {Aequationes mathematicae}\ }\textbf {\bibinfo {volume} {4}},\ \bibinfo {pages} {127} (\bibinfo {year} {1970})}\BibitemShut {NoStop}%
\bibitem [{\citenamefont {Lovelock}(1971)}]{lovelock1971einstein}%
  \BibitemOpen
  \bibfield  {author} {\bibinfo {author} {\bibfnamefont {D.}~\bibnamefont {Lovelock}},\ }\bibfield  {title} {\bibinfo {title} {The einstein tensor and its generalizations},\ }\href@noop {} {\bibfield  {journal} {\bibinfo  {journal} {Journal of Mathematical Physics}\ }\textbf {\bibinfo {volume} {12}},\ \bibinfo {pages} {498} (\bibinfo {year} {1971})}\BibitemShut {NoStop}%
\bibitem [{\citenamefont {Mignemi}\ and\ \citenamefont {Stewart}(1993)}]{mignemi1993charged}%
  \BibitemOpen
  \bibfield  {author} {\bibinfo {author} {\bibfnamefont {S.}~\bibnamefont {Mignemi}}\ and\ \bibinfo {author} {\bibfnamefont {N.}~\bibnamefont {Stewart}},\ }\bibfield  {title} {\bibinfo {title} {Charged black holes in effective string theory},\ }\href@noop {} {\bibfield  {journal} {\bibinfo  {journal} {Physical Review D}\ }\textbf {\bibinfo {volume} {47}},\ \bibinfo {pages} {5259} (\bibinfo {year} {1993})}\BibitemShut {NoStop}%
\bibitem [{\citenamefont {Kanti}\ \emph {et~al.}(1996)\citenamefont {Kanti}, \citenamefont {Mavromatos}, \citenamefont {Rizos}, \citenamefont {Tamvakis},\ and\ \citenamefont {Winstanley}}]{kanti1996dilatonic}%
  \BibitemOpen
  \bibfield  {author} {\bibinfo {author} {\bibfnamefont {P.}~\bibnamefont {Kanti}}, \bibinfo {author} {\bibfnamefont {N.~E.}\ \bibnamefont {Mavromatos}}, \bibinfo {author} {\bibfnamefont {J.}~\bibnamefont {Rizos}}, \bibinfo {author} {\bibfnamefont {K.}~\bibnamefont {Tamvakis}},\ and\ \bibinfo {author} {\bibfnamefont {E.}~\bibnamefont {Winstanley}},\ }\bibfield  {title} {\bibinfo {title} {Dilatonic black holes in higher curvature string gravity},\ }\href@noop {} {\bibfield  {journal} {\bibinfo  {journal} {Physical Review D}\ }\textbf {\bibinfo {volume} {54}},\ \bibinfo {pages} {5049} (\bibinfo {year} {1996})}\BibitemShut {NoStop}%
\bibitem [{\citenamefont {Antoniou}\ \emph {et~al.}(2018{\natexlab{a}})\citenamefont {Antoniou}, \citenamefont {Bakopoulos},\ and\ \citenamefont {Kanti}}]{antoniou2018evasion}%
  \BibitemOpen
  \bibfield  {author} {\bibinfo {author} {\bibfnamefont {G.}~\bibnamefont {Antoniou}}, \bibinfo {author} {\bibfnamefont {A.}~\bibnamefont {Bakopoulos}},\ and\ \bibinfo {author} {\bibfnamefont {P.}~\bibnamefont {Kanti}},\ }\bibfield  {title} {\bibinfo {title} {Evasion of no-hair theorems and novel black-hole solutions in gauss-bonnet theories},\ }\href@noop {} {\bibfield  {journal} {\bibinfo  {journal} {Physical review letters}\ }\textbf {\bibinfo {volume} {120}},\ \bibinfo {pages} {131102} (\bibinfo {year} {2018}{\natexlab{a}})}\BibitemShut {NoStop}%
\bibitem [{\citenamefont {Antoniou}\ \emph {et~al.}(2018{\natexlab{b}})\citenamefont {Antoniou}, \citenamefont {Bakopoulos},\ and\ \citenamefont {Kanti}}]{antoniou2018black}%
  \BibitemOpen
  \bibfield  {author} {\bibinfo {author} {\bibfnamefont {G.}~\bibnamefont {Antoniou}}, \bibinfo {author} {\bibfnamefont {A.}~\bibnamefont {Bakopoulos}},\ and\ \bibinfo {author} {\bibfnamefont {P.}~\bibnamefont {Kanti}},\ }\bibfield  {title} {\bibinfo {title} {Black-hole solutions with scalar hair in einstein-scalar-gauss-bonnet theories},\ }\href@noop {} {\bibfield  {journal} {\bibinfo  {journal} {Physical Review D}\ }\textbf {\bibinfo {volume} {97}},\ \bibinfo {pages} {084037} (\bibinfo {year} {2018}{\natexlab{b}})}\BibitemShut {NoStop}%
\bibitem [{\citenamefont {Gibbons}\ and\ \citenamefont {Maeda}(1988)}]{gibbons1988black}%
  \BibitemOpen
  \bibfield  {author} {\bibinfo {author} {\bibfnamefont {G.~W.}\ \bibnamefont {Gibbons}}\ and\ \bibinfo {author} {\bibfnamefont {K.-i.}\ \bibnamefont {Maeda}},\ }\bibfield  {title} {\bibinfo {title} {Black holes and membranes in higher-dimensional theories with dilaton fields},\ }\href@noop {} {\bibfield  {journal} {\bibinfo  {journal} {Nuclear Physics B}\ }\textbf {\bibinfo {volume} {298}},\ \bibinfo {pages} {741} (\bibinfo {year} {1988})}\BibitemShut {NoStop}%
\bibitem [{\citenamefont {Garfinkle}\ \emph {et~al.}(1991)\citenamefont {Garfinkle}, \citenamefont {Horowitz},\ and\ \citenamefont {Strominger}}]{garfinkle1991charged}%
  \BibitemOpen
  \bibfield  {author} {\bibinfo {author} {\bibfnamefont {D.}~\bibnamefont {Garfinkle}}, \bibinfo {author} {\bibfnamefont {G.~T.}\ \bibnamefont {Horowitz}},\ and\ \bibinfo {author} {\bibfnamefont {A.}~\bibnamefont {Strominger}},\ }\bibfield  {title} {\bibinfo {title} {Charged black holes in string theory},\ }\href@noop {} {\bibfield  {journal} {\bibinfo  {journal} {Physical Review D}\ }\textbf {\bibinfo {volume} {43}},\ \bibinfo {pages} {3140} (\bibinfo {year} {1991})}\BibitemShut {NoStop}%
\bibitem [{\citenamefont {Doneva}\ \emph {et~al.}(2024{\natexlab{a}})\citenamefont {Doneva}, \citenamefont {Ramazano{\u{g}}lu}, \citenamefont {Silva}, \citenamefont {Sotiriou},\ and\ \citenamefont {Yazadjiev}}]{doneva2024spontaneous}%
  \BibitemOpen
  \bibfield  {author} {\bibinfo {author} {\bibfnamefont {D.~D.}\ \bibnamefont {Doneva}}, \bibinfo {author} {\bibfnamefont {F.~M.}\ \bibnamefont {Ramazano{\u{g}}lu}}, \bibinfo {author} {\bibfnamefont {H.~O.}\ \bibnamefont {Silva}}, \bibinfo {author} {\bibfnamefont {T.~P.}\ \bibnamefont {Sotiriou}},\ and\ \bibinfo {author} {\bibfnamefont {S.~S.}\ \bibnamefont {Yazadjiev}},\ }\bibfield  {title} {\bibinfo {title} {Spontaneous scalarization},\ }\href@noop {} {\bibfield  {journal} {\bibinfo  {journal} {Reviews of Modern Physics}\ }\textbf {\bibinfo {volume} {96}},\ \bibinfo {pages} {015004} (\bibinfo {year} {2024}{\natexlab{a}})}\BibitemShut {NoStop}%
\bibitem [{\citenamefont {Damour}\ and\ \citenamefont {Esposito-Farese}(1993)}]{damour1993nonperturbative}%
  \BibitemOpen
  \bibfield  {author} {\bibinfo {author} {\bibfnamefont {T.}~\bibnamefont {Damour}}\ and\ \bibinfo {author} {\bibfnamefont {G.}~\bibnamefont {Esposito-Farese}},\ }\bibfield  {title} {\bibinfo {title} {Nonperturbative strong-field effects in tensor-scalar theories of gravitation},\ }\href@noop {} {\bibfield  {journal} {\bibinfo  {journal} {Physical Review Letters}\ }\textbf {\bibinfo {volume} {70}},\ \bibinfo {pages} {2220} (\bibinfo {year} {1993})}\BibitemShut {NoStop}%
\bibitem [{\citenamefont {Doneva}\ and\ \citenamefont {Yazadjiev}(2018)}]{doneva2018new}%
  \BibitemOpen
  \bibfield  {author} {\bibinfo {author} {\bibfnamefont {D.~D.}\ \bibnamefont {Doneva}}\ and\ \bibinfo {author} {\bibfnamefont {S.~S.}\ \bibnamefont {Yazadjiev}},\ }\bibfield  {title} {\bibinfo {title} {New gauss-bonnet black holes with curvature-induced scalarization in extended scalar-tensor theories},\ }\href@noop {} {\bibfield  {journal} {\bibinfo  {journal} {Physical review letters}\ }\textbf {\bibinfo {volume} {120}},\ \bibinfo {pages} {131103} (\bibinfo {year} {2018})}\BibitemShut {NoStop}%
\bibitem [{\citenamefont {Silva}\ \emph {et~al.}(2018)\citenamefont {Silva}, \citenamefont {Sakstein}, \citenamefont {Gualtieri}, \citenamefont {Sotiriou},\ and\ \citenamefont {Berti}}]{Silva2018SpontaneousCoupling}%
  \BibitemOpen
  \bibfield  {author} {\bibinfo {author} {\bibfnamefont {H.~O.}\ \bibnamefont {Silva}}, \bibinfo {author} {\bibfnamefont {J.}~\bibnamefont {Sakstein}}, \bibinfo {author} {\bibfnamefont {L.}~\bibnamefont {Gualtieri}}, \bibinfo {author} {\bibfnamefont {T.~P.}\ \bibnamefont {Sotiriou}},\ and\ \bibinfo {author} {\bibfnamefont {E.}~\bibnamefont {Berti}},\ }\bibfield  {title} {\bibinfo {title} {{Spontaneous scalarization of black holes and compact stars from a Gauss-Bonnet coupling}},\ }\bibfield  {journal} {\bibinfo  {journal} {Physical Review Letters}\ }\textbf {\bibinfo {volume} {120}},\ \href {https://doi.org/10.1103/PhysRevLett.120.131104} {10.1103/PhysRevLett.120.131104} (\bibinfo {year} {2018})\BibitemShut {NoStop}%
\bibitem [{\citenamefont {Herdeiro}\ \emph {et~al.}(2018)\citenamefont {Herdeiro}, \citenamefont {Radu}, \citenamefont {Sanchis-Gual},\ and\ \citenamefont {Font}}]{herdeiro2018spontaneous}%
  \BibitemOpen
  \bibfield  {author} {\bibinfo {author} {\bibfnamefont {C.~A.}\ \bibnamefont {Herdeiro}}, \bibinfo {author} {\bibfnamefont {E.}~\bibnamefont {Radu}}, \bibinfo {author} {\bibfnamefont {N.}~\bibnamefont {Sanchis-Gual}},\ and\ \bibinfo {author} {\bibfnamefont {J.~A.}\ \bibnamefont {Font}},\ }\bibfield  {title} {\bibinfo {title} {Spontaneous scalarization of charged black holes},\ }\href@noop {} {\bibfield  {journal} {\bibinfo  {journal} {Physical review letters}\ }\textbf {\bibinfo {volume} {121}},\ \bibinfo {pages} {101102} (\bibinfo {year} {2018})}\BibitemShut {NoStop}%
\bibitem [{\citenamefont {Myung}\ and\ \citenamefont {Zou}(2019{\natexlab{a}})}]{myung2019instability}%
  \BibitemOpen
  \bibfield  {author} {\bibinfo {author} {\bibfnamefont {Y.~S.}\ \bibnamefont {Myung}}\ and\ \bibinfo {author} {\bibfnamefont {D.-C.}\ \bibnamefont {Zou}},\ }\bibfield  {title} {\bibinfo {title} {Instability of reissner--nordstr{\"o}m black hole in einstein-maxwell-scalar theory},\ }\href@noop {} {\bibfield  {journal} {\bibinfo  {journal} {The European Physical Journal C}\ }\textbf {\bibinfo {volume} {79}},\ \bibinfo {pages} {1} (\bibinfo {year} {2019}{\natexlab{a}})}\BibitemShut {NoStop}%
\bibitem [{\citenamefont {Doneva}\ and\ \citenamefont {Yazadjiev}(2021)}]{doneva2021spontaneously}%
  \BibitemOpen
  \bibfield  {author} {\bibinfo {author} {\bibfnamefont {D.~D.}\ \bibnamefont {Doneva}}\ and\ \bibinfo {author} {\bibfnamefont {S.~S.}\ \bibnamefont {Yazadjiev}},\ }\bibfield  {title} {\bibinfo {title} {Spontaneously scalarized black holes in dynamical chern-simons gravity: dynamics and equilibrium solutions},\ }\href@noop {} {\bibfield  {journal} {\bibinfo  {journal} {Physical Review D}\ }\textbf {\bibinfo {volume} {103}},\ \bibinfo {pages} {083007} (\bibinfo {year} {2021})}\BibitemShut {NoStop}%
\bibitem [{\citenamefont {Gao}\ \emph {et~al.}(2019)\citenamefont {Gao}, \citenamefont {Huang},\ and\ \citenamefont {Liu}}]{gao2019scalar}%
  \BibitemOpen
  \bibfield  {author} {\bibinfo {author} {\bibfnamefont {Y.-X.}\ \bibnamefont {Gao}}, \bibinfo {author} {\bibfnamefont {Y.}~\bibnamefont {Huang}},\ and\ \bibinfo {author} {\bibfnamefont {D.-J.}\ \bibnamefont {Liu}},\ }\bibfield  {title} {\bibinfo {title} {Scalar perturbations on the background of kerr black holes in the quadratic dynamical chern-simons gravity},\ }\href@noop {} {\bibfield  {journal} {\bibinfo  {journal} {Physical Review D}\ }\textbf {\bibinfo {volume} {99}},\ \bibinfo {pages} {044020} (\bibinfo {year} {2019})}\BibitemShut {NoStop}%
\bibitem [{\citenamefont {Myung}\ and\ \citenamefont {Zou}(2021)}]{myung2021onset}%
  \BibitemOpen
  \bibfield  {author} {\bibinfo {author} {\bibfnamefont {Y.~S.}\ \bibnamefont {Myung}}\ and\ \bibinfo {author} {\bibfnamefont {D.-C.}\ \bibnamefont {Zou}},\ }\bibfield  {title} {\bibinfo {title} {Onset of rotating scalarized black holes in einstein-chern-simons-scalar theory},\ }\href@noop {} {\bibfield  {journal} {\bibinfo  {journal} {Physics Letters B}\ }\textbf {\bibinfo {volume} {814}},\ \bibinfo {pages} {136081} (\bibinfo {year} {2021})}\BibitemShut {NoStop}%
\bibitem [{\citenamefont {Minamitsuji}\ and\ \citenamefont {Ikeda}(2019{\natexlab{a}})}]{minamitsuji2019scalarized}%
  \BibitemOpen
  \bibfield  {author} {\bibinfo {author} {\bibfnamefont {M.}~\bibnamefont {Minamitsuji}}\ and\ \bibinfo {author} {\bibfnamefont {T.}~\bibnamefont {Ikeda}},\ }\bibfield  {title} {\bibinfo {title} {Scalarized black holes in the presence of the coupling to gauss-bonnet gravity},\ }\href@noop {} {\bibfield  {journal} {\bibinfo  {journal} {Physical Review D}\ }\textbf {\bibinfo {volume} {99}},\ \bibinfo {pages} {044017} (\bibinfo {year} {2019}{\natexlab{a}})}\BibitemShut {NoStop}%
\bibitem [{\citenamefont {Silva}\ \emph {et~al.}(2021)\citenamefont {Silva}, \citenamefont {Witek}, \citenamefont {Elley},\ and\ \citenamefont {Yunes}}]{silva2021dynamical}%
  \BibitemOpen
  \bibfield  {author} {\bibinfo {author} {\bibfnamefont {H.~O.}\ \bibnamefont {Silva}}, \bibinfo {author} {\bibfnamefont {H.}~\bibnamefont {Witek}}, \bibinfo {author} {\bibfnamefont {M.}~\bibnamefont {Elley}},\ and\ \bibinfo {author} {\bibfnamefont {N.}~\bibnamefont {Yunes}},\ }\bibfield  {title} {\bibinfo {title} {Dynamical descalarization in binary black hole mergers},\ }\href@noop {} {\bibfield  {journal} {\bibinfo  {journal} {Physical review letters}\ }\textbf {\bibinfo {volume} {127}},\ \bibinfo {pages} {031101} (\bibinfo {year} {2021})}\BibitemShut {NoStop}%
\bibitem [{\citenamefont {Andreou}\ \emph {et~al.}(2019)\citenamefont {Andreou}, \citenamefont {Franchini}, \citenamefont {Ventagli},\ and\ \citenamefont {Sotiriou}}]{andreou2019spontaneous}%
  \BibitemOpen
  \bibfield  {author} {\bibinfo {author} {\bibfnamefont {N.}~\bibnamefont {Andreou}}, \bibinfo {author} {\bibfnamefont {N.}~\bibnamefont {Franchini}}, \bibinfo {author} {\bibfnamefont {G.}~\bibnamefont {Ventagli}},\ and\ \bibinfo {author} {\bibfnamefont {T.~P.}\ \bibnamefont {Sotiriou}},\ }\bibfield  {title} {\bibinfo {title} {Spontaneous scalarization in generalized scalar-tensor theory},\ }\href@noop {} {\bibfield  {journal} {\bibinfo  {journal} {Physical Review D}\ }\textbf {\bibinfo {volume} {99}},\ \bibinfo {pages} {124022} (\bibinfo {year} {2019})}\BibitemShut {NoStop}%
\bibitem [{\citenamefont {Doneva}\ \emph {et~al.}(2019)\citenamefont {Doneva}, \citenamefont {Staykov},\ and\ \citenamefont {Yazadjiev}}]{doneva2019gauss}%
  \BibitemOpen
  \bibfield  {author} {\bibinfo {author} {\bibfnamefont {D.~D.}\ \bibnamefont {Doneva}}, \bibinfo {author} {\bibfnamefont {K.~V.}\ \bibnamefont {Staykov}},\ and\ \bibinfo {author} {\bibfnamefont {S.~S.}\ \bibnamefont {Yazadjiev}},\ }\bibfield  {title} {\bibinfo {title} {Gauss-bonnet black holes with a massive scalar field},\ }\href@noop {} {\bibfield  {journal} {\bibinfo  {journal} {Physical Review D}\ }\textbf {\bibinfo {volume} {99}},\ \bibinfo {pages} {104045} (\bibinfo {year} {2019})}\BibitemShut {NoStop}%
\bibitem [{\citenamefont {Bl{\'a}zquez-Salcedo}\ \emph {et~al.}(2020{\natexlab{a}})\citenamefont {Bl{\'a}zquez-Salcedo}, \citenamefont {Doneva}, \citenamefont {Kahlen}, \citenamefont {Kunz}, \citenamefont {Nedkova},\ and\ \citenamefont {Yazadjiev}}]{blazquez2020axial}%
  \BibitemOpen
  \bibfield  {author} {\bibinfo {author} {\bibfnamefont {J.~L.}\ \bibnamefont {Bl{\'a}zquez-Salcedo}}, \bibinfo {author} {\bibfnamefont {D.~D.}\ \bibnamefont {Doneva}}, \bibinfo {author} {\bibfnamefont {S.}~\bibnamefont {Kahlen}}, \bibinfo {author} {\bibfnamefont {J.}~\bibnamefont {Kunz}}, \bibinfo {author} {\bibfnamefont {P.}~\bibnamefont {Nedkova}},\ and\ \bibinfo {author} {\bibfnamefont {S.~S.}\ \bibnamefont {Yazadjiev}},\ }\bibfield  {title} {\bibinfo {title} {Axial perturbations of the scalarized einstein-gauss-bonnet black holes},\ }\href@noop {} {\bibfield  {journal} {\bibinfo  {journal} {Physical Review D}\ }\textbf {\bibinfo {volume} {101}},\ \bibinfo {pages} {104006} (\bibinfo {year} {2020}{\natexlab{a}})}\BibitemShut {NoStop}%
\bibitem [{\citenamefont {Bl{\'a}zquez-Salcedo}\ \emph {et~al.}(2020{\natexlab{b}})\citenamefont {Bl{\'a}zquez-Salcedo}, \citenamefont {Doneva}, \citenamefont {Kahlen}, \citenamefont {Kunz}, \citenamefont {Nedkova},\ and\ \citenamefont {Yazadjiev}}]{blazquez2020polar}%
  \BibitemOpen
  \bibfield  {author} {\bibinfo {author} {\bibfnamefont {J.~L.}\ \bibnamefont {Bl{\'a}zquez-Salcedo}}, \bibinfo {author} {\bibfnamefont {D.~D.}\ \bibnamefont {Doneva}}, \bibinfo {author} {\bibfnamefont {S.}~\bibnamefont {Kahlen}}, \bibinfo {author} {\bibfnamefont {J.}~\bibnamefont {Kunz}}, \bibinfo {author} {\bibfnamefont {P.}~\bibnamefont {Nedkova}},\ and\ \bibinfo {author} {\bibfnamefont {S.~S.}\ \bibnamefont {Yazadjiev}},\ }\bibfield  {title} {\bibinfo {title} {Polar quasinormal modes of the scalarized einstein-gauss-bonnet black holes},\ }\href@noop {} {\bibfield  {journal} {\bibinfo  {journal} {Physical Review D}\ }\textbf {\bibinfo {volume} {102}},\ \bibinfo {pages} {024086} (\bibinfo {year} {2020}{\natexlab{b}})}\BibitemShut {NoStop}%
\bibitem [{\citenamefont {Brihaye}\ and\ \citenamefont {Ducobu}(2019)}]{brihaye2019hairy}%
  \BibitemOpen
  \bibfield  {author} {\bibinfo {author} {\bibfnamefont {Y.}~\bibnamefont {Brihaye}}\ and\ \bibinfo {author} {\bibfnamefont {L.}~\bibnamefont {Ducobu}},\ }\bibfield  {title} {\bibinfo {title} {Hairy black holes, boson stars and non-minimal coupling to curvature invariants},\ }\href@noop {} {\bibfield  {journal} {\bibinfo  {journal} {Physics Letters B}\ }\textbf {\bibinfo {volume} {795}},\ \bibinfo {pages} {135} (\bibinfo {year} {2019})}\BibitemShut {NoStop}%
\bibitem [{\citenamefont {Juli{\'e}}\ \emph {et~al.}(2022)\citenamefont {Juli{\'e}}, \citenamefont {Silva}, \citenamefont {Berti},\ and\ \citenamefont {Yunes}}]{julie2022black}%
  \BibitemOpen
  \bibfield  {author} {\bibinfo {author} {\bibfnamefont {F.-L.}\ \bibnamefont {Juli{\'e}}}, \bibinfo {author} {\bibfnamefont {H.~O.}\ \bibnamefont {Silva}}, \bibinfo {author} {\bibfnamefont {E.}~\bibnamefont {Berti}},\ and\ \bibinfo {author} {\bibfnamefont {N.}~\bibnamefont {Yunes}},\ }\bibfield  {title} {\bibinfo {title} {Black hole sensitivities in einstein-scalar-gauss-bonnet gravity},\ }\href@noop {} {\bibfield  {journal} {\bibinfo  {journal} {Physical Review D}\ }\textbf {\bibinfo {volume} {105}},\ \bibinfo {pages} {124031} (\bibinfo {year} {2022})}\BibitemShut {NoStop}%
\bibitem [{\citenamefont {Doneva}\ \emph {et~al.}(2022)\citenamefont {Doneva}, \citenamefont {Va{\~n}{\'o}-Vi{\~n}uales},\ and\ \citenamefont {Yazadjiev}}]{doneva2022dynamical}%
  \BibitemOpen
  \bibfield  {author} {\bibinfo {author} {\bibfnamefont {D.~D.}\ \bibnamefont {Doneva}}, \bibinfo {author} {\bibfnamefont {A.}~\bibnamefont {Va{\~n}{\'o}-Vi{\~n}uales}},\ and\ \bibinfo {author} {\bibfnamefont {S.~S.}\ \bibnamefont {Yazadjiev}},\ }\bibfield  {title} {\bibinfo {title} {Dynamical descalarization with a jump during a black hole merger},\ }\href@noop {} {\bibfield  {journal} {\bibinfo  {journal} {Physical Review D}\ }\textbf {\bibinfo {volume} {106}},\ \bibinfo {pages} {L061502} (\bibinfo {year} {2022})}\BibitemShut {NoStop}%
\bibitem [{\citenamefont {Minamitsuji}\ and\ \citenamefont {Ikeda}(2019{\natexlab{b}})}]{minamitsuji2019spontaneous}%
  \BibitemOpen
  \bibfield  {author} {\bibinfo {author} {\bibfnamefont {M.}~\bibnamefont {Minamitsuji}}\ and\ \bibinfo {author} {\bibfnamefont {T.}~\bibnamefont {Ikeda}},\ }\bibfield  {title} {\bibinfo {title} {Spontaneous scalarization of black holes in the horndeski theory},\ }\href@noop {} {\bibfield  {journal} {\bibinfo  {journal} {Physical Review D}\ }\textbf {\bibinfo {volume} {99}},\ \bibinfo {pages} {104069} (\bibinfo {year} {2019}{\natexlab{b}})}\BibitemShut {NoStop}%
\bibitem [{\citenamefont {Wong}\ \emph {et~al.}(2022)\citenamefont {Wong}, \citenamefont {Herdeiro},\ and\ \citenamefont {Radu}}]{wong2022constraining}%
  \BibitemOpen
  \bibfield  {author} {\bibinfo {author} {\bibfnamefont {L.~K.}\ \bibnamefont {Wong}}, \bibinfo {author} {\bibfnamefont {C.~A.}\ \bibnamefont {Herdeiro}},\ and\ \bibinfo {author} {\bibfnamefont {E.}~\bibnamefont {Radu}},\ }\bibfield  {title} {\bibinfo {title} {Constraining spontaneous black hole scalarization in scalar-tensor-gauss-bonnet theories with current gravitational-wave data},\ }\href@noop {} {\bibfield  {journal} {\bibinfo  {journal} {Physical Review D}\ }\textbf {\bibinfo {volume} {106}},\ \bibinfo {pages} {024008} (\bibinfo {year} {2022})}\BibitemShut {NoStop}%
\bibitem [{\citenamefont {Brihaye}\ \emph {et~al.}(2020)\citenamefont {Brihaye}, \citenamefont {Herdeiro},\ and\ \citenamefont {Radu}}]{brihaye2020black}%
  \BibitemOpen
  \bibfield  {author} {\bibinfo {author} {\bibfnamefont {Y.}~\bibnamefont {Brihaye}}, \bibinfo {author} {\bibfnamefont {C.}~\bibnamefont {Herdeiro}},\ and\ \bibinfo {author} {\bibfnamefont {E.}~\bibnamefont {Radu}},\ }\bibfield  {title} {\bibinfo {title} {Black hole spontaneous scalarisation with a positive cosmological constant},\ }\href@noop {} {\bibfield  {journal} {\bibinfo  {journal} {Physics Letters B}\ }\textbf {\bibinfo {volume} {802}},\ \bibinfo {pages} {135269} (\bibinfo {year} {2020})}\BibitemShut {NoStop}%
\bibitem [{\citenamefont {Doneva}\ and\ \citenamefont {Yazadjiev}(2022)}]{doneva2022beyond}%
  \BibitemOpen
  \bibfield  {author} {\bibinfo {author} {\bibfnamefont {D.~D.}\ \bibnamefont {Doneva}}\ and\ \bibinfo {author} {\bibfnamefont {S.~S.}\ \bibnamefont {Yazadjiev}},\ }\bibfield  {title} {\bibinfo {title} {Beyond the spontaneous scalarization: New fully nonlinear mechanism for the formation of scalarized black holes and its dynamical development},\ }\href@noop {} {\bibfield  {journal} {\bibinfo  {journal} {Physical Review D}\ }\textbf {\bibinfo {volume} {105}},\ \bibinfo {pages} {L041502} (\bibinfo {year} {2022})}\BibitemShut {NoStop}%
\bibitem [{\citenamefont {Antoniou}\ \emph {et~al.}(2021{\natexlab{a}})\citenamefont {Antoniou}, \citenamefont {Bordin},\ and\ \citenamefont {Sotiriou}}]{antoniou2021compact}%
  \BibitemOpen
  \bibfield  {author} {\bibinfo {author} {\bibfnamefont {G.}~\bibnamefont {Antoniou}}, \bibinfo {author} {\bibfnamefont {L.}~\bibnamefont {Bordin}},\ and\ \bibinfo {author} {\bibfnamefont {T.~P.}\ \bibnamefont {Sotiriou}},\ }\bibfield  {title} {\bibinfo {title} {Compact object scalarization with general relativity as a cosmic attractor},\ }\href@noop {} {\bibfield  {journal} {\bibinfo  {journal} {Physical Review D}\ }\textbf {\bibinfo {volume} {103}},\ \bibinfo {pages} {024012} (\bibinfo {year} {2021}{\natexlab{a}})}\BibitemShut {NoStop}%
\bibitem [{\citenamefont {Ventagli}\ \emph {et~al.}(2020)\citenamefont {Ventagli}, \citenamefont {Leh{\'e}bel},\ and\ \citenamefont {Sotiriou}}]{ventagli2020onset}%
  \BibitemOpen
  \bibfield  {author} {\bibinfo {author} {\bibfnamefont {G.}~\bibnamefont {Ventagli}}, \bibinfo {author} {\bibfnamefont {A.}~\bibnamefont {Leh{\'e}bel}},\ and\ \bibinfo {author} {\bibfnamefont {T.~P.}\ \bibnamefont {Sotiriou}},\ }\bibfield  {title} {\bibinfo {title} {Onset of spontaneous scalarization in generalized scalar-tensor theories},\ }\href@noop {} {\bibfield  {journal} {\bibinfo  {journal} {Physical Review D}\ }\textbf {\bibinfo {volume} {102}},\ \bibinfo {pages} {024050} (\bibinfo {year} {2020})}\BibitemShut {NoStop}%
\bibitem [{\citenamefont {Cunha}\ \emph {et~al.}(2019)\citenamefont {Cunha}, \citenamefont {Herdeiro},\ and\ \citenamefont {Radu}}]{cunha2019spontaneously}%
  \BibitemOpen
  \bibfield  {author} {\bibinfo {author} {\bibfnamefont {P.~V.}\ \bibnamefont {Cunha}}, \bibinfo {author} {\bibfnamefont {C.~A.}\ \bibnamefont {Herdeiro}},\ and\ \bibinfo {author} {\bibfnamefont {E.}~\bibnamefont {Radu}},\ }\bibfield  {title} {\bibinfo {title} {Spontaneously scalarized kerr black holes in extended scalar-tensor--gauss-bonnet gravity},\ }\href@noop {} {\bibfield  {journal} {\bibinfo  {journal} {Physical Review Letters}\ }\textbf {\bibinfo {volume} {123}},\ \bibinfo {pages} {011101} (\bibinfo {year} {2019})}\BibitemShut {NoStop}%
\bibitem [{\citenamefont {Collodel}\ \emph {et~al.}(2020)\citenamefont {Collodel}, \citenamefont {Kleihaus}, \citenamefont {Kunz},\ and\ \citenamefont {Berti}}]{collodel2020spinning}%
  \BibitemOpen
  \bibfield  {author} {\bibinfo {author} {\bibfnamefont {L.~G.}\ \bibnamefont {Collodel}}, \bibinfo {author} {\bibfnamefont {B.}~\bibnamefont {Kleihaus}}, \bibinfo {author} {\bibfnamefont {J.}~\bibnamefont {Kunz}},\ and\ \bibinfo {author} {\bibfnamefont {E.}~\bibnamefont {Berti}},\ }\bibfield  {title} {\bibinfo {title} {Spinning and excited black holes in einstein-scalar-gauss--bonnet theory},\ }\href@noop {} {\bibfield  {journal} {\bibinfo  {journal} {Classical and Quantum Gravity}\ }\textbf {\bibinfo {volume} {37}},\ \bibinfo {pages} {075018} (\bibinfo {year} {2020})}\BibitemShut {NoStop}%
\bibitem [{\citenamefont {Kuan}\ \emph {et~al.}(2021)\citenamefont {Kuan}, \citenamefont {Doneva},\ and\ \citenamefont {Yazadjiev}}]{kuan2021dynamical}%
  \BibitemOpen
  \bibfield  {author} {\bibinfo {author} {\bibfnamefont {H.-J.}\ \bibnamefont {Kuan}}, \bibinfo {author} {\bibfnamefont {D.~D.}\ \bibnamefont {Doneva}},\ and\ \bibinfo {author} {\bibfnamefont {S.~S.}\ \bibnamefont {Yazadjiev}},\ }\bibfield  {title} {\bibinfo {title} {Dynamical formation of scalarized black holes and neutron stars through stellar core collapse},\ }\href@noop {} {\bibfield  {journal} {\bibinfo  {journal} {Physical Review Letters}\ }\textbf {\bibinfo {volume} {127}},\ \bibinfo {pages} {161103} (\bibinfo {year} {2021})}\BibitemShut {NoStop}%
\bibitem [{\citenamefont {Danchev}\ \emph {et~al.}(2022)\citenamefont {Danchev}, \citenamefont {Doneva},\ and\ \citenamefont {Yazadjiev}}]{danchev2022constraining}%
  \BibitemOpen
  \bibfield  {author} {\bibinfo {author} {\bibfnamefont {V.~I.}\ \bibnamefont {Danchev}}, \bibinfo {author} {\bibfnamefont {D.~D.}\ \bibnamefont {Doneva}},\ and\ \bibinfo {author} {\bibfnamefont {S.~S.}\ \bibnamefont {Yazadjiev}},\ }\bibfield  {title} {\bibinfo {title} {Constraining scalarization in scalar-gauss-bonnet gravity through binary pulsars},\ }\href@noop {} {\bibfield  {journal} {\bibinfo  {journal} {Physical Review D}\ }\textbf {\bibinfo {volume} {106}},\ \bibinfo {pages} {124001} (\bibinfo {year} {2022})}\BibitemShut {NoStop}%
\bibitem [{\citenamefont {Fernandes}\ \emph {et~al.}(2019)\citenamefont {Fernandes}, \citenamefont {Herdeiro}, \citenamefont {Pombo}, \citenamefont {Radu},\ and\ \citenamefont {Sanchis-Gual}}]{fernandes2019spontaneous}%
  \BibitemOpen
  \bibfield  {author} {\bibinfo {author} {\bibfnamefont {P.~G.}\ \bibnamefont {Fernandes}}, \bibinfo {author} {\bibfnamefont {C.~A.}\ \bibnamefont {Herdeiro}}, \bibinfo {author} {\bibfnamefont {A.~M.}\ \bibnamefont {Pombo}}, \bibinfo {author} {\bibfnamefont {E.}~\bibnamefont {Radu}},\ and\ \bibinfo {author} {\bibfnamefont {N.}~\bibnamefont {Sanchis-Gual}},\ }\bibfield  {title} {\bibinfo {title} {Spontaneous scalarisation of charged black holes: coupling dependence and dynamical features},\ }\href@noop {} {\bibfield  {journal} {\bibinfo  {journal} {Classical and Quantum Gravity}\ }\textbf {\bibinfo {volume} {36}},\ \bibinfo {pages} {134002} (\bibinfo {year} {2019})}\BibitemShut {NoStop}%
\bibitem [{\citenamefont {Astefanesei}\ \emph {et~al.}(2019)\citenamefont {Astefanesei}, \citenamefont {Herdeiro}, \citenamefont {Pombo},\ and\ \citenamefont {Radu}}]{astefanesei2019einstein}%
  \BibitemOpen
  \bibfield  {author} {\bibinfo {author} {\bibfnamefont {D.}~\bibnamefont {Astefanesei}}, \bibinfo {author} {\bibfnamefont {C.}~\bibnamefont {Herdeiro}}, \bibinfo {author} {\bibfnamefont {A.}~\bibnamefont {Pombo}},\ and\ \bibinfo {author} {\bibfnamefont {E.}~\bibnamefont {Radu}},\ }\bibfield  {title} {\bibinfo {title} {Einstein-maxwell-scalar black holes: classes of solutions, dyons and extremality},\ }\href@noop {} {\bibfield  {journal} {\bibinfo  {journal} {Journal of High Energy Physics}\ }\textbf {\bibinfo {volume} {2019}},\ \bibinfo {pages} {1} (\bibinfo {year} {2019})}\BibitemShut {NoStop}%
\bibitem [{\citenamefont {Belkhadria}\ and\ \citenamefont {Pombo}(2024)}]{belkhadria2024mixed}%
  \BibitemOpen
  \bibfield  {author} {\bibinfo {author} {\bibfnamefont {Z.}~\bibnamefont {Belkhadria}}\ and\ \bibinfo {author} {\bibfnamefont {A.~M.}\ \bibnamefont {Pombo}},\ }\bibfield  {title} {\bibinfo {title} {Mixed scalarization of charged black holes: From spontaneous to nonlinear scalarization},\ }\href@noop {} {\bibfield  {journal} {\bibinfo  {journal} {Physical Review D}\ }\textbf {\bibinfo {volume} {110}},\ \bibinfo {pages} {044014} (\bibinfo {year} {2024})}\BibitemShut {NoStop}%
\bibitem [{\citenamefont {Konoplya}\ and\ \citenamefont {Zhidenko}(2019)}]{konoplya2019analytical}%
  \BibitemOpen
  \bibfield  {author} {\bibinfo {author} {\bibfnamefont {R.}~\bibnamefont {Konoplya}}\ and\ \bibinfo {author} {\bibfnamefont {A.}~\bibnamefont {Zhidenko}},\ }\bibfield  {title} {\bibinfo {title} {Analytical representation for metrics of scalarized einstein-maxwell black holes and their shadows},\ }\href@noop {} {\bibfield  {journal} {\bibinfo  {journal} {Physical Review D}\ }\textbf {\bibinfo {volume} {100}},\ \bibinfo {pages} {044015} (\bibinfo {year} {2019})}\BibitemShut {NoStop}%
\bibitem [{\citenamefont {Gan}\ \emph {et~al.}(2021)\citenamefont {Gan}, \citenamefont {Wang}, \citenamefont {Wu},\ and\ \citenamefont {Yang}}]{gan2021photon}%
  \BibitemOpen
  \bibfield  {author} {\bibinfo {author} {\bibfnamefont {Q.}~\bibnamefont {Gan}}, \bibinfo {author} {\bibfnamefont {P.}~\bibnamefont {Wang}}, \bibinfo {author} {\bibfnamefont {H.}~\bibnamefont {Wu}},\ and\ \bibinfo {author} {\bibfnamefont {H.}~\bibnamefont {Yang}},\ }\bibfield  {title} {\bibinfo {title} {Photon ring and observational appearance of a hairy black hole},\ }\href@noop {} {\bibfield  {journal} {\bibinfo  {journal} {Physical Review D}\ }\textbf {\bibinfo {volume} {104}},\ \bibinfo {pages} {044049} (\bibinfo {year} {2021})}\BibitemShut {NoStop}%
\bibitem [{\citenamefont {Myung}\ and\ \citenamefont {Zou}(2019{\natexlab{b}})}]{myung2019quasinormal}%
  \BibitemOpen
  \bibfield  {author} {\bibinfo {author} {\bibfnamefont {Y.~S.}\ \bibnamefont {Myung}}\ and\ \bibinfo {author} {\bibfnamefont {D.-C.}\ \bibnamefont {Zou}},\ }\bibfield  {title} {\bibinfo {title} {Quasinormal modes of scalarized black holes in the einstein--maxwell--scalar theory},\ }\href@noop {} {\bibfield  {journal} {\bibinfo  {journal} {Physics Letters B}\ }\textbf {\bibinfo {volume} {790}},\ \bibinfo {pages} {400} (\bibinfo {year} {2019}{\natexlab{b}})}\BibitemShut {NoStop}%
\bibitem [{\citenamefont {Myung}\ and\ \citenamefont {Zou}(2019{\natexlab{c}})}]{myung2019stability}%
  \BibitemOpen
  \bibfield  {author} {\bibinfo {author} {\bibfnamefont {Y.~S.}\ \bibnamefont {Myung}}\ and\ \bibinfo {author} {\bibfnamefont {D.-C.}\ \bibnamefont {Zou}},\ }\bibfield  {title} {\bibinfo {title} {Stability of scalarized charged black holes in the einstein--maxwell--scalar theory},\ }\href@noop {} {\bibfield  {journal} {\bibinfo  {journal} {The European Physical Journal C}\ }\textbf {\bibinfo {volume} {79}},\ \bibinfo {pages} {1} (\bibinfo {year} {2019}{\natexlab{c}})}\BibitemShut {NoStop}%
\bibitem [{\citenamefont {Zou}\ and\ \citenamefont {Myung}(2019)}]{zou2019scalarized}%
  \BibitemOpen
  \bibfield  {author} {\bibinfo {author} {\bibfnamefont {D.-C.}\ \bibnamefont {Zou}}\ and\ \bibinfo {author} {\bibfnamefont {Y.~S.}\ \bibnamefont {Myung}},\ }\bibfield  {title} {\bibinfo {title} {Scalarized charged black holes with scalar mass term},\ }\href@noop {} {\bibfield  {journal} {\bibinfo  {journal} {Physical Review D}\ }\textbf {\bibinfo {volume} {100}},\ \bibinfo {pages} {124055} (\bibinfo {year} {2019})}\BibitemShut {NoStop}%
\bibitem [{\citenamefont {Bl{\'a}zquez-Salcedo}\ \emph {et~al.}(2020{\natexlab{c}})\citenamefont {Bl{\'a}zquez-Salcedo}, \citenamefont {Herdeiro}, \citenamefont {Kunz}, \citenamefont {Pombo},\ and\ \citenamefont {Radu}}]{blazquez2020einstein}%
  \BibitemOpen
  \bibfield  {author} {\bibinfo {author} {\bibfnamefont {J.~L.}\ \bibnamefont {Bl{\'a}zquez-Salcedo}}, \bibinfo {author} {\bibfnamefont {C.~A.}\ \bibnamefont {Herdeiro}}, \bibinfo {author} {\bibfnamefont {J.}~\bibnamefont {Kunz}}, \bibinfo {author} {\bibfnamefont {A.~M.}\ \bibnamefont {Pombo}},\ and\ \bibinfo {author} {\bibfnamefont {E.}~\bibnamefont {Radu}},\ }\bibfield  {title} {\bibinfo {title} {Einstein-maxwell-scalar black holes: the hot, the cold and the bald},\ }\href@noop {} {\bibfield  {journal} {\bibinfo  {journal} {Physics Letters B}\ }\textbf {\bibinfo {volume} {806}},\ \bibinfo {pages} {135493} (\bibinfo {year} {2020}{\natexlab{c}})}\BibitemShut {NoStop}%
\bibitem [{\citenamefont {Bl{\'a}zquez-Salcedo}\ \emph {et~al.}(2021)\citenamefont {Bl{\'a}zquez-Salcedo}, \citenamefont {Herdeiro}, \citenamefont {Kahlen}, \citenamefont {Kunz}, \citenamefont {Pombo},\ and\ \citenamefont {Radu}}]{blazquez2021quasinormal}%
  \BibitemOpen
  \bibfield  {author} {\bibinfo {author} {\bibfnamefont {J.~L.}\ \bibnamefont {Bl{\'a}zquez-Salcedo}}, \bibinfo {author} {\bibfnamefont {C.~A.}\ \bibnamefont {Herdeiro}}, \bibinfo {author} {\bibfnamefont {S.}~\bibnamefont {Kahlen}}, \bibinfo {author} {\bibfnamefont {J.}~\bibnamefont {Kunz}}, \bibinfo {author} {\bibfnamefont {A.~M.}\ \bibnamefont {Pombo}},\ and\ \bibinfo {author} {\bibfnamefont {E.}~\bibnamefont {Radu}},\ }\bibfield  {title} {\bibinfo {title} {Quasinormal modes of hot, cold and bald einstein--maxwell-scalar black holes},\ }\href@noop {} {\bibfield  {journal} {\bibinfo  {journal} {The European Physical Journal C}\ }\textbf {\bibinfo {volume} {81}},\ \bibinfo {pages} {1} (\bibinfo {year} {2021})}\BibitemShut {NoStop}%
\bibitem [{\citenamefont {Guo}\ \emph {et~al.}(2022)\citenamefont {Guo}, \citenamefont {Wang}, \citenamefont {Wu},\ and\ \citenamefont {Yang}}]{guo2022quasinormal}%
  \BibitemOpen
  \bibfield  {author} {\bibinfo {author} {\bibfnamefont {G.}~\bibnamefont {Guo}}, \bibinfo {author} {\bibfnamefont {P.}~\bibnamefont {Wang}}, \bibinfo {author} {\bibfnamefont {H.}~\bibnamefont {Wu}},\ and\ \bibinfo {author} {\bibfnamefont {H.}~\bibnamefont {Yang}},\ }\bibfield  {title} {\bibinfo {title} {Quasinormal modes of black holes with multiple photon spheres},\ }\href@noop {} {\bibfield  {journal} {\bibinfo  {journal} {Journal of High Energy Physics}\ }\textbf {\bibinfo {volume} {2022}},\ \bibinfo {pages} {1} (\bibinfo {year} {2022})}\BibitemShut {NoStop}%
\bibitem [{\citenamefont {Fernandes}(2020)}]{fernandes2020einstein}%
  \BibitemOpen
  \bibfield  {author} {\bibinfo {author} {\bibfnamefont {P.~G.}\ \bibnamefont {Fernandes}},\ }\bibfield  {title} {\bibinfo {title} {Einstein--maxwell-scalar black holes with massive and self-interacting scalar hair},\ }\href@noop {} {\bibfield  {journal} {\bibinfo  {journal} {Physics of the Dark Universe}\ }\textbf {\bibinfo {volume} {30}},\ \bibinfo {pages} {100716} (\bibinfo {year} {2020})}\BibitemShut {NoStop}%
\bibitem [{\citenamefont {Zhang}\ \emph {et~al.}(2022)\citenamefont {Zhang}, \citenamefont {Chen}, \citenamefont {Liu}, \citenamefont {Luo}, \citenamefont {Tian},\ and\ \citenamefont {Wang}}]{zhang2022critical}%
  \BibitemOpen
  \bibfield  {author} {\bibinfo {author} {\bibfnamefont {C.-Y.}\ \bibnamefont {Zhang}}, \bibinfo {author} {\bibfnamefont {Q.}~\bibnamefont {Chen}}, \bibinfo {author} {\bibfnamefont {Y.}~\bibnamefont {Liu}}, \bibinfo {author} {\bibfnamefont {W.-K.}\ \bibnamefont {Luo}}, \bibinfo {author} {\bibfnamefont {Y.}~\bibnamefont {Tian}},\ and\ \bibinfo {author} {\bibfnamefont {B.}~\bibnamefont {Wang}},\ }\bibfield  {title} {\bibinfo {title} {Critical phenomena in dynamical scalarization of charged black holes},\ }\href@noop {} {\bibfield  {journal} {\bibinfo  {journal} {Physical Review Letters}\ }\textbf {\bibinfo {volume} {128}},\ \bibinfo {pages} {161105} (\bibinfo {year} {2022})}\BibitemShut {NoStop}%
\bibitem [{\citenamefont {Zhang}\ \emph {et~al.}(2021)\citenamefont {Zhang}, \citenamefont {Liu}, \citenamefont {Liu}, \citenamefont {Niu},\ and\ \citenamefont {Wang}}]{zhang2021dynamical}%
  \BibitemOpen
  \bibfield  {author} {\bibinfo {author} {\bibfnamefont {C.-Y.}\ \bibnamefont {Zhang}}, \bibinfo {author} {\bibfnamefont {P.}~\bibnamefont {Liu}}, \bibinfo {author} {\bibfnamefont {Y.-Q.}\ \bibnamefont {Liu}}, \bibinfo {author} {\bibfnamefont {C.}~\bibnamefont {Niu}},\ and\ \bibinfo {author} {\bibfnamefont {B.}~\bibnamefont {Wang}},\ }\bibfield  {title} {\bibinfo {title} {Dynamical charged black hole spontaneous scalarization in anti--de sitter spacetimes},\ }\href@noop {} {\bibfield  {journal} {\bibinfo  {journal} {Physical Review D}\ }\textbf {\bibinfo {volume} {104}},\ \bibinfo {pages} {084089} (\bibinfo {year} {2021})}\BibitemShut {NoStop}%
\bibitem [{\citenamefont {Khalil}\ \emph {et~al.}(2019)\citenamefont {Khalil}, \citenamefont {Sennett}, \citenamefont {Steinhoff},\ and\ \citenamefont {Buonanno}}]{khalil2019theory}%
  \BibitemOpen
  \bibfield  {author} {\bibinfo {author} {\bibfnamefont {M.}~\bibnamefont {Khalil}}, \bibinfo {author} {\bibfnamefont {N.}~\bibnamefont {Sennett}}, \bibinfo {author} {\bibfnamefont {J.}~\bibnamefont {Steinhoff}},\ and\ \bibinfo {author} {\bibfnamefont {A.}~\bibnamefont {Buonanno}},\ }\bibfield  {title} {\bibinfo {title} {Theory-agnostic framework for dynamical scalarization of compact binaries},\ }\href@noop {} {\bibfield  {journal} {\bibinfo  {journal} {Physical Review D}\ }\textbf {\bibinfo {volume} {100}},\ \bibinfo {pages} {124013} (\bibinfo {year} {2019})}\BibitemShut {NoStop}%
\bibitem [{\citenamefont {Doneva}\ \emph {et~al.}(2018)\citenamefont {Doneva}, \citenamefont {Kiorpelidi}, \citenamefont {Nedkova}, \citenamefont {Papantonopoulos},\ and\ \citenamefont {Yazadjiev}}]{doneva2018charged}%
  \BibitemOpen
  \bibfield  {author} {\bibinfo {author} {\bibfnamefont {D.~D.}\ \bibnamefont {Doneva}}, \bibinfo {author} {\bibfnamefont {S.}~\bibnamefont {Kiorpelidi}}, \bibinfo {author} {\bibfnamefont {P.~G.}\ \bibnamefont {Nedkova}}, \bibinfo {author} {\bibfnamefont {E.}~\bibnamefont {Papantonopoulos}},\ and\ \bibinfo {author} {\bibfnamefont {S.~S.}\ \bibnamefont {Yazadjiev}},\ }\bibfield  {title} {\bibinfo {title} {Charged gauss-bonnet black holes with curvature induced scalarization in the extended scalar-tensor theories},\ }\href@noop {} {\bibfield  {journal} {\bibinfo  {journal} {Physical Review D}\ }\textbf {\bibinfo {volume} {98}},\ \bibinfo {pages} {104056} (\bibinfo {year} {2018})}\BibitemShut {NoStop}%
\bibitem [{\citenamefont {Brihaye}\ and\ \citenamefont {Hartmann}(2019)}]{brihaye2019spontaneous}%
  \BibitemOpen
  \bibfield  {author} {\bibinfo {author} {\bibfnamefont {Y.}~\bibnamefont {Brihaye}}\ and\ \bibinfo {author} {\bibfnamefont {B.}~\bibnamefont {Hartmann}},\ }\bibfield  {title} {\bibinfo {title} {Spontaneous scalarization of charged black holes at the approach to extremality},\ }\href@noop {} {\bibfield  {journal} {\bibinfo  {journal} {Physics Letters B}\ }\textbf {\bibinfo {volume} {792}},\ \bibinfo {pages} {244} (\bibinfo {year} {2019})}\BibitemShut {NoStop}%
\bibitem [{\citenamefont {Herdeiro}\ \emph {et~al.}(2021{\natexlab{a}})\citenamefont {Herdeiro}, \citenamefont {Pombo},\ and\ \citenamefont {Radu}}]{herdeiro2021aspects}%
  \BibitemOpen
  \bibfield  {author} {\bibinfo {author} {\bibfnamefont {C.~A.}\ \bibnamefont {Herdeiro}}, \bibinfo {author} {\bibfnamefont {A.~M.}\ \bibnamefont {Pombo}},\ and\ \bibinfo {author} {\bibfnamefont {E.}~\bibnamefont {Radu}},\ }\bibfield  {title} {\bibinfo {title} {Aspects of gauss-bonnet scalarisation of charged black holes},\ }\href@noop {} {\bibfield  {journal} {\bibinfo  {journal} {Universe}\ }\textbf {\bibinfo {volume} {7}},\ \bibinfo {pages} {483} (\bibinfo {year} {2021}{\natexlab{a}})}\BibitemShut {NoStop}%
\bibitem [{\citenamefont {Bl{\'a}zquez-Salcedo}\ \emph {et~al.}(2024)\citenamefont {Bl{\'a}zquez-Salcedo}, \citenamefont {Kleihaus},\ and\ \citenamefont {Kunz}}]{blazquez2024instabilities}%
  \BibitemOpen
  \bibfield  {author} {\bibinfo {author} {\bibfnamefont {J.~L.}\ \bibnamefont {Bl{\'a}zquez-Salcedo}}, \bibinfo {author} {\bibfnamefont {B.}~\bibnamefont {Kleihaus}},\ and\ \bibinfo {author} {\bibfnamefont {J.}~\bibnamefont {Kunz}},\ }\bibfield  {title} {\bibinfo {title} {Instabilities of black holes in einstein-scalar--gauss--bonnet theories},\ }\href@noop {} {\bibfield  {journal} {\bibinfo  {journal} {General Relativity and Gravitation}\ }\textbf {\bibinfo {volume} {56}},\ \bibinfo {pages} {99} (\bibinfo {year} {2024})}\BibitemShut {NoStop}%
\bibitem [{\citenamefont {Bl{\'a}zquez-Salcedo}\ \emph {et~al.}(2018)\citenamefont {Bl{\'a}zquez-Salcedo}, \citenamefont {Doneva}, \citenamefont {Kunz},\ and\ \citenamefont {Yazadjiev}}]{blazquez2018radial}%
  \BibitemOpen
  \bibfield  {author} {\bibinfo {author} {\bibfnamefont {J.~L.}\ \bibnamefont {Bl{\'a}zquez-Salcedo}}, \bibinfo {author} {\bibfnamefont {D.~D.}\ \bibnamefont {Doneva}}, \bibinfo {author} {\bibfnamefont {J.}~\bibnamefont {Kunz}},\ and\ \bibinfo {author} {\bibfnamefont {S.~S.}\ \bibnamefont {Yazadjiev}},\ }\bibfield  {title} {\bibinfo {title} {Radial perturbations of the scalarized einstein-gauss-bonnet black holes},\ }\href@noop {} {\bibfield  {journal} {\bibinfo  {journal} {Physical Review D}\ }\textbf {\bibinfo {volume} {98}},\ \bibinfo {pages} {084011} (\bibinfo {year} {2018})}\BibitemShut {NoStop}%
\bibitem [{\citenamefont {Silva}\ \emph {et~al.}(2019)\citenamefont {Silva}, \citenamefont {Macedo}, \citenamefont {Sotiriou}, \citenamefont {Gualtieri}, \citenamefont {Sakstein},\ and\ \citenamefont {Berti}}]{silva2019stability}%
  \BibitemOpen
  \bibfield  {author} {\bibinfo {author} {\bibfnamefont {H.~O.}\ \bibnamefont {Silva}}, \bibinfo {author} {\bibfnamefont {C.~F.}\ \bibnamefont {Macedo}}, \bibinfo {author} {\bibfnamefont {T.~P.}\ \bibnamefont {Sotiriou}}, \bibinfo {author} {\bibfnamefont {L.}~\bibnamefont {Gualtieri}}, \bibinfo {author} {\bibfnamefont {J.}~\bibnamefont {Sakstein}},\ and\ \bibinfo {author} {\bibfnamefont {E.}~\bibnamefont {Berti}},\ }\bibfield  {title} {\bibinfo {title} {Stability of scalarized black hole solutions in scalar-gauss-bonnet gravity},\ }\href@noop {} {\bibfield  {journal} {\bibinfo  {journal} {Physical Review D}\ }\textbf {\bibinfo {volume} {99}},\ \bibinfo {pages} {064011} (\bibinfo {year} {2019})}\BibitemShut {NoStop}%
\bibitem [{\citenamefont {Macedo}\ \emph {et~al.}(2019)\citenamefont {Macedo}, \citenamefont {Sakstein}, \citenamefont {Berti}, \citenamefont {Gualtieri}, \citenamefont {Silva},\ and\ \citenamefont {Sotiriou}}]{macedo2019self}%
  \BibitemOpen
  \bibfield  {author} {\bibinfo {author} {\bibfnamefont {C.~F.}\ \bibnamefont {Macedo}}, \bibinfo {author} {\bibfnamefont {J.}~\bibnamefont {Sakstein}}, \bibinfo {author} {\bibfnamefont {E.}~\bibnamefont {Berti}}, \bibinfo {author} {\bibfnamefont {L.}~\bibnamefont {Gualtieri}}, \bibinfo {author} {\bibfnamefont {H.~O.}\ \bibnamefont {Silva}},\ and\ \bibinfo {author} {\bibfnamefont {T.~P.}\ \bibnamefont {Sotiriou}},\ }\bibfield  {title} {\bibinfo {title} {Self-interactions and spontaneous black hole scalarization},\ }\href@noop {} {\bibfield  {journal} {\bibinfo  {journal} {Physical Review D}\ }\textbf {\bibinfo {volume} {99}},\ \bibinfo {pages} {104041} (\bibinfo {year} {2019})}\BibitemShut {NoStop}%
\bibitem [{\citenamefont {Antoniou}\ \emph {et~al.}(2021{\natexlab{b}})\citenamefont {Antoniou}, \citenamefont {Leh{\'e}bel}, \citenamefont {Ventagli},\ and\ \citenamefont {Sotiriou}}]{antoniou2021black}%
  \BibitemOpen
  \bibfield  {author} {\bibinfo {author} {\bibfnamefont {G.}~\bibnamefont {Antoniou}}, \bibinfo {author} {\bibfnamefont {A.}~\bibnamefont {Leh{\'e}bel}}, \bibinfo {author} {\bibfnamefont {G.}~\bibnamefont {Ventagli}},\ and\ \bibinfo {author} {\bibfnamefont {T.~P.}\ \bibnamefont {Sotiriou}},\ }\bibfield  {title} {\bibinfo {title} {Black hole scalarization with gauss-bonnet and ricci scalar couplings},\ }\href@noop {} {\bibfield  {journal} {\bibinfo  {journal} {Physical Review D}\ }\textbf {\bibinfo {volume} {104}},\ \bibinfo {pages} {044002} (\bibinfo {year} {2021}{\natexlab{b}})}\BibitemShut {NoStop}%
\bibitem [{\citenamefont {Antoniou}\ \emph {et~al.}(2022)\citenamefont {Antoniou}, \citenamefont {Macedo}, \citenamefont {McManus},\ and\ \citenamefont {Sotiriou}}]{antoniou2022stable}%
  \BibitemOpen
  \bibfield  {author} {\bibinfo {author} {\bibfnamefont {G.}~\bibnamefont {Antoniou}}, \bibinfo {author} {\bibfnamefont {C.~F.}\ \bibnamefont {Macedo}}, \bibinfo {author} {\bibfnamefont {R.}~\bibnamefont {McManus}},\ and\ \bibinfo {author} {\bibfnamefont {T.~P.}\ \bibnamefont {Sotiriou}},\ }\bibfield  {title} {\bibinfo {title} {Stable spontaneously-scalarized black holes in generalized scalar-tensor theories},\ }\href@noop {} {\bibfield  {journal} {\bibinfo  {journal} {Physical Review D}\ }\textbf {\bibinfo {volume} {106}},\ \bibinfo {pages} {024029} (\bibinfo {year} {2022})}\BibitemShut {NoStop}%
\bibitem [{\citenamefont {Kleihaus}\ \emph {et~al.}(2023)\citenamefont {Kleihaus}, \citenamefont {Kunz}, \citenamefont {Uterm{\"o}hlen},\ and\ \citenamefont {Berti}}]{kleihaus2023quadrupole}%
  \BibitemOpen
  \bibfield  {author} {\bibinfo {author} {\bibfnamefont {B.}~\bibnamefont {Kleihaus}}, \bibinfo {author} {\bibfnamefont {J.}~\bibnamefont {Kunz}}, \bibinfo {author} {\bibfnamefont {T.}~\bibnamefont {Uterm{\"o}hlen}},\ and\ \bibinfo {author} {\bibfnamefont {E.}~\bibnamefont {Berti}},\ }\bibfield  {title} {\bibinfo {title} {Quadrupole instability of static scalarized black holes},\ }\href@noop {} {\bibfield  {journal} {\bibinfo  {journal} {Physical Review D}\ }\textbf {\bibinfo {volume} {107}},\ \bibinfo {pages} {L081501} (\bibinfo {year} {2023})}\BibitemShut {NoStop}%
\bibitem [{\citenamefont {Minamitsuji}\ and\ \citenamefont {Mukohyama}(2023)}]{minamitsuji2023instability}%
  \BibitemOpen
  \bibfield  {author} {\bibinfo {author} {\bibfnamefont {M.}~\bibnamefont {Minamitsuji}}\ and\ \bibinfo {author} {\bibfnamefont {S.}~\bibnamefont {Mukohyama}},\ }\bibfield  {title} {\bibinfo {title} {Instability of scalarized compact objects in einstein-scalar-gauss-bonnet theories},\ }\href@noop {} {\bibfield  {journal} {\bibinfo  {journal} {Physical Review D}\ }\textbf {\bibinfo {volume} {108}},\ \bibinfo {pages} {024029} (\bibinfo {year} {2023})}\BibitemShut {NoStop}%
\bibitem [{\citenamefont {Minamitsuji}\ \emph {et~al.}(2024)\citenamefont {Minamitsuji}, \citenamefont {Mukohyama},\ and\ \citenamefont {Tsujikawa}}]{minamitsuji2024angular}%
  \BibitemOpen
  \bibfield  {author} {\bibinfo {author} {\bibfnamefont {M.}~\bibnamefont {Minamitsuji}}, \bibinfo {author} {\bibfnamefont {S.}~\bibnamefont {Mukohyama}},\ and\ \bibinfo {author} {\bibfnamefont {S.}~\bibnamefont {Tsujikawa}},\ }\bibfield  {title} {\bibinfo {title} {Angular and radial stabilities of spontaneously scalarized black holes in the presence of scalar-gauss-bonnet couplings},\ }\href@noop {} {\bibfield  {journal} {\bibinfo  {journal} {Physical Review D}\ }\textbf {\bibinfo {volume} {109}},\ \bibinfo {pages} {104057} (\bibinfo {year} {2024})}\BibitemShut {NoStop}%
\bibitem [{\citenamefont {East}\ and\ \citenamefont {Ripley}(2021)}]{east2021dynamics}%
  \BibitemOpen
  \bibfield  {author} {\bibinfo {author} {\bibfnamefont {W.~E.}\ \bibnamefont {East}}\ and\ \bibinfo {author} {\bibfnamefont {J.~L.}\ \bibnamefont {Ripley}},\ }\bibfield  {title} {\bibinfo {title} {Dynamics of spontaneous black hole scalarization and mergers in einstein-scalar-gauss-bonnet gravity},\ }\href@noop {} {\bibfield  {journal} {\bibinfo  {journal} {Physical Review Letters}\ }\textbf {\bibinfo {volume} {127}},\ \bibinfo {pages} {101102} (\bibinfo {year} {2021})}\BibitemShut {NoStop}%
\bibitem [{\citenamefont {Doneva}\ \emph {et~al.}(2023)\citenamefont {Doneva}, \citenamefont {Sal{\'o}}, \citenamefont {Clough}, \citenamefont {Figueras},\ and\ \citenamefont {Yazadjiev}}]{doneva2023testing}%
  \BibitemOpen
  \bibfield  {author} {\bibinfo {author} {\bibfnamefont {D.~D.}\ \bibnamefont {Doneva}}, \bibinfo {author} {\bibfnamefont {L.~A.}\ \bibnamefont {Sal{\'o}}}, \bibinfo {author} {\bibfnamefont {K.}~\bibnamefont {Clough}}, \bibinfo {author} {\bibfnamefont {P.}~\bibnamefont {Figueras}},\ and\ \bibinfo {author} {\bibfnamefont {S.~S.}\ \bibnamefont {Yazadjiev}},\ }\bibfield  {title} {\bibinfo {title} {Testing the limits of scalar-gauss-bonnet gravity through nonlinear evolutions of spin-induced scalarization},\ }\href@noop {} {\bibfield  {journal} {\bibinfo  {journal} {Physical Review D}\ }\textbf {\bibinfo {volume} {108}},\ \bibinfo {pages} {084017} (\bibinfo {year} {2023})}\BibitemShut {NoStop}%
\bibitem [{\citenamefont {Doneva}\ \emph {et~al.}(2024{\natexlab{b}})\citenamefont {Doneva}, \citenamefont {Sal{\'o}},\ and\ \citenamefont {Yazadjiev}}]{doneva20243+}%
  \BibitemOpen
  \bibfield  {author} {\bibinfo {author} {\bibfnamefont {D.~D.}\ \bibnamefont {Doneva}}, \bibinfo {author} {\bibfnamefont {L.~A.}\ \bibnamefont {Sal{\'o}}},\ and\ \bibinfo {author} {\bibfnamefont {S.~S.}\ \bibnamefont {Yazadjiev}},\ }\bibfield  {title} {\bibinfo {title} {3+ 1 nonlinear evolution of ricci-coupled scalar-gauss-bonnet gravity},\ }\href@noop {} {\bibfield  {journal} {\bibinfo  {journal} {Physical Review D}\ }\textbf {\bibinfo {volume} {110}},\ \bibinfo {pages} {024040} (\bibinfo {year} {2024}{\natexlab{b}})}\BibitemShut {NoStop}%
\bibitem [{\citenamefont {Hegade~KR}\ \emph {et~al.}(2023)\citenamefont {Hegade~KR}, \citenamefont {Ripley},\ and\ \citenamefont {Yunes}}]{hegade2023and}%
  \BibitemOpen
  \bibfield  {author} {\bibinfo {author} {\bibfnamefont {A.}~\bibnamefont {Hegade~KR}}, \bibinfo {author} {\bibfnamefont {J.~L.}\ \bibnamefont {Ripley}},\ and\ \bibinfo {author} {\bibfnamefont {N.}~\bibnamefont {Yunes}},\ }\bibfield  {title} {\bibinfo {title} {Where and why does einstein-scalar-gauss-bonnet theory break down?},\ }\href@noop {} {\bibfield  {journal} {\bibinfo  {journal} {Physical Review D}\ }\textbf {\bibinfo {volume} {107}},\ \bibinfo {pages} {044044} (\bibinfo {year} {2023})}\BibitemShut {NoStop}%
\bibitem [{\citenamefont {Thaalba}\ \emph {et~al.}(2024)\citenamefont {Thaalba}, \citenamefont {Bezares}, \citenamefont {Franchini},\ and\ \citenamefont {Sotiriou}}]{thaalba2024spherical}%
  \BibitemOpen
  \bibfield  {author} {\bibinfo {author} {\bibfnamefont {F.}~\bibnamefont {Thaalba}}, \bibinfo {author} {\bibfnamefont {M.}~\bibnamefont {Bezares}}, \bibinfo {author} {\bibfnamefont {N.}~\bibnamefont {Franchini}},\ and\ \bibinfo {author} {\bibfnamefont {T.~P.}\ \bibnamefont {Sotiriou}},\ }\bibfield  {title} {\bibinfo {title} {Spherical collapse in scalar-gauss-bonnet gravity: Taming ill-posedness with a ricci coupling},\ }\href@noop {} {\bibfield  {journal} {\bibinfo  {journal} {Physical Review D}\ }\textbf {\bibinfo {volume} {109}},\ \bibinfo {pages} {L041503} (\bibinfo {year} {2024})}\BibitemShut {NoStop}%
\bibitem [{\citenamefont {Franchini}\ \emph {et~al.}(2022)\citenamefont {Franchini}, \citenamefont {Bezares}, \citenamefont {Barausse},\ and\ \citenamefont {Lehner}}]{franchini2022fixing}%
  \BibitemOpen
  \bibfield  {author} {\bibinfo {author} {\bibfnamefont {N.}~\bibnamefont {Franchini}}, \bibinfo {author} {\bibfnamefont {M.}~\bibnamefont {Bezares}}, \bibinfo {author} {\bibfnamefont {E.}~\bibnamefont {Barausse}},\ and\ \bibinfo {author} {\bibfnamefont {L.}~\bibnamefont {Lehner}},\ }\bibfield  {title} {\bibinfo {title} {Fixing the dynamical evolution in scalar-gauss-bonnet gravity},\ }\href@noop {} {\bibfield  {journal} {\bibinfo  {journal} {Physical Review D}\ }\textbf {\bibinfo {volume} {106}},\ \bibinfo {pages} {064061} (\bibinfo {year} {2022})}\BibitemShut {NoStop}%
\bibitem [{\citenamefont {Mignemi}(1995)}]{mignemi1995dyonic}%
  \BibitemOpen
  \bibfield  {author} {\bibinfo {author} {\bibfnamefont {S.}~\bibnamefont {Mignemi}},\ }\bibfield  {title} {\bibinfo {title} {Dyonic black holes in effective string theory},\ }\href@noop {} {\bibfield  {journal} {\bibinfo  {journal} {Physical Review D}\ }\textbf {\bibinfo {volume} {51}},\ \bibinfo {pages} {934} (\bibinfo {year} {1995})}\BibitemShut {NoStop}%
\bibitem [{\citenamefont {Torii}\ \emph {et~al.}(1997)\citenamefont {Torii}, \citenamefont {Yajima},\ and\ \citenamefont {Maeda}}]{torii1997dilatonic}%
  \BibitemOpen
  \bibfield  {author} {\bibinfo {author} {\bibfnamefont {T.}~\bibnamefont {Torii}}, \bibinfo {author} {\bibfnamefont {H.}~\bibnamefont {Yajima}},\ and\ \bibinfo {author} {\bibfnamefont {K.-i.}\ \bibnamefont {Maeda}},\ }\bibfield  {title} {\bibinfo {title} {Dilatonic black holes with a gauss-bonnet term},\ }\href@noop {} {\bibfield  {journal} {\bibinfo  {journal} {Physical Review D}\ }\textbf {\bibinfo {volume} {55}},\ \bibinfo {pages} {739} (\bibinfo {year} {1997})}\BibitemShut {NoStop}%
\bibitem [{\citenamefont {Alexeyev}\ and\ \citenamefont {Pomazanov}(1997)}]{alexeyev1997singular}%
  \BibitemOpen
  \bibfield  {author} {\bibinfo {author} {\bibfnamefont {S.}~\bibnamefont {Alexeyev}}\ and\ \bibinfo {author} {\bibfnamefont {M.}~\bibnamefont {Pomazanov}},\ }\bibfield  {title} {\bibinfo {title} {Singular regions in black hole solutions in higher order curvature gravity},\ }\href@noop {} {\bibfield  {journal} {\bibinfo  {journal} {arXiv preprint gr-qc/9706066}\ } (\bibinfo {year} {1997})}\BibitemShut {NoStop}%
\bibitem [{\citenamefont {Torii}\ and\ \citenamefont {Maeda}(1998)}]{torii1998stability}%
  \BibitemOpen
  \bibfield  {author} {\bibinfo {author} {\bibfnamefont {T.}~\bibnamefont {Torii}}\ and\ \bibinfo {author} {\bibfnamefont {K.-i.}\ \bibnamefont {Maeda}},\ }\bibfield  {title} {\bibinfo {title} {Stability of a dilatonic black hole with a gauss-bonnet term},\ }\href@noop {} {\bibfield  {journal} {\bibinfo  {journal} {Physical Review D}\ }\textbf {\bibinfo {volume} {58}},\ \bibinfo {pages} {084004} (\bibinfo {year} {1998})}\BibitemShut {NoStop}%
\bibitem [{\citenamefont {Kase}\ and\ \citenamefont {Tsujikawa}(2023)}]{kase2023black}%
  \BibitemOpen
  \bibfield  {author} {\bibinfo {author} {\bibfnamefont {R.}~\bibnamefont {Kase}}\ and\ \bibinfo {author} {\bibfnamefont {S.}~\bibnamefont {Tsujikawa}},\ }\bibfield  {title} {\bibinfo {title} {Black hole perturbations in maxwell-horndeski theories},\ }\href@noop {} {\bibfield  {journal} {\bibinfo  {journal} {Physical Review D}\ }\textbf {\bibinfo {volume} {107}},\ \bibinfo {pages} {104045} (\bibinfo {year} {2023})}\BibitemShut {NoStop}%
\bibitem [{\citenamefont {Bai}\ \emph {et~al.}(2020)\citenamefont {Bai}, \citenamefont {Berger}, \citenamefont {Korwar},\ and\ \citenamefont {Orlofsky}}]{bai2020phenomenology}%
  \BibitemOpen
  \bibfield  {author} {\bibinfo {author} {\bibfnamefont {Y.}~\bibnamefont {Bai}}, \bibinfo {author} {\bibfnamefont {J.}~\bibnamefont {Berger}}, \bibinfo {author} {\bibfnamefont {M.}~\bibnamefont {Korwar}},\ and\ \bibinfo {author} {\bibfnamefont {N.}~\bibnamefont {Orlofsky}},\ }\bibfield  {title} {\bibinfo {title} {Phenomenology of magnetic black holes with electroweak-symmetric coronas},\ }\href@noop {} {\bibfield  {journal} {\bibinfo  {journal} {Journal of High Energy Physics}\ }\textbf {\bibinfo {volume} {2020}},\ \bibinfo {pages} {1} (\bibinfo {year} {2020})}\BibitemShut {NoStop}%
\bibitem [{\citenamefont {Ghosh}\ \emph {et~al.}(2021)\citenamefont {Ghosh}, \citenamefont {Thalapillil},\ and\ \citenamefont {Ullah}}]{ghosh2021astrophysical}%
  \BibitemOpen
  \bibfield  {author} {\bibinfo {author} {\bibfnamefont {D.}~\bibnamefont {Ghosh}}, \bibinfo {author} {\bibfnamefont {A.}~\bibnamefont {Thalapillil}},\ and\ \bibinfo {author} {\bibfnamefont {F.}~\bibnamefont {Ullah}},\ }\bibfield  {title} {\bibinfo {title} {Astrophysical hints for magnetic black holes},\ }\href@noop {} {\bibfield  {journal} {\bibinfo  {journal} {Physical Review D}\ }\textbf {\bibinfo {volume} {103}},\ \bibinfo {pages} {023006} (\bibinfo {year} {2021})}\BibitemShut {NoStop}%
\bibitem [{\citenamefont {Diamond}\ and\ \citenamefont {Kaplan}(2022)}]{diamond2022constraints}%
  \BibitemOpen
  \bibfield  {author} {\bibinfo {author} {\bibfnamefont {M.~D.}\ \bibnamefont {Diamond}}\ and\ \bibinfo {author} {\bibfnamefont {D.~E.}\ \bibnamefont {Kaplan}},\ }\bibfield  {title} {\bibinfo {title} {Constraints on relic magnetic black holes},\ }\href@noop {} {\bibfield  {journal} {\bibinfo  {journal} {Journal of High Energy Physics}\ }\textbf {\bibinfo {volume} {2022}},\ \bibinfo {pages} {1} (\bibinfo {year} {2022})}\BibitemShut {NoStop}%
\bibitem [{\citenamefont {Gervalle}\ and\ \citenamefont {Volkov}(2024)}]{gervalle2024black}%
  \BibitemOpen
  \bibfield  {author} {\bibinfo {author} {\bibfnamefont {R.}~\bibnamefont {Gervalle}}\ and\ \bibinfo {author} {\bibfnamefont {M.~S.}\ \bibnamefont {Volkov}},\ }\bibfield  {title} {\bibinfo {title} {Black holes with electroweak hair},\ }\href@noop {} {\bibfield  {journal} {\bibinfo  {journal} {Physical Review Letters}\ }\textbf {\bibinfo {volume} {133}},\ \bibinfo {pages} {171402} (\bibinfo {year} {2024})}\BibitemShut {NoStop}%
\bibitem [{\citenamefont {Pere{\~n}iguez}\ \emph {et~al.}(2024)\citenamefont {Pere{\~n}iguez}, \citenamefont {de~Amicis}, \citenamefont {Brito},\ and\ \citenamefont {Macedo}}]{pereniguez2024superradiant}%
  \BibitemOpen
  \bibfield  {author} {\bibinfo {author} {\bibfnamefont {D.}~\bibnamefont {Pere{\~n}iguez}}, \bibinfo {author} {\bibfnamefont {M.}~\bibnamefont {de~Amicis}}, \bibinfo {author} {\bibfnamefont {R.}~\bibnamefont {Brito}},\ and\ \bibinfo {author} {\bibfnamefont {R.~P.}\ \bibnamefont {Macedo}},\ }\bibfield  {title} {\bibinfo {title} {Superradiant instability of magnetic black holes},\ }\href@noop {} {\bibfield  {journal} {\bibinfo  {journal} {Physical Review D}\ }\textbf {\bibinfo {volume} {110}},\ \bibinfo {pages} {104001} (\bibinfo {year} {2024})}\BibitemShut {NoStop}%
\bibitem [{\citenamefont {Dyson}\ and\ \citenamefont {Pere{\~n}iguez}(2023)}]{dyson2023magnetic}%
  \BibitemOpen
  \bibfield  {author} {\bibinfo {author} {\bibfnamefont {C.}~\bibnamefont {Dyson}}\ and\ \bibinfo {author} {\bibfnamefont {D.}~\bibnamefont {Pere{\~n}iguez}},\ }\bibfield  {title} {\bibinfo {title} {Magnetic black holes: From thomson dipoles to the penrose process and cosmic censorship},\ }\href@noop {} {\bibfield  {journal} {\bibinfo  {journal} {Physical Review D}\ }\textbf {\bibinfo {volume} {108}},\ \bibinfo {pages} {084064} (\bibinfo {year} {2023})}\BibitemShut {NoStop}%
\bibitem [{\citenamefont {Maldacena}(2021)}]{maldacena2021comments}%
  \BibitemOpen
  \bibfield  {author} {\bibinfo {author} {\bibfnamefont {J.}~\bibnamefont {Maldacena}},\ }\bibfield  {title} {\bibinfo {title} {Comments on magnetic black holes},\ }\href@noop {} {\bibfield  {journal} {\bibinfo  {journal} {Journal of High Energy Physics}\ }\textbf {\bibinfo {volume} {2021}},\ \bibinfo {pages} {1} (\bibinfo {year} {2021})}\BibitemShut {NoStop}%
\bibitem [{\citenamefont {Alexander}\ \emph {et~al.}(2016)\citenamefont {Alexander}, \citenamefont {Battaglieri}, \citenamefont {Echenard}, \citenamefont {Essig}, \citenamefont {Graham}, \citenamefont {Izaguirre}, \citenamefont {Jaros}, \citenamefont {Krnjaic}, \citenamefont {Mardon}, \citenamefont {Morrissey} \emph {et~al.}}]{alexander2016dark}%
  \BibitemOpen
  \bibfield  {author} {\bibinfo {author} {\bibfnamefont {J.}~\bibnamefont {Alexander}}, \bibinfo {author} {\bibfnamefont {M.}~\bibnamefont {Battaglieri}}, \bibinfo {author} {\bibfnamefont {B.}~\bibnamefont {Echenard}}, \bibinfo {author} {\bibfnamefont {R.}~\bibnamefont {Essig}}, \bibinfo {author} {\bibfnamefont {M.}~\bibnamefont {Graham}}, \bibinfo {author} {\bibfnamefont {E.}~\bibnamefont {Izaguirre}}, \bibinfo {author} {\bibfnamefont {J.}~\bibnamefont {Jaros}}, \bibinfo {author} {\bibfnamefont {G.}~\bibnamefont {Krnjaic}}, \bibinfo {author} {\bibfnamefont {J.}~\bibnamefont {Mardon}}, \bibinfo {author} {\bibfnamefont {D.}~\bibnamefont {Morrissey}}, \emph {et~al.},\ }\bibfield  {title} {\bibinfo {title} {Dark sectors 2016 workshop: community report},\ }\href@noop {} {\bibfield  {journal} {\bibinfo  {journal} {arXiv preprint arXiv:1608.08632}\ } (\bibinfo {year} {2016})}\BibitemShut {NoStop}%
\bibitem [{\citenamefont {Ackerman}\ \emph {et~al.}(2009)\citenamefont {Ackerman}, \citenamefont {Buckley}, \citenamefont {Carroll},\ and\ \citenamefont {Kamionkowski}}]{ackerman2009dark}%
  \BibitemOpen
  \bibfield  {author} {\bibinfo {author} {\bibfnamefont {L.}~\bibnamefont {Ackerman}}, \bibinfo {author} {\bibfnamefont {M.~R.}\ \bibnamefont {Buckley}}, \bibinfo {author} {\bibfnamefont {S.~M.}\ \bibnamefont {Carroll}},\ and\ \bibinfo {author} {\bibfnamefont {M.}~\bibnamefont {Kamionkowski}},\ }\bibfield  {title} {\bibinfo {title} {Dark matter and dark radiation},\ }\href@noop {} {\bibfield  {journal} {\bibinfo  {journal} {Physical Review D—Particles, Fields, Gravitation, and Cosmology}\ }\textbf {\bibinfo {volume} {79}},\ \bibinfo {pages} {023519} (\bibinfo {year} {2009})}\BibitemShut {NoStop}%
\bibitem [{\citenamefont {Nelson}\ and\ \citenamefont {Scholtz}(2011)}]{nelson2011dark}%
  \BibitemOpen
  \bibfield  {author} {\bibinfo {author} {\bibfnamefont {A.~E.}\ \bibnamefont {Nelson}}\ and\ \bibinfo {author} {\bibfnamefont {J.}~\bibnamefont {Scholtz}},\ }\bibfield  {title} {\bibinfo {title} {Dark light, dark matter, and the misalignment mechanism},\ }\href@noop {} {\bibfield  {journal} {\bibinfo  {journal} {Physical Review D—Particles, Fields, Gravitation, and Cosmology}\ }\textbf {\bibinfo {volume} {84}},\ \bibinfo {pages} {103501} (\bibinfo {year} {2011})}\BibitemShut {NoStop}%
\bibitem [{\citenamefont {Fabbrichesi}\ \emph {et~al.}(2021)\citenamefont {Fabbrichesi}, \citenamefont {Gabrielli},\ and\ \citenamefont {Lanfranchi}}]{fabbrichesi2021physics}%
  \BibitemOpen
  \bibfield  {author} {\bibinfo {author} {\bibfnamefont {M.}~\bibnamefont {Fabbrichesi}}, \bibinfo {author} {\bibfnamefont {E.}~\bibnamefont {Gabrielli}},\ and\ \bibinfo {author} {\bibfnamefont {G.}~\bibnamefont {Lanfranchi}},\ }\href@noop {} {\emph {\bibinfo {title} {The physics of the dark photon: a primer}}}\ (\bibinfo  {publisher} {Springer},\ \bibinfo {year} {2021})\BibitemShut {NoStop}%
\bibitem [{\citenamefont {Curtin}\ \emph {et~al.}(2015)\citenamefont {Curtin}, \citenamefont {Essig}, \citenamefont {Gori},\ and\ \citenamefont {Shelton}}]{curtin2015illuminating}%
  \BibitemOpen
  \bibfield  {author} {\bibinfo {author} {\bibfnamefont {D.}~\bibnamefont {Curtin}}, \bibinfo {author} {\bibfnamefont {R.}~\bibnamefont {Essig}}, \bibinfo {author} {\bibfnamefont {S.}~\bibnamefont {Gori}},\ and\ \bibinfo {author} {\bibfnamefont {J.}~\bibnamefont {Shelton}},\ }\bibfield  {title} {\bibinfo {title} {Illuminating dark photons with high-energy colliders},\ }\href@noop {} {\bibfield  {journal} {\bibinfo  {journal} {Journal of High Energy Physics}\ }\textbf {\bibinfo {volume} {2015}},\ \bibinfo {pages} {1} (\bibinfo {year} {2015})}\BibitemShut {NoStop}%
\bibitem [{\citenamefont {Abel}\ \emph {et~al.}(2008)\citenamefont {Abel}, \citenamefont {Goodsell}, \citenamefont {Jaeckel}, \citenamefont {Khoze},\ and\ \citenamefont {Ringwald}}]{abel2008kinetic}%
  \BibitemOpen
  \bibfield  {author} {\bibinfo {author} {\bibfnamefont {S.~A.}\ \bibnamefont {Abel}}, \bibinfo {author} {\bibfnamefont {M.~D.}\ \bibnamefont {Goodsell}}, \bibinfo {author} {\bibfnamefont {J.}~\bibnamefont {Jaeckel}}, \bibinfo {author} {\bibfnamefont {V.}~\bibnamefont {Khoze}},\ and\ \bibinfo {author} {\bibfnamefont {A.}~\bibnamefont {Ringwald}},\ }\bibfield  {title} {\bibinfo {title} {Kinetic mixing of the photon with hidden u (1) s in string phenomenology},\ }\href@noop {} {\bibfield  {journal} {\bibinfo  {journal} {Journal of High Energy Physics}\ }\textbf {\bibinfo {volume} {2008}},\ \bibinfo {pages} {124} (\bibinfo {year} {2008})}\BibitemShut {NoStop}%
\bibitem [{\citenamefont {Foot}\ and\ \citenamefont {Vagnozzi}(2015)}]{foot2015dissipative}%
  \BibitemOpen
  \bibfield  {author} {\bibinfo {author} {\bibfnamefont {R.}~\bibnamefont {Foot}}\ and\ \bibinfo {author} {\bibfnamefont {S.}~\bibnamefont {Vagnozzi}},\ }\bibfield  {title} {\bibinfo {title} {Dissipative hidden sector dark matter},\ }\href@noop {} {\bibfield  {journal} {\bibinfo  {journal} {Physical Review D}\ }\textbf {\bibinfo {volume} {91}},\ \bibinfo {pages} {023512} (\bibinfo {year} {2015})}\BibitemShut {NoStop}%
\bibitem [{\citenamefont {An}\ \emph {et~al.}(2013{\natexlab{a}})\citenamefont {An}, \citenamefont {Pospelov},\ and\ \citenamefont {Pradler}}]{an2013new}%
  \BibitemOpen
  \bibfield  {author} {\bibinfo {author} {\bibfnamefont {H.}~\bibnamefont {An}}, \bibinfo {author} {\bibfnamefont {M.}~\bibnamefont {Pospelov}},\ and\ \bibinfo {author} {\bibfnamefont {J.}~\bibnamefont {Pradler}},\ }\bibfield  {title} {\bibinfo {title} {New stellar constraints on dark photons},\ }\href@noop {} {\bibfield  {journal} {\bibinfo  {journal} {Physics Letters B}\ }\textbf {\bibinfo {volume} {725}},\ \bibinfo {pages} {190} (\bibinfo {year} {2013}{\natexlab{a}})}\BibitemShut {NoStop}%
\bibitem [{\citenamefont {Caputo}\ \emph {et~al.}(2021{\natexlab{a}})\citenamefont {Caputo}, \citenamefont {Millar}, \citenamefont {O’Hare},\ and\ \citenamefont {Vitagliano}}]{caputo2021dark}%
  \BibitemOpen
  \bibfield  {author} {\bibinfo {author} {\bibfnamefont {A.}~\bibnamefont {Caputo}}, \bibinfo {author} {\bibfnamefont {A.~J.}\ \bibnamefont {Millar}}, \bibinfo {author} {\bibfnamefont {C.~A.}\ \bibnamefont {O’Hare}},\ and\ \bibinfo {author} {\bibfnamefont {E.}~\bibnamefont {Vitagliano}},\ }\bibfield  {title} {\bibinfo {title} {Dark photon limits: A handbook},\ }\href@noop {} {\bibfield  {journal} {\bibinfo  {journal} {Physical Review D}\ }\textbf {\bibinfo {volume} {104}},\ \bibinfo {pages} {095029} (\bibinfo {year} {2021}{\natexlab{a}})}\BibitemShut {NoStop}%
\bibitem [{\citenamefont {Agrawal}\ \emph {et~al.}(2020)\citenamefont {Agrawal}, \citenamefont {Kitajima}, \citenamefont {Reece}, \citenamefont {Sekiguchi},\ and\ \citenamefont {Takahashi}}]{agrawal2020relic}%
  \BibitemOpen
  \bibfield  {author} {\bibinfo {author} {\bibfnamefont {P.}~\bibnamefont {Agrawal}}, \bibinfo {author} {\bibfnamefont {N.}~\bibnamefont {Kitajima}}, \bibinfo {author} {\bibfnamefont {M.}~\bibnamefont {Reece}}, \bibinfo {author} {\bibfnamefont {T.}~\bibnamefont {Sekiguchi}},\ and\ \bibinfo {author} {\bibfnamefont {F.}~\bibnamefont {Takahashi}},\ }\bibfield  {title} {\bibinfo {title} {Relic abundance of dark photon dark matter},\ }\href@noop {} {\bibfield  {journal} {\bibinfo  {journal} {Physics Letters B}\ }\textbf {\bibinfo {volume} {801}},\ \bibinfo {pages} {135136} (\bibinfo {year} {2020})}\BibitemShut {NoStop}%
\bibitem [{\citenamefont {Chaudhuri}\ \emph {et~al.}(2015)\citenamefont {Chaudhuri}, \citenamefont {Graham}, \citenamefont {Irwin}, \citenamefont {Mardon}, \citenamefont {Rajendran},\ and\ \citenamefont {Zhao}}]{chaudhuri2015radio}%
  \BibitemOpen
  \bibfield  {author} {\bibinfo {author} {\bibfnamefont {S.}~\bibnamefont {Chaudhuri}}, \bibinfo {author} {\bibfnamefont {P.~W.}\ \bibnamefont {Graham}}, \bibinfo {author} {\bibfnamefont {K.}~\bibnamefont {Irwin}}, \bibinfo {author} {\bibfnamefont {J.}~\bibnamefont {Mardon}}, \bibinfo {author} {\bibfnamefont {S.}~\bibnamefont {Rajendran}},\ and\ \bibinfo {author} {\bibfnamefont {Y.}~\bibnamefont {Zhao}},\ }\bibfield  {title} {\bibinfo {title} {Radio for hidden-photon dark matter detection},\ }\href@noop {} {\bibfield  {journal} {\bibinfo  {journal} {Physical Review D}\ }\textbf {\bibinfo {volume} {92}},\ \bibinfo {pages} {075012} (\bibinfo {year} {2015})}\BibitemShut {NoStop}%
\bibitem [{\citenamefont {Mirizzi}\ \emph {et~al.}(2009)\citenamefont {Mirizzi}, \citenamefont {Redondo},\ and\ \citenamefont {Sigl}}]{mirizzi2009microwave}%
  \BibitemOpen
  \bibfield  {author} {\bibinfo {author} {\bibfnamefont {A.}~\bibnamefont {Mirizzi}}, \bibinfo {author} {\bibfnamefont {J.}~\bibnamefont {Redondo}},\ and\ \bibinfo {author} {\bibfnamefont {G.}~\bibnamefont {Sigl}},\ }\bibfield  {title} {\bibinfo {title} {Microwave background constraints on mixing of photons with hidden photons},\ }\href@noop {} {\bibfield  {journal} {\bibinfo  {journal} {Journal of Cosmology and Astroparticle Physics}\ }\textbf {\bibinfo {volume} {2009}}\bibinfo  {number} { (03)},\ \bibinfo {pages} {026}}\BibitemShut {NoStop}%
\bibitem [{\citenamefont {An}\ \emph {et~al.}(2013{\natexlab{b}})\citenamefont {An}, \citenamefont {Pospelov},\ and\ \citenamefont {Pradler}}]{an2013dark}%
  \BibitemOpen
\bibfield  {number} {  }\bibfield  {author} {\bibinfo {author} {\bibfnamefont {H.}~\bibnamefont {An}}, \bibinfo {author} {\bibfnamefont {M.}~\bibnamefont {Pospelov}},\ and\ \bibinfo {author} {\bibfnamefont {J.}~\bibnamefont {Pradler}},\ }\bibfield  {title} {\bibinfo {title} {Dark matter detectors as dark photon helioscopes},\ }\href@noop {} {\bibfield  {journal} {\bibinfo  {journal} {Physical review letters}\ }\textbf {\bibinfo {volume} {111}},\ \bibinfo {pages} {041302} (\bibinfo {year} {2013}{\natexlab{b}})}\BibitemShut {NoStop}%
\bibitem [{\citenamefont {Pierce}\ \emph {et~al.}(2018)\citenamefont {Pierce}, \citenamefont {Riles},\ and\ \citenamefont {Zhao}}]{pierce2018searching}%
  \BibitemOpen
  \bibfield  {author} {\bibinfo {author} {\bibfnamefont {A.}~\bibnamefont {Pierce}}, \bibinfo {author} {\bibfnamefont {K.}~\bibnamefont {Riles}},\ and\ \bibinfo {author} {\bibfnamefont {Y.}~\bibnamefont {Zhao}},\ }\bibfield  {title} {\bibinfo {title} {Searching for dark photon dark matter with gravitational-wave detectors},\ }\href@noop {} {\bibfield  {journal} {\bibinfo  {journal} {Physical review letters}\ }\textbf {\bibinfo {volume} {121}},\ \bibinfo {pages} {061102} (\bibinfo {year} {2018})}\BibitemShut {NoStop}%
\bibitem [{\citenamefont {Caputo}\ \emph {et~al.}(2020)\citenamefont {Caputo}, \citenamefont {Liu}, \citenamefont {Mishra-Sharma},\ and\ \citenamefont {Ruderman}}]{caputo2020dark}%
  \BibitemOpen
  \bibfield  {author} {\bibinfo {author} {\bibfnamefont {A.}~\bibnamefont {Caputo}}, \bibinfo {author} {\bibfnamefont {H.}~\bibnamefont {Liu}}, \bibinfo {author} {\bibfnamefont {S.}~\bibnamefont {Mishra-Sharma}},\ and\ \bibinfo {author} {\bibfnamefont {J.~T.}\ \bibnamefont {Ruderman}},\ }\bibfield  {title} {\bibinfo {title} {Dark photon oscillations in our inhomogeneous universe},\ }\href@noop {} {\bibfield  {journal} {\bibinfo  {journal} {Physical Review Letters}\ }\textbf {\bibinfo {volume} {125}},\ \bibinfo {pages} {221303} (\bibinfo {year} {2020})}\BibitemShut {NoStop}%
\bibitem [{\citenamefont {Feng}\ \emph {et~al.}(2016)\citenamefont {Feng}, \citenamefont {Smolinsky},\ and\ \citenamefont {Tanedo}}]{feng2016detecting}%
  \BibitemOpen
  \bibfield  {author} {\bibinfo {author} {\bibfnamefont {J.~L.}\ \bibnamefont {Feng}}, \bibinfo {author} {\bibfnamefont {J.}~\bibnamefont {Smolinsky}},\ and\ \bibinfo {author} {\bibfnamefont {P.}~\bibnamefont {Tanedo}},\ }\bibfield  {title} {\bibinfo {title} {Detecting dark matter through dark photons from the sun: Charged particle signatures},\ }\href@noop {} {\bibfield  {journal} {\bibinfo  {journal} {Physical Review D}\ }\textbf {\bibinfo {volume} {93}},\ \bibinfo {pages} {115036} (\bibinfo {year} {2016})}\BibitemShut {NoStop}%
\bibitem [{\citenamefont {Villas-Boas}\ \emph {et~al.}(2025)\citenamefont {Villas-Boas}, \citenamefont {M{\'a}ximo}, \citenamefont {Paulino}, \citenamefont {Bachelard},\ and\ \citenamefont {Rempe}}]{villas2025bright}%
  \BibitemOpen
  \bibfield  {author} {\bibinfo {author} {\bibfnamefont {C.~J.}\ \bibnamefont {Villas-Boas}}, \bibinfo {author} {\bibfnamefont {C.~E.}\ \bibnamefont {M{\'a}ximo}}, \bibinfo {author} {\bibfnamefont {P.~J.}\ \bibnamefont {Paulino}}, \bibinfo {author} {\bibfnamefont {R.~P.}\ \bibnamefont {Bachelard}},\ and\ \bibinfo {author} {\bibfnamefont {G.}~\bibnamefont {Rempe}},\ }\bibfield  {title} {\bibinfo {title} {Bright and dark states of light: The quantum origin of classical interference},\ }\href@noop {} {\bibfield  {journal} {\bibinfo  {journal} {Physical Review Letters}\ }\textbf {\bibinfo {volume} {134}},\ \bibinfo {pages} {133603} (\bibinfo {year} {2025})}\BibitemShut {NoStop}%
\bibitem [{\citenamefont {An}\ \emph {et~al.}(2025)\citenamefont {An}, \citenamefont {Ge}, \citenamefont {Liu},\ and\ \citenamefont {Liu}}]{an2025situ}%
  \BibitemOpen
  \bibfield  {author} {\bibinfo {author} {\bibfnamefont {H.}~\bibnamefont {An}}, \bibinfo {author} {\bibfnamefont {S.}~\bibnamefont {Ge}}, \bibinfo {author} {\bibfnamefont {J.}~\bibnamefont {Liu}},\ and\ \bibinfo {author} {\bibfnamefont {M.}~\bibnamefont {Liu}},\ }\bibfield  {title} {\bibinfo {title} {In situ measurements of dark photon dark matter using parker solar probe: Going beyond the radio window},\ }\href@noop {} {\bibfield  {journal} {\bibinfo  {journal} {Physical Review Letters}\ }\textbf {\bibinfo {volume} {134}},\ \bibinfo {pages} {171001} (\bibinfo {year} {2025})}\BibitemShut {NoStop}%
\bibitem [{\citenamefont {Long}\ and\ \citenamefont {Wang}(2019)}]{long2019dark}%
  \BibitemOpen
  \bibfield  {author} {\bibinfo {author} {\bibfnamefont {A.~J.}\ \bibnamefont {Long}}\ and\ \bibinfo {author} {\bibfnamefont {L.-T.}\ \bibnamefont {Wang}},\ }\bibfield  {title} {\bibinfo {title} {Dark photon dark matter from a network of cosmic strings},\ }\href@noop {} {\bibfield  {journal} {\bibinfo  {journal} {Physical Review D}\ }\textbf {\bibinfo {volume} {99}},\ \bibinfo {pages} {063529} (\bibinfo {year} {2019})}\BibitemShut {NoStop}%
\bibitem [{\citenamefont {Cardoso}\ \emph {et~al.}(2016)\citenamefont {Cardoso}, \citenamefont {Macedo}, \citenamefont {Pani},\ and\ \citenamefont {Ferrari}}]{cardoso2016black}%
  \BibitemOpen
  \bibfield  {author} {\bibinfo {author} {\bibfnamefont {V.}~\bibnamefont {Cardoso}}, \bibinfo {author} {\bibfnamefont {C.~F.}\ \bibnamefont {Macedo}}, \bibinfo {author} {\bibfnamefont {P.}~\bibnamefont {Pani}},\ and\ \bibinfo {author} {\bibfnamefont {V.}~\bibnamefont {Ferrari}},\ }\bibfield  {title} {\bibinfo {title} {Black holes and gravitational waves in models of minicharged dark matter},\ }\href@noop {} {\bibfield  {journal} {\bibinfo  {journal} {Journal of Cosmology and Astroparticle Physics}\ }\textbf {\bibinfo {volume} {2016}}\bibinfo  {number} { (05)},\ \bibinfo {pages} {054}}\BibitemShut {NoStop}%
\bibitem [{\citenamefont {Cardoso}\ \emph {et~al.}(2017)\citenamefont {Cardoso}, \citenamefont {Pani},\ and\ \citenamefont {Yu}}]{cardoso2017superradiance}%
  \BibitemOpen
\bibfield  {number} {  }\bibfield  {author} {\bibinfo {author} {\bibfnamefont {V.}~\bibnamefont {Cardoso}}, \bibinfo {author} {\bibfnamefont {P.}~\bibnamefont {Pani}},\ and\ \bibinfo {author} {\bibfnamefont {T.-T.}\ \bibnamefont {Yu}},\ }\bibfield  {title} {\bibinfo {title} {Superradiance in rotating stars and pulsar-timing constraints on dark photons},\ }\href@noop {} {\bibfield  {journal} {\bibinfo  {journal} {Physical Review D}\ }\textbf {\bibinfo {volume} {95}},\ \bibinfo {pages} {124056} (\bibinfo {year} {2017})}\BibitemShut {NoStop}%
\bibitem [{\citenamefont {Cardoso}\ \emph {et~al.}(2018)\citenamefont {Cardoso}, \citenamefont {Dias}, \citenamefont {Hartnett}, \citenamefont {Middleton}, \citenamefont {Pani},\ and\ \citenamefont {Santos}}]{cardoso2018constraining}%
  \BibitemOpen
  \bibfield  {author} {\bibinfo {author} {\bibfnamefont {V.}~\bibnamefont {Cardoso}}, \bibinfo {author} {\bibfnamefont {{\'O}.~J.}\ \bibnamefont {Dias}}, \bibinfo {author} {\bibfnamefont {G.~S.}\ \bibnamefont {Hartnett}}, \bibinfo {author} {\bibfnamefont {M.}~\bibnamefont {Middleton}}, \bibinfo {author} {\bibfnamefont {P.}~\bibnamefont {Pani}},\ and\ \bibinfo {author} {\bibfnamefont {J.~E.}\ \bibnamefont {Santos}},\ }\bibfield  {title} {\bibinfo {title} {Constraining the mass of dark photons and axion-like particles through black-hole superradiance},\ }\href@noop {} {\bibfield  {journal} {\bibinfo  {journal} {Journal of Cosmology and Astroparticle Physics}\ }\textbf {\bibinfo {volume} {2018}}\bibinfo  {number} { (03)},\ \bibinfo {pages} {043}}\BibitemShut {NoStop}%
\bibitem [{\citenamefont {Alexander}\ \emph {et~al.}(2018)\citenamefont {Alexander}, \citenamefont {McDonough}, \citenamefont {Sims},\ and\ \citenamefont {Yunes}}]{alexander2018hidden}%
  \BibitemOpen
\bibfield  {number} {  }\bibfield  {author} {\bibinfo {author} {\bibfnamefont {S.}~\bibnamefont {Alexander}}, \bibinfo {author} {\bibfnamefont {E.}~\bibnamefont {McDonough}}, \bibinfo {author} {\bibfnamefont {R.}~\bibnamefont {Sims}},\ and\ \bibinfo {author} {\bibfnamefont {N.}~\bibnamefont {Yunes}},\ }\bibfield  {title} {\bibinfo {title} {Hidden-sector modifications to gravitational waves from binary inspirals},\ }\href@noop {} {\bibfield  {journal} {\bibinfo  {journal} {Classical and Quantum Gravity}\ }\textbf {\bibinfo {volume} {35}},\ \bibinfo {pages} {235012} (\bibinfo {year} {2018})}\BibitemShut {NoStop}%
\bibitem [{\citenamefont {Caputo}\ \emph {et~al.}(2021{\natexlab{b}})\citenamefont {Caputo}, \citenamefont {Witte}, \citenamefont {Blas},\ and\ \citenamefont {Pani}}]{caputo2021electromagnetic}%
  \BibitemOpen
  \bibfield  {author} {\bibinfo {author} {\bibfnamefont {A.}~\bibnamefont {Caputo}}, \bibinfo {author} {\bibfnamefont {S.~J.}\ \bibnamefont {Witte}}, \bibinfo {author} {\bibfnamefont {D.}~\bibnamefont {Blas}},\ and\ \bibinfo {author} {\bibfnamefont {P.}~\bibnamefont {Pani}},\ }\bibfield  {title} {\bibinfo {title} {Electromagnetic signatures of dark photon superradiance},\ }\href@noop {} {\bibfield  {journal} {\bibinfo  {journal} {Physical Review D}\ }\textbf {\bibinfo {volume} {104}},\ \bibinfo {pages} {043006} (\bibinfo {year} {2021}{\natexlab{b}})}\BibitemShut {NoStop}%
\bibitem [{\citenamefont {Gupta}\ \emph {et~al.}(2021)\citenamefont {Gupta}, \citenamefont {Spieksma}, \citenamefont {Pang}, \citenamefont {Koekoek},\ and\ \citenamefont {Van Den~Broeck}}]{gupta2021bounding}%
  \BibitemOpen
  \bibfield  {author} {\bibinfo {author} {\bibfnamefont {P.~K.}\ \bibnamefont {Gupta}}, \bibinfo {author} {\bibfnamefont {T.~F.}\ \bibnamefont {Spieksma}}, \bibinfo {author} {\bibfnamefont {P.~T.}\ \bibnamefont {Pang}}, \bibinfo {author} {\bibfnamefont {G.}~\bibnamefont {Koekoek}},\ and\ \bibinfo {author} {\bibfnamefont {C.}~\bibnamefont {Van Den~Broeck}},\ }\bibfield  {title} {\bibinfo {title} {Bounding dark charges on binary black holes using gravitational waves},\ }\href@noop {} {\bibfield  {journal} {\bibinfo  {journal} {Physical Review D}\ }\textbf {\bibinfo {volume} {104}},\ \bibinfo {pages} {063041} (\bibinfo {year} {2021})}\BibitemShut {NoStop}%
\bibitem [{\citenamefont {East}\ and\ \citenamefont {Huang}(2022)}]{east2022dark}%
  \BibitemOpen
  \bibfield  {author} {\bibinfo {author} {\bibfnamefont {W.~E.}\ \bibnamefont {East}}\ and\ \bibinfo {author} {\bibfnamefont {J.}~\bibnamefont {Huang}},\ }\bibfield  {title} {\bibinfo {title} {Dark photon vortex formation and dynamics},\ }\href@noop {} {\bibfield  {journal} {\bibinfo  {journal} {Journal of High Energy Physics}\ }\textbf {\bibinfo {volume} {2022}},\ \bibinfo {pages} {1} (\bibinfo {year} {2022})}\BibitemShut {NoStop}%
\bibitem [{\citenamefont {Cannizzaro}\ \emph {et~al.}(2022)\citenamefont {Cannizzaro}, \citenamefont {Sberna}, \citenamefont {Caputo},\ and\ \citenamefont {Pani}}]{cannizzaro2022dark}%
  \BibitemOpen
  \bibfield  {author} {\bibinfo {author} {\bibfnamefont {E.}~\bibnamefont {Cannizzaro}}, \bibinfo {author} {\bibfnamefont {L.}~\bibnamefont {Sberna}}, \bibinfo {author} {\bibfnamefont {A.}~\bibnamefont {Caputo}},\ and\ \bibinfo {author} {\bibfnamefont {P.}~\bibnamefont {Pani}},\ }\bibfield  {title} {\bibinfo {title} {Dark photon superradiance quenched by dark matter},\ }\href@noop {} {\bibfield  {journal} {\bibinfo  {journal} {Physical Review D}\ }\textbf {\bibinfo {volume} {106}},\ \bibinfo {pages} {083019} (\bibinfo {year} {2022})}\BibitemShut {NoStop}%
\bibitem [{\citenamefont {East}(2022)}]{east2022vortex}%
  \BibitemOpen
  \bibfield  {author} {\bibinfo {author} {\bibfnamefont {W.~E.}\ \bibnamefont {East}},\ }\bibfield  {title} {\bibinfo {title} {Vortex string formation in black hole superradiance of a dark photon with the higgs mechanism},\ }\href@noop {} {\bibfield  {journal} {\bibinfo  {journal} {Physical Review Letters}\ }\textbf {\bibinfo {volume} {129}},\ \bibinfo {pages} {141103} (\bibinfo {year} {2022})}\BibitemShut {NoStop}%
\bibitem [{\citenamefont {Siemonsen}\ \emph {et~al.}(2023)\citenamefont {Siemonsen}, \citenamefont {Mondino}, \citenamefont {Egana-Ugrinovic}, \citenamefont {Huang}, \citenamefont {Baryakhtar},\ and\ \citenamefont {East}}]{siemonsen2023dark}%
  \BibitemOpen
  \bibfield  {author} {\bibinfo {author} {\bibfnamefont {N.}~\bibnamefont {Siemonsen}}, \bibinfo {author} {\bibfnamefont {C.}~\bibnamefont {Mondino}}, \bibinfo {author} {\bibfnamefont {D.}~\bibnamefont {Egana-Ugrinovic}}, \bibinfo {author} {\bibfnamefont {J.}~\bibnamefont {Huang}}, \bibinfo {author} {\bibfnamefont {M.}~\bibnamefont {Baryakhtar}},\ and\ \bibinfo {author} {\bibfnamefont {W.~E.}\ \bibnamefont {East}},\ }\bibfield  {title} {\bibinfo {title} {Dark photon superradiance: Electrodynamics and multimessenger signals},\ }\href@noop {} {\bibfield  {journal} {\bibinfo  {journal} {Physical Review D}\ }\textbf {\bibinfo {volume} {107}},\ \bibinfo {pages} {075025} (\bibinfo {year} {2023})}\BibitemShut {NoStop}%
\bibitem [{\citenamefont {Xin}\ and\ \citenamefont {Most}(2025)}]{xin2025dark}%
  \BibitemOpen
  \bibfield  {author} {\bibinfo {author} {\bibfnamefont {S.}~\bibnamefont {Xin}}\ and\ \bibinfo {author} {\bibfnamefont {E.~R.}\ \bibnamefont {Most}},\ }\bibfield  {title} {\bibinfo {title} {Dark magnetohydrodynamics: Black hole accretion in superradiant dark photon clouds},\ }\href@noop {} {\bibfield  {journal} {\bibinfo  {journal} {Physical Review D}\ }\textbf {\bibinfo {volume} {111}},\ \bibinfo {pages} {063050} (\bibinfo {year} {2025})}\BibitemShut {NoStop}%
\bibitem [{\citenamefont {{\"O}vg{\"u}n}\ and\ \citenamefont {Pantig}(2025)}]{ovgun2025black}%
  \BibitemOpen
  \bibfield  {author} {\bibinfo {author} {\bibfnamefont {A.}~\bibnamefont {{\"O}vg{\"u}n}}\ and\ \bibinfo {author} {\bibfnamefont {R.~C.}\ \bibnamefont {Pantig}},\ }\bibfield  {title} {\bibinfo {title} {Black hole solutions in dark photon models with higher order corrections},\ }\href@noop {} {\bibfield  {journal} {\bibinfo  {journal} {arXiv preprint arXiv:2505.01649}\ } (\bibinfo {year} {2025})}\BibitemShut {NoStop}%
\bibitem [{\citenamefont {Bhattacharyya}\ \emph {et~al.}(2023)\citenamefont {Bhattacharyya}, \citenamefont {Ghosh},\ and\ \citenamefont {Pal}}]{bhattacharyya2023worldline}%
  \BibitemOpen
  \bibfield  {author} {\bibinfo {author} {\bibfnamefont {A.}~\bibnamefont {Bhattacharyya}}, \bibinfo {author} {\bibfnamefont {S.}~\bibnamefont {Ghosh}},\ and\ \bibinfo {author} {\bibfnamefont {S.}~\bibnamefont {Pal}},\ }\bibfield  {title} {\bibinfo {title} {Worldline effective field theory of inspiralling black hole binaries in presence of dark photon and axionic dark matter},\ }\href@noop {} {\bibfield  {journal} {\bibinfo  {journal} {Journal of High Energy Physics}\ }\textbf {\bibinfo {volume} {2023}},\ \bibinfo {pages} {1} (\bibinfo {year} {2023})}\BibitemShut {NoStop}%
\bibitem [{\citenamefont {Zilhao}\ \emph {et~al.}(2012)\citenamefont {Zilhao}, \citenamefont {Cardoso}, \citenamefont {Herdeiro}, \citenamefont {Lehner},\ and\ \citenamefont {Sperhake}}]{zilhao2012collisions}%
  \BibitemOpen
  \bibfield  {author} {\bibinfo {author} {\bibfnamefont {M.}~\bibnamefont {Zilhao}}, \bibinfo {author} {\bibfnamefont {V.}~\bibnamefont {Cardoso}}, \bibinfo {author} {\bibfnamefont {C.}~\bibnamefont {Herdeiro}}, \bibinfo {author} {\bibfnamefont {L.}~\bibnamefont {Lehner}},\ and\ \bibinfo {author} {\bibfnamefont {U.}~\bibnamefont {Sperhake}},\ }\bibfield  {title} {\bibinfo {title} {Collisions of charged black holes},\ }\href@noop {} {\bibfield  {journal} {\bibinfo  {journal} {Physical Review D—Particles, Fields, Gravitation, and Cosmology}\ }\textbf {\bibinfo {volume} {85}},\ \bibinfo {pages} {124062} (\bibinfo {year} {2012})}\BibitemShut {NoStop}%
\bibitem [{\citenamefont {Majumdar}(1947)}]{majumdar1947class}%
  \BibitemOpen
  \bibfield  {author} {\bibinfo {author} {\bibfnamefont {S.~D.}\ \bibnamefont {Majumdar}},\ }\bibfield  {title} {\bibinfo {title} {A class of exact solutions of einstein's field equations},\ }\href@noop {} {\bibfield  {journal} {\bibinfo  {journal} {Physical Review}\ }\textbf {\bibinfo {volume} {72}},\ \bibinfo {pages} {390} (\bibinfo {year} {1947})}\BibitemShut {NoStop}%
\bibitem [{\citenamefont {Papapetrou}(1948)}]{papapetrou1948einstein}%
  \BibitemOpen
  \bibfield  {author} {\bibinfo {author} {\bibfnamefont {A.}~\bibnamefont {Papapetrou}},\ }\bibfield  {title} {\bibinfo {title} {Einstein's theory of gravitation and flat space},\ }in\ \href@noop {} {\emph {\bibinfo {booktitle} {Proceedings of the Royal Irish Academy. Section A: Mathematical and Physical Sciences}}},\ Vol.~\bibinfo {volume} {52}\ (\bibinfo {organization} {JSTOR},\ \bibinfo {year} {1948})\ pp.\ \bibinfo {pages} {11--23}\BibitemShut {NoStop}%
\bibitem [{\citenamefont {Gibbons}\ and\ \citenamefont {Hull}(1982)}]{gibbons1982bogomolny}%
  \BibitemOpen
  \bibfield  {author} {\bibinfo {author} {\bibfnamefont {G.}~\bibnamefont {Gibbons}}\ and\ \bibinfo {author} {\bibfnamefont {C.}~\bibnamefont {Hull}},\ }\bibfield  {title} {\bibinfo {title} {A bogomolny bound for general relativity and solitons in n= 2 supergravity},\ }\href@noop {} {\bibfield  {journal} {\bibinfo  {journal} {Physics Letters B}\ }\textbf {\bibinfo {volume} {109}},\ \bibinfo {pages} {190} (\bibinfo {year} {1982})}\BibitemShut {NoStop}%
\bibitem [{\citenamefont {Tod}(1983)}]{tod1983all}%
  \BibitemOpen
  \bibfield  {author} {\bibinfo {author} {\bibfnamefont {K.}~\bibnamefont {Tod}},\ }\bibfield  {title} {\bibinfo {title} {All metrics admitting super-covariantly constant spinors},\ }\href@noop {} {\bibfield  {journal} {\bibinfo  {journal} {Physics Letters B}\ }\textbf {\bibinfo {volume} {121}},\ \bibinfo {pages} {241} (\bibinfo {year} {1983})}\BibitemShut {NoStop}%
\bibitem [{\citenamefont {Misner}\ and\ \citenamefont {Sharp}(1964)}]{misner1964relativistic}%
  \BibitemOpen
  \bibfield  {author} {\bibinfo {author} {\bibfnamefont {C.~W.}\ \bibnamefont {Misner}}\ and\ \bibinfo {author} {\bibfnamefont {D.~H.}\ \bibnamefont {Sharp}},\ }\bibfield  {title} {\bibinfo {title} {Relativistic equations for adiabatic, spherically symmetric gravitational collapse},\ }\href@noop {} {\bibfield  {journal} {\bibinfo  {journal} {Physical Review}\ }\textbf {\bibinfo {volume} {136}},\ \bibinfo {pages} {B571} (\bibinfo {year} {1964})}\BibitemShut {NoStop}%
\bibitem [{\citenamefont {Dima}\ \emph {et~al.}(2020)\citenamefont {Dima}, \citenamefont {Barausse}, \citenamefont {Franchini},\ and\ \citenamefont {Sotiriou}}]{dima2020spin}%
  \BibitemOpen
  \bibfield  {author} {\bibinfo {author} {\bibfnamefont {A.}~\bibnamefont {Dima}}, \bibinfo {author} {\bibfnamefont {E.}~\bibnamefont {Barausse}}, \bibinfo {author} {\bibfnamefont {N.}~\bibnamefont {Franchini}},\ and\ \bibinfo {author} {\bibfnamefont {T.~P.}\ \bibnamefont {Sotiriou}},\ }\bibfield  {title} {\bibinfo {title} {Spin-induced black hole spontaneous scalarization},\ }\href@noop {} {\bibfield  {journal} {\bibinfo  {journal} {Physical Review Letters}\ }\textbf {\bibinfo {volume} {125}},\ \bibinfo {pages} {231101} (\bibinfo {year} {2020})}\BibitemShut {NoStop}%
\bibitem [{\citenamefont {Herdeiro}\ \emph {et~al.}(2021{\natexlab{b}})\citenamefont {Herdeiro}, \citenamefont {Radu}, \citenamefont {Silva}, \citenamefont {Sotiriou},\ and\ \citenamefont {Yunes}}]{herdeiro2021spin}%
  \BibitemOpen
  \bibfield  {author} {\bibinfo {author} {\bibfnamefont {C.~A.}\ \bibnamefont {Herdeiro}}, \bibinfo {author} {\bibfnamefont {E.}~\bibnamefont {Radu}}, \bibinfo {author} {\bibfnamefont {H.~O.}\ \bibnamefont {Silva}}, \bibinfo {author} {\bibfnamefont {T.~P.}\ \bibnamefont {Sotiriou}},\ and\ \bibinfo {author} {\bibfnamefont {N.}~\bibnamefont {Yunes}},\ }\bibfield  {title} {\bibinfo {title} {Spin-induced scalarized black holes},\ }\href@noop {} {\bibfield  {journal} {\bibinfo  {journal} {Physical review letters}\ }\textbf {\bibinfo {volume} {126}},\ \bibinfo {pages} {011103} (\bibinfo {year} {2021}{\natexlab{b}})}\BibitemShut {NoStop}%
\bibitem [{\citenamefont {Berti}\ \emph {et~al.}(2021)\citenamefont {Berti}, \citenamefont {Collodel}, \citenamefont {Kleihaus},\ and\ \citenamefont {Kunz}}]{berti2021spin}%
  \BibitemOpen
  \bibfield  {author} {\bibinfo {author} {\bibfnamefont {E.}~\bibnamefont {Berti}}, \bibinfo {author} {\bibfnamefont {L.~G.}\ \bibnamefont {Collodel}}, \bibinfo {author} {\bibfnamefont {B.}~\bibnamefont {Kleihaus}},\ and\ \bibinfo {author} {\bibfnamefont {J.}~\bibnamefont {Kunz}},\ }\bibfield  {title} {\bibinfo {title} {Spin-induced black hole scalarization in einstein-scalar-gauss-bonnet theory},\ }\href@noop {} {\bibfield  {journal} {\bibinfo  {journal} {Physical Review Letters}\ }\textbf {\bibinfo {volume} {126}},\ \bibinfo {pages} {011104} (\bibinfo {year} {2021})}\BibitemShut {NoStop}%
\bibitem [{\citenamefont {Wald}(1993)}]{wald1993black}%
  \BibitemOpen
  \bibfield  {author} {\bibinfo {author} {\bibfnamefont {R.~M.}\ \bibnamefont {Wald}},\ }\bibfield  {title} {\bibinfo {title} {Black hole entropy is the noether charge},\ }\href@noop {} {\bibfield  {journal} {\bibinfo  {journal} {Physical Review D}\ }\textbf {\bibinfo {volume} {48}},\ \bibinfo {pages} {R3427} (\bibinfo {year} {1993})}\BibitemShut {NoStop}%
\bibitem [{\citenamefont {Iyer}\ and\ \citenamefont {Wald}(1994)}]{iyer1994some}%
  \BibitemOpen
  \bibfield  {author} {\bibinfo {author} {\bibfnamefont {V.}~\bibnamefont {Iyer}}\ and\ \bibinfo {author} {\bibfnamefont {R.~M.}\ \bibnamefont {Wald}},\ }\bibfield  {title} {\bibinfo {title} {Some properties of the noether charge and a proposal for dynamical black hole entropy},\ }\href@noop {} {\bibfield  {journal} {\bibinfo  {journal} {Physical review D}\ }\textbf {\bibinfo {volume} {50}},\ \bibinfo {pages} {846} (\bibinfo {year} {1994})}\BibitemShut {NoStop}%
\bibitem [{\citenamefont {Liberati}\ and\ \citenamefont {Pacilio}(2016)}]{liberati2016smarr}%
  \BibitemOpen
  \bibfield  {author} {\bibinfo {author} {\bibfnamefont {S.}~\bibnamefont {Liberati}}\ and\ \bibinfo {author} {\bibfnamefont {C.}~\bibnamefont {Pacilio}},\ }\bibfield  {title} {\bibinfo {title} {Smarr formula for lovelock black holes: A lagrangian approach},\ }\href@noop {} {\bibfield  {journal} {\bibinfo  {journal} {Physical Review D}\ }\textbf {\bibinfo {volume} {93}},\ \bibinfo {pages} {084044} (\bibinfo {year} {2016})}\BibitemShut {NoStop}%
\bibitem [{\citenamefont {Fernandes}\ \emph {et~al.}(2022)\citenamefont {Fernandes}, \citenamefont {Mulryne},\ and\ \citenamefont {Delgado}}]{fernandes2022exploring}%
  \BibitemOpen
  \bibfield  {author} {\bibinfo {author} {\bibfnamefont {P.~G.}\ \bibnamefont {Fernandes}}, \bibinfo {author} {\bibfnamefont {D.~J.}\ \bibnamefont {Mulryne}},\ and\ \bibinfo {author} {\bibfnamefont {J.~F.}\ \bibnamefont {Delgado}},\ }\bibfield  {title} {\bibinfo {title} {Exploring the small mass limit of stationary black holes in theories with gauss--bonnet terms},\ }\href@noop {} {\bibfield  {journal} {\bibinfo  {journal} {Classical and Quantum Gravity}\ }\textbf {\bibinfo {volume} {39}},\ \bibinfo {pages} {235015} (\bibinfo {year} {2022})}\BibitemShut {NoStop}%
\bibitem [{\citenamefont {Abbott}\ \emph {et~al.}(2017)\citenamefont {Abbott}, \citenamefont {Abbott}, \citenamefont {Abbott}, \citenamefont {Acernese}, \citenamefont {Ackley}, \citenamefont {Adams}, \citenamefont {Adams}, \citenamefont {Addesso}, \citenamefont {Adhikari}, \citenamefont {Adya} \emph {et~al.}}]{abbott2017gw170817}%
  \BibitemOpen
  \bibfield  {author} {\bibinfo {author} {\bibfnamefont {B.~P.}\ \bibnamefont {Abbott}}, \bibinfo {author} {\bibfnamefont {R.}~\bibnamefont {Abbott}}, \bibinfo {author} {\bibfnamefont {T.}~\bibnamefont {Abbott}}, \bibinfo {author} {\bibfnamefont {F.}~\bibnamefont {Acernese}}, \bibinfo {author} {\bibfnamefont {K.}~\bibnamefont {Ackley}}, \bibinfo {author} {\bibfnamefont {C.}~\bibnamefont {Adams}}, \bibinfo {author} {\bibfnamefont {T.}~\bibnamefont {Adams}}, \bibinfo {author} {\bibfnamefont {P.}~\bibnamefont {Addesso}}, \bibinfo {author} {\bibfnamefont {R.~X.}\ \bibnamefont {Adhikari}}, \bibinfo {author} {\bibfnamefont {V.~B.}\ \bibnamefont {Adya}}, \emph {et~al.},\ }\bibfield  {title} {\bibinfo {title} {Gw170817: observation of gravitational waves from a binary neutron star inspiral},\ }\href@noop {} {\bibfield  {journal} {\bibinfo  {journal} {Physical review letters}\ }\textbf {\bibinfo {volume} {119}},\ \bibinfo {pages} {161101} (\bibinfo {year} {2017})}\BibitemShut {NoStop}%
\bibitem [{\citenamefont {Ezquiaga}\ and\ \citenamefont {Zumalac{\'a}rregui}(2017)}]{ezquiaga2017dark}%
  \BibitemOpen
  \bibfield  {author} {\bibinfo {author} {\bibfnamefont {J.~M.}\ \bibnamefont {Ezquiaga}}\ and\ \bibinfo {author} {\bibfnamefont {M.}~\bibnamefont {Zumalac{\'a}rregui}},\ }\bibfield  {title} {\bibinfo {title} {Dark energy after gw170817: dead ends and the road ahead},\ }\href@noop {} {\bibfield  {journal} {\bibinfo  {journal} {Physical review letters}\ }\textbf {\bibinfo {volume} {119}},\ \bibinfo {pages} {251304} (\bibinfo {year} {2017})}\BibitemShut {NoStop}%
\bibitem [{\citenamefont {Creminelli}\ and\ \citenamefont {Vernizzi}(2017)}]{creminelli2017dark}%
  \BibitemOpen
  \bibfield  {author} {\bibinfo {author} {\bibfnamefont {P.}~\bibnamefont {Creminelli}}\ and\ \bibinfo {author} {\bibfnamefont {F.}~\bibnamefont {Vernizzi}},\ }\bibfield  {title} {\bibinfo {title} {Dark energy after gw170817 and grb170817a},\ }\href@noop {} {\bibfield  {journal} {\bibinfo  {journal} {Physical review letters}\ }\textbf {\bibinfo {volume} {119}},\ \bibinfo {pages} {251302} (\bibinfo {year} {2017})}\BibitemShut {NoStop}%
\bibitem [{\citenamefont {Baker}\ \emph {et~al.}(2017)\citenamefont {Baker}, \citenamefont {Bellini}, \citenamefont {Ferreira}, \citenamefont {Lagos}, \citenamefont {Noller},\ and\ \citenamefont {Sawicki}}]{baker2017strong}%
  \BibitemOpen
  \bibfield  {author} {\bibinfo {author} {\bibfnamefont {T.}~\bibnamefont {Baker}}, \bibinfo {author} {\bibfnamefont {E.}~\bibnamefont {Bellini}}, \bibinfo {author} {\bibfnamefont {P.~G.}\ \bibnamefont {Ferreira}}, \bibinfo {author} {\bibfnamefont {M.}~\bibnamefont {Lagos}}, \bibinfo {author} {\bibfnamefont {J.}~\bibnamefont {Noller}},\ and\ \bibinfo {author} {\bibfnamefont {I.}~\bibnamefont {Sawicki}},\ }\bibfield  {title} {\bibinfo {title} {Strong constraints on cosmological gravity from gw170817 and grb 170817a},\ }\href@noop {} {\bibfield  {journal} {\bibinfo  {journal} {Physical review letters}\ }\textbf {\bibinfo {volume} {119}},\ \bibinfo {pages} {251301} (\bibinfo {year} {2017})}\BibitemShut {NoStop}%
\bibitem [{\citenamefont {Sakstein}\ and\ \citenamefont {Jain}(2017)}]{sakstein2017implications}%
  \BibitemOpen
  \bibfield  {author} {\bibinfo {author} {\bibfnamefont {J.}~\bibnamefont {Sakstein}}\ and\ \bibinfo {author} {\bibfnamefont {B.}~\bibnamefont {Jain}},\ }\bibfield  {title} {\bibinfo {title} {Implications of the neutron star merger gw170817 for cosmological scalar-tensor theories},\ }\href@noop {} {\bibfield  {journal} {\bibinfo  {journal} {Physical review letters}\ }\textbf {\bibinfo {volume} {119}},\ \bibinfo {pages} {251303} (\bibinfo {year} {2017})}\BibitemShut {NoStop}%
\end{thebibliography}%

\end{document}